\documentclass[reprint,aps,pra,showpacs,superscriptaddress]{revtex4-1} 
\usepackage{subfiles}
\usepackage{url}
\usepackage{amsmath,amssymb}
\usepackage{mathrsfs}
\usepackage{bbm}
\usepackage[dvipdfmx]{graphicx, color}
\usepackage{dcolumn}
\usepackage{bm}
\usepackage{ascmac}
\usepackage{enumerate}
\usepackage{theorem}
\newtheorem{theorem}{Theorem}
\newtheorem{lemma}{Lemma}
{}

\makeatletter
\def\loweq@align#1#2{\lower.6ex\vbox{\baselineskip\z@skip\lineskip\z@
     \ialign{$\m@th#1\hfil##\hfil$\crcr#2\crcr=\crcr}}}
\def\lowsim@align#1#2{\lower.6ex\vbox{\baselineskip\z@skip\lineskip\z@
     \ialign{$\m@th#1\hfil##\hfil$\crcr#2\crcr\sim\crcr}}}
\def\geqq{\mathrel{\mathpalette\loweq@align >}}
\def\leqq{\mathrel{\mathpalette\loweq@align <}}
\def\grsim{\mathrel{\mathpalette\lowsim@align >}}
\def\lesssim{\mathrel{\mathpalette\lowsim@align <}}
\def\gsim{\mathrel{\mathpalette\lowsim@align >}}
\def\lsim{\mathrel{\mathpalette\lowsim@align <}}

\def\argmin{\mathop{\mathrm{argmin}}}

\def\Trace{\mathop{\mathrm{Tr}}}
\def\est{\mathop{\mathrm{est}}}

\newcommand{\Expectation}{\mathbbm{E}}

\newcommand{\norm}[1]{\left\| #1  \right\| }

\newcommand{\as}{\mathrm{a.s.}}

\newcommand{\qoset}{\bm{s}}
\newcommand{\sest}{\qoset^{\textrm{est}}}
\newcommand{\strue}{\qoset^{\textrm{true}}}
\newcommand{\starget}{\qoset^{\textrm{target}}}

\newcommand{\numGates}{n_{\mathrm{g}}}

\newcommand{\state}{\rho}
\newcommand{\povm}{\bm{\Pi}}
\newcommand{\outcome}{x}

\newcommand{\povmelement}{\Pi_{\outcome}}
\newcommand{\gate}{\mathcal{G}}

\newcommand{\qosetsphysical}{\mathcal{S}}

\newcommand{\indexSequence}{\bm{i}}
\newcommand{\setIndexSequence}{{\bf Id}}

\newcommand{\dlangle}{\langle\!\langle}
\newcommand{\drangle}{\rangle\!\rangle}

\newcommand{\HSrep}{\mathrm{HS}}

\newcommand{\CJrep}{\mathrm{CJ}}

\newcommand{\mapAdjoint}{\dagger}

\newcommand{\Loss}{{\rm L}}
\newcommand{\Regu}{{\rm R}}

\definecolor{dgreen}{rgb}{0.0, 0.5, 0.0}
\newcommand{\dgreen}{\color{dgreen}}
\begin{document}

\title{Reliable Characterization for Improving and Validating Accurate Quantum Operations}

\author{Takanori Sugiyama*}
\email{sugiyama@qc.rcast.u-tokyo.ac.jp}
\affiliation{Research Center for Advanced Science and Technology, The University of Tokyo, 4--6--1 Komaba Meguro-ku, Tokyo Japan, 153--8904.}

\author{Shinpei Imori}
\email{imori@hiroshima-u.ac.jp}
\affiliation{Graduate School of Advanced Science and Engineering, Hiroshima University, 1--3--1 Kagamiyama Higashi-Hiroshima-shi, Hiroshima Japan, 739-8526.}

\author{Fuyuhiko Tanaka}
\email{ftanaka@sigmath.es.osaka-u.ac.jp}
\affiliation{Department of Systems Innovation, Graduate School of Engineering Science, Osaka University,
1--3 Machikaneyama-chou Toyonaka-shi, Osaka Japan, 560--8531.}
\affiliation{Quantum Information and Quantum Biology Division,
Institute for Open and Transdisciplinary Research Initiatives, Osaka University,
1--3 Machikaneyama-chou Toyonaka-shi, Osaka Japan, 560--8531.}

\date{\today}
\begin{abstract}
A reliable method for characterizing quantum operations that is suitable for improving and validating their accuracies is indispensable for realizing a practical quantum computer.
Known methods are still not sufficient because they lack reliability or are not suitable for use in the improvement and validation steps.
Here we propose a reliable characterization method that is suitable for the accuracy validation step.
First, we introduce a new self-consistent estimator with regularization and physicality constraints that are designed for improvement and validation.
Second, we mathematically prove that the method provides estimation results that are stringently physical and converge to the gauge-equivalence class of the quantum operations of interest at the limit of data size going to infinity.
The asymptotic convergence guarantees the reliability of the method, and the physical and regularized results ensure the suitability to the validation task.
We also derive the asymptotic convergence rate, which would be optimal.
Finally, we show numerical results on 1-qubit system, which confirm the theoretical results and prove that the method proposed is practical.
\end{abstract}

\maketitle

\section{Introduction}\label{sec:introduction}

   As error rates of elemental quantum operations implemented in recent experiments approach a fault-tolerant threshold of a surface code \cite{Barends2014}, it becomes more important to develop more reliable methods for characterizing their accuracies.
   In particular, in order to realize a practical quantum computer, it is necessary to further improve the quantum operations as much as possible beyond the threshold towards successful work of a quantum error correction in practice. 
   The results of characterization experiments are employed to validate the implemented operations of quantum information processing (QIP) or to improve the accuracies of these operations.  
   The design of the characterization methods should be guided by the suitability of their results for such improvement and validation steps. 
   Reliability of a characterization method, which is defined as a closeness between true object characterized and result of characterization, is an important and fundamental suitability, and we focus on it in the paper.  
   
   Standard randomized benchmarking (RB) \cite{RB_Emerson2005,RB_Emerson2007,RB_Emerson2008,RB_Standard_Magesan2011,RB_Standard_Magesan2012} and the relatives \cite{RB_interleaved_Magesan2012,RB_simultaneous_Gambetta2012,RB_leakage_Chasseur2015,RB_leakage_Wallan2016,RB_purity_Wallman2015,RB_purity_Scheldon2016,RB_loss_Wallman2015,RB_nonClliford_Cross2016} are efficient methods specified for estimating an accuracy parameter like the average gate fidelity, except for a tomographic RB protocol \cite{RB_unital_Kimmel} for multi-parameter estimation.
   Although they are frequently used in current experiments, recent numerical work revealed that a non-negligible bias can exist in the estimation results in realistic experimental settings \cite{RB_Reliability_Epstein2014, RB_Reliability_Proctor2017, RB_Reliability_Qi2018}.
   Standard quantum tomography (QT) \cite{QT_State_Fano1957,QT_State_Smithey1993,QT_State_Hradil1997,QT_State_Banaszek1999,QT_Process_Poyatos1997,QT_Process_Chuang1997,QT_POVM_Luis1999,Text_QStateEstimation_Paris2004} are methods for estimating full information of state preparations, measurements, or gates.    
   They are also popular in experiments but have two disadvantages, exponentially growing costs of implementation and inevitable biases caused by unknown imperfections in experiments.
   When we restrict the use of QT to small subsystems like a few qubits, the high implementation costs do not pose a problem.
   If the biases stem from finiteness of data size, we can make their effects as small as necessary by increasing the size.
   However, the biases in estimation results of RB and QT survive even at the limit of data size going to infinity. 
   Hence, the low reliability of QT and RB is crucial because the purpose of quantum characterization is to reliably characterize super-accurate operations beyond the fault-tolerance threshold.
   
    \begin{table*}[bt]
       \centering
       \begin{tabular}{|l||c|c|}
       \hline
            & GST & RSCQT \\
        \hline
        \hline
        Data-fitting optimization & Nonlinear & Nonlinear \\
        \hline    
        Gauge fixing method & Additional gauge optimization & Regularization at data-fitting \\
        \hline
        Physicality constraints & Not fully taken into account & Fully taken into account \\
        \hline
        Long gate sequences & Implemented & Not implemented \\
        \hline
       \end{tabular}
       \caption{Comparison of GST and RSCQT. GST is a current representative tomographic method in the self-consistent approach. RSCQT is the method proposed in the paper. Both suffer from nonlinearity of data-fitting, which can cause numerical instability. There are two possible advantages of RSCQT compared to GST. One is that RSCQT does not need additional gauge-optimization, which reduces numerical costs of data-processing. The other is that RSCQT fully takes physicality constraints into account, which is suitable to the accuracy validation step. On the other hand, long gate sequences, which have been implemented in GST to amplify effects of tiny physical errors, have not been implemented in RSCQT yet.}
       \label{table:Comparison_GST_RSCQT}
   \end{table*}

   Self-consistent quantum tomography (SCQT) \cite{SCQT_Process_Merkel2013,SCQT_GramSchmidt_Stark2014,SCQT_GST_BlumeKohout2013} is an approach towards overcoming the low reliability of standard QT by treating all quantum operations used in a characterization experiment as unknown objects to be estimated.
   In the SCQT approach, unknown imperfections are treated as unknown objects, in contrast to standard QTs that model them by some known ones, and we can avoid biases caused from our pre-knowledge discrepancy between the true unknown objects and assumed models.
   On the other hand, the approach causes a problem that we cannot uniquely determine the set of quantum operations only from experimental data even if we have infinite amount of data.
   This is because there exist experimentally undetectable gauge degrees of freedom \cite{SCQT_GramSchmidt_Stark2014}.
   In order to obtain estimates of quantum operations in the setting of SCQT, we have to choose how to fix the gauge.
   Gate-set tomography (GST) \cite{SCQT_GST_BlumeKohout2013} is a current representative method in SCQT.
   A software package for performing GST, named pyGSTi, is provided \cite{pyGSTi}.
   For the gauge-fixing, GST uses an optimization with respect to a norm over the gauge degrees of freedom \cite{SCQT_GST_BlumeKohout2017}.
   GST has superior features, e.g., it is self-consistent and free from the pre-knowledge errors, there is a method for testing the existence of time-dependent errors with data for the GST experiment, and so on.
   However, it has at least two problems.
   First, the data-processing procedure in GST is very complicated, and it becomes hard to theoretically evaluate the estimation error caused by finiteness of data.
   Then, it is unclear how to choose the amount of data to be taken for guaranteeing a precision to start the experiment.
   Second, the optimization is a nonlinear problem, and its numerical implementation suffers from high numerical cost, low numerical stability, and hardness of taking into account physicality constraints. 
  Actually, the current version of pyGSTi can ensure physicality of gates, but physicality of state preparation and measurement (SPAM) are not guaranteed \cite{pyGSTi_FAQ}.
  Such a gauge-fixing method with possibly unphysical results is not suitable for use in the validation step.

   Here, we propose a new method based on SCQT with regularization.
   We call the method Regularized Self-Consistent Quantum Tomography (RSCQT).
   Comparisons of RSCQT and GST are summarized in Table~\ref{table:Comparison_GST_RSCQT}. 
   In this paper, first, we mathematically prove its asymptotic convergence and derive the convergence rate, which have been proven for the first time in SCQT methods.
   The theoretical results make it possible to decide the amount of data for guaranteeing a required precision, which ensures the theoretical reliability of the method.
   Second, we show results of numerical experiments on the method for 1-qubit system, which confirm the theoretical results and indicate the possible suitability of the method for use in the  improvement step.
   We also describe the suitability of the SCQT approach for use in the validation step.
   In Sec.~\ref{sec:settingAndNotation}, settings and notation are explained.
   In Sec.~\ref{sec:theoreticalResults}, we introduce the SCQT method with regularization, show theoretical results, and describe their meanings and importance.
   Proofs of theorems in Sec.~\ref{sec:theoreticalResults} are given in Appendix~\ref{sec:proofOfTheorem1} and \ref{sec:proofOfTheorem2}.
   We also explain how to use estimation results of the method at the improvement step in Sec.~\ref{sec:theoreticalResults}, whose technical details are described in Appendix~\ref{sec:dynamicsGeneratorAnalysis}.  
   We performed numerical experiments for 1-qubit system.
   The numerical results are reported in Sec.~\ref{sec:numericalResults}.
   Details of the numerical experiments are described in Appendix~\ref{sec:numericalExperiments}.
   Sec.~\ref{sec:discussions} is devoted for discussions. 
   In particular, we discuss effects of biases caused by regularization and suitability of our gauge-fixing method for use in the validation step.
   We conclude the main text in Sec.~\ref{sec:conclusion}. 
   Roles of a quantum characterization method in a QIP experiment and requirements on it are described and discussed in Appendix~\ref{sec:roleInExperiment} and \ref{sec:requiredConditions}, respectively.
   We also explain relations between results in the paper and requirements there.
   Supplemental information on a SCQT approach is explained in Appendix~\ref{sec:parametrization} and \ref{sec:gaugeDegreesOfFreedom}.
   Brief explanations of statistical methods mentioned in the main text are given in Appendix~\ref{sec:regularization} and \ref{sec:crossValidation}.

\section{Setting and Notation}\label{sec:settingAndNotation}
  
   We consider a characterization problem of quantum operations on a quantum system.
   Let $d$ denote the dimension of the system.
   We assume that $d$ is finite and known.
   The purpose is to know mathematical representations of a set of unknown state preparations, measurements, and gates that are implemented in a QIP protocol.
   We use notations $\rho$, $\bm{\Pi}$, and $\mathcal{G}$ for a density matrix for a state preparation, a positive operator-valued measure (POVM) for a measurement, and a linear trace-preserving and completely-positive (TPCP) map for a gate. 
   For simplicity, we consider a set of quantum operations, $\bm{s}$, consisting of single state preparation, single measurement, and multiple gates.
   Generalizations of theoretical results in Sec.~\ref{sec:theoreticalResults} to cases of multiple state preparations and measurements are straightforward.
   Let $n_{\mathrm{g}}$ denote the number of gates in $\bm{s}$.
  
   A set of quantum operation can be parametrized with a real Euclidean vector.
   We identify the set $\bm{s}$ and the parametrization vector.
   Details of the parametrization are explained in Appendix~\ref{sec:parametrization}.
   Let $\mathcal{S}$ denote the physical region in the Euclidean space.
   Let $\bm{s}^{\mathrm{target}} \in \mathcal{S}$ denote the ideal, noiseless, and known set of quantum operations that we aim to implement in a lab.
   An implemented set, say $\bm{s}^{\mathrm{true}} \in \mathcal{S}$, is unknown, noisy, and different from $\bm{s}^{\mathrm{target}}$ because of imperfections on experimental devices.
   The information of the discrepancy is used to further improve the accuracy of $\strue$ or to validate the suitability for the QIP protocol.
   
\begin{figure}[bt]
   \begin{center}
      \includegraphics[width=\linewidth]{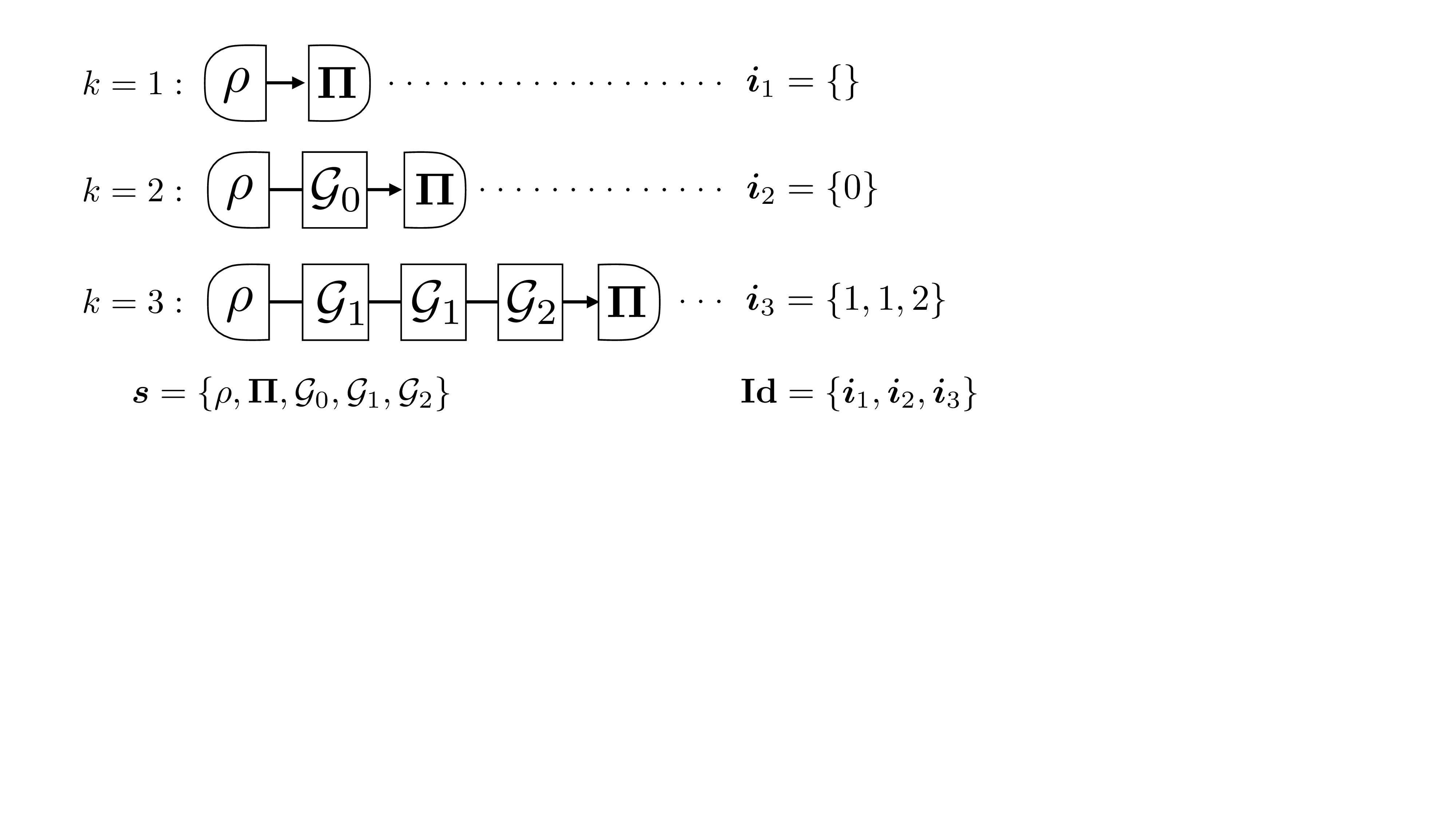}
      \caption{Example of an experimental schedule. A set of quantum operations $\bm{s}$ consists of $\rho$, $\bm{\Pi}$, $\mathcal{G}_0$, $\mathcal{G}_1$, and $\mathcal{G}_2$. In this case, the experimental schedule $\bm{\mathrm{Id}}$ consists of three index sequences, $\bm{i}_1$, $\bm{i}_2$, and $\bm{i}_3$.}
      \label{fig:example_experimental_schedule}
   \end{center}
\end{figure}

   We perform a set of experiments for estimating $\bm{s}^{\mathrm{true}}$, which consists of many different combinations of a state preparation, gate sequences, and a measurement.
   Each combination is specified with an index sequence of gates, say $\bm{i}$, and a set of index sequences is denoted by $\setIndexSequence$, called experimental schedule.
   An example of an experimental schedule is shown in Fig.~\ref{fig:example_experimental_schedule}.
   
   Let $\bm{p}^{\bm{i}}(\bm{s})$ denote the probability distribution of measurement outcomes of the $\bm{i}$-th experiment with a set of quantum operations $\bm{s}\in\mathcal{S}$.
   We define $\bm{p}(\setIndexSequence, \bm{s}) := \{ \bm{p}^{\bm{i}}(\bm{s}) \}_{\bm{i} \in \setIndexSequence}$.
   We repeat the experiment $N$ times for each $\bm{i} \in \setIndexSequence$.
   We assume that the actions of quantum operations are independent of the timing in any sequences and identical for different sequences and repetition orders during the whole experiment.
   Let $\bm{f}^{\bm{i}}_{N}$ denote an empirical distribution calculated from data obtained in the $N$ repetitions of a sequence $\bm{i}$ and $\bm{f}_N(\setIndexSequence) := \{ \bm{f}^{\bm{i}}_{N} \}_{\bm{i} \in \setIndexSequence}$.
   The total amount of data is $N |\setIndexSequence|$.
   
   For any $\bm{s}^{\mathrm{true}}$, there exist sets of quantum operations $\tilde{\bm{s}}\in\mathcal{S}$ satisfying $\bm{p}^{\bm{i}} (\bm{s}^{\mathrm{true}}) = \bm{p}^{\bm{i}}(\tilde{\bm{s}})$ for any $\bm{i}$ in arbitrary $\setIndexSequence$ \cite{SCQT_GST_BlumeKohout2013}.
   We call such $\tilde{\bm{s}}$ gauge-equivalent to $\bm{s}^{\mathrm{true}}$.
   Let $[\bm{s}^{\mathrm{true}}]$ denote the gauge-equivalence class of $\bm{s}^{\mathrm{true}}$, i.e., the set of all $\tilde{\bm{s}}$ gauge-equivalent to $\bm{s}^{\mathrm{true}}$.
   Any difference in the gauge degrees of freedom is superficial and experimentally undetectable.
   In order to obtain an estimate of $\bm{s}^{\mathrm{true}}$ from experimental data, we have to choose how to fix the gauge.
   Details of the gauge degrees of freedom is explained in Appendix~\ref{sec:gaugeDegreesOfFreedom}.
\section{Theoretical Results}\label{sec:theoreticalResults}  

  In this section, we show our theoretical results.
  In Sec. \ref{subsec:InformationalCompletenessAndGaugeEquivalence}, we introduce a concept of informational completeness for the self-consistent approach, which is an expansion of informational completeness in the standard quantum tomography \cite{Text_QStateEstimation_Paris2004}.
  We prove that the expanded informational completeness is a sufficient condition on experiments for estimating all parameters of a set except for the gauge degrees of freedom (Theorem \ref{theorem:GaugeEquivalence-SCIC}).
  In Sec. \ref{subsec:AsymptoricallyGauge-EquivalentEstimator}, we propose a data-processing method, called an estimator in statistics, with regularization for estimating the unknown set of quantum operations of interest from data.
  Let $N$ denote an amount of data. 
  We prove that, by tuning a regularization parameter appropriately, an estimate sequence of the estimator $\bm{s}^{\mathrm{est}}_N$ converges into the gauge-equivalent set of the true set $[\bm{s}^{\mathrm{true}}]$ at the limit of data going to infinity, assuming that an experiment satisfies the informational completeness.
  We also prove that the estimation error of the probability distributions calculated from $\bm{s}^{\mathrm{est}}_{N}$ converges to the true probability distributions with convergence rate equivalent to or faster than that of empirical distributions does, which would be optimal.
  These results guarantees the reliability of the proposed method for sufficiently large data.
  In Sec. \ref{subsec:DynamicsGeneratorAnalysis}, we give formulae for extracting information of dynamics generators such as Hamiltonian and Dissipator from the estimates.
  The formulae would be useful for the improvement step in QIP experiments.

  \subsection{Informational completeness and gauge equivalence}\label{subsec:InformationalCompletenessAndGaugeEquivalence}
  
   We derive a sufficient condition on an experimental schedule $\setIndexSequence$ to self-consistently characterize quantum operations.
   Under the condition, we can know full information of $\bm{s}^{\mathrm{true}}$ except for the gauge degrees of freedom.

   We introduce informational completeness in the context of SCQT. 
  Let ${\setIndexSequence} = 
\{ (i_{\mathrm{g}1_{k}}, \ldots , i_{\mathrm{g}L_{k}} ):\ k=1, \ldots \} $ 
denote a set of index vectors, where $k$ is a index for gate sequences and $i_{\mathrm{g}}$ is a index for gates.
   We call an experimental schedule ${\setIndexSequence}$ {\it state-informationally complete} if a set of density matrix 
   \begin{eqnarray}
   \{ \rho^{\indexSequence} := \mathcal{G}_{i_{\mathrm{g}L_{k}}} \circ \cdots \circ \mathcal{G}_{i_{\mathrm{g}1_{k}}} (\rho) \}_{\indexSequence \in {\setIndexSequence}}
   \end{eqnarray}
   is a (possibly over-complete) basis of $d \times d$ matrix space.    
   We call an experimental schedule ${\setIndexSequence}$ {\it POVM-informationally complete} if a set of POVMs 
   \begin{eqnarray}
   \{ \povm^{\indexSequence} := \mathcal{G}_{i_{\mathrm{g}1_{k}}}^{\mapAdjoint} \circ \cdots \circ \mathcal{G}_{i_{\mathrm{g}L_{k}}}^{\mapAdjoint} (\povm) \}_{\indexSequence \in {\setIndexSequence}}
   \end{eqnarray} 
   is a (possibly over-complete) basis of the space.
   Let $\indexSequence \cup \indexSequence^{\prime}  $ denote the direct union of two index vectors, i.e., 
  $ \indexSequence \cup \indexSequence^{\prime} = (i_{1}, \dots, i_{L}, i^{\prime}_{1}, \dots, i^{\prime}_{L^{\prime}} )$.
   We call $\setIndexSequence$ {\it self-consistently informationally complete} (SCIC) if it includes
   \begin{eqnarray}
      \big\{ 
             \indexSequence_{\mathrm{s}} \cup  \indexSequence_{\mathrm{p}} \  
      \big| \ 
      \indexSequence_{\mathrm{s}} \in {\setIndexSequence}_{\mathrm{s}},
      \indexSequence_{\mathrm{p}} \in {\setIndexSequence}_{\mathrm{p}} 
      \big\} \label{eq:SCIC_state_povm}
   \end{eqnarray}   
   and
   \begin{eqnarray}
      \big\{ 
       \indexSequence_{\mathrm{s}} \cup i_{\mathrm{g}} \cup \indexSequence_{\mathrm{p}} \ 
      \big| \ 
      \indexSequence_{\mathrm{s}} \in {\setIndexSequence}_{\mathrm{s}},
      i_{\mathrm{g}} \in \{  1, \ldots, \numGates \},
      \indexSequence_{\mathrm{p}} \in {\setIndexSequence}_{\mathrm{p}} 
      \big\} \label{eq:SCIC_state_gate_povm}
   \end{eqnarray}
   as subsets where ${\setIndexSequence}_{\mathrm{s}}$ and ${\setIndexSequence}_{\mathrm{p}}$ are state-informationally and POVM-informationally complete sets of gate index sequences, respectively. 
   Eq.~(\ref{eq:SCIC_state_gate_povm}) means that when $\indexSequence$ is SCIC, it includes quantum process tomography (QPT) experiment for all gates in $\bm{s}$, and Eq.~(\ref{eq:SCIC_state_povm}) means that it includes quantum state tomography (QST) experiment for state preparations and POVM tomography (POVMT) experiment for measurements used in the QPT experiment.
   The SCIC condition implies that ${\setIndexSequence}$ includes QST,POVMT, and QPT  experiments allowing duplication of index sequences.
   Hence, we expect that we can obtain full information of $\bm{s}$ except for the gauge degrees of freedom from experimental data of a SCIC $\setIndexSequence$.
   
   \begin{theorem}\label{theorem:GaugeEquivalence-SCIC}
      Suppose that $\setIndexSequence$ is SCIC and inverse maps $\gate_{i_{\mathrm{g}}}^{-1}$ exists for $i_{\mathrm{g}} = 1, \ldots , \numGates$.
      Then, for any $\qoset ,\ \tilde{\qoset} \in \qosetsphysical $, the following two statements are equivalent:
      \begin{enumerate}
         \item $\tilde{\qoset} \in [\qoset]$.
         \item $\bm{p}({\setIndexSequence}, \tilde{\qoset}) = \bm{p}({\setIndexSequence}, \qoset)$.
      \end{enumerate}
   \end{theorem}
   Proof of Theorem \ref{theorem:GaugeEquivalence-SCIC} is given at Appendix \ref{sec:proofOfTheorem1}. 
   Note that the inverse maps mentioned in Theorem \ref{theorem:GaugeEquivalence-SCIC} are not required to be TPCP.
   The inverse map always exists if a quantum gate is implemented with dynamics obeying a time-dependent GKLS master equation \cite{GKS_1976,Lindblad_1976,Text_BreuerPetruccione2002}, the time period is finite, and the dissipator of the dynamics is bounded \cite{HornJohnson_textbook_1991} (see Appendix~\ref{sec:dynamicsGeneratorAnalysis}.~1).
   These conditions are considered as natural in usual settings of QIP experiments.
   Hence the condition that the inverses of all gates exist can be considered as satisfied in experiments.
   
   Theorem \ref{theorem:GaugeEquivalence-SCIC} indicates that the experimental indistinguishability implies the gauge equivalence when the set of gate index sequences is SCIC.
   By taking contraposition of Theorem \ref{theorem:GaugeEquivalence-SCIC}, we have
   \begin{eqnarray}
      \tilde{\qoset} \notin [\qoset] 
      \Leftrightarrow
      \bm{p}({\setIndexSequence}, \tilde{\qoset}) \neq \bm{p}({\setIndexSequence}, \qoset).  
   \end{eqnarray}   
   This means that we can distinguish gauge-inequivalent sets of quantum operations from probability distributions of experiments satisfying the SCIC condition.
   Therefore the SCIC condition is a sufficient condition.

\subsection{Asymptotically Gauge-Equivalent Estimator}\label{subsec:AsymptoricallyGauge-EquivalentEstimator}

   We propose an estimator with regularization.
   The estimator is formulated with three parts, loss function, regularization function, and regularization parameter.
   Here we specify the classes of loss and regularization functions into squared errors for simplicity.
   Results in this subsection hold for wider classes, which is mentioned in Appendix \ref{sec:proofOfTheorem2}. 
   
   We define a loss function and regularization function as
   \begin{eqnarray}
      \hspace{-2mm}
      \Loss \big( \bm{p}\left( {\setIndexSequence}, \qoset ), \bm{f}_{N}({\setIndexSequence}\right) \big) 
      := \frac{1}{|\setIndexSequence|} \sum_{\indexSequence \in \setIndexSequence} 
           \frac{1}{2}\left\| \bm{p}^{\indexSequence}(\qoset) - \bm{f}^{\indexSequence}_{N} \right\|_{2}^{\ 2}, \label{def_loss_squared}
   \end{eqnarray}
   and 
   \begin{eqnarray}        
      \Regu (\qoset , \qoset^{\prime}) &:=& 
      \frac{1}{2} \left\| \state - \state^{\prime} \right\|_{2}^{\ 2} \notag \\
      & &+ \frac{1}{|\mathcal{X}|} \sum_{x \in \mathcal{X}} \frac{1}{2} \left\| \povmelement - \povmelement^{\prime} \right\|_{2}^{\ 2} \notag \\
      & &+ \sum_{i_{\mathrm{g}}=1}^{\numGates} \frac{1}{2 d^2}
       \left\| \HSrep \left( \mathcal{G}_{i_\mathrm{g}} \right) - \HSrep \left( \mathcal{G}_{i_\mathrm{g}}^{\prime} \right) \right\|_{2}^{\ 2},\label{def_reg_squared}
   \end{eqnarray}
   where $\HSrep (\mathcal{G})$ denote a Hilbert-Schmidt matrix representation of a TPCP map $\mathcal{G}$.

   We propose the following estimator:
   \begin{eqnarray}
      \sest_{N} 
      &:=& \argmin_{\qoset \in \qosetsphysical}  
         \left\{
         \Loss \big( \bm{p}({\setIndexSequence}, \qoset ), \bm{f}_{N}({\setIndexSequence}) \big) \right.\notag \\
      & & \hspace{12mm} \left.  + r_{N} \Regu (\qoset, \starget)
         \right\}, \label{eq:def_estimate}
   \end{eqnarray}
   where $r_N$ is a positive number, called regularization parameter.
   It is a user-tunable parameter and it can depend not only on $N$, but also on data. 
   The regularization term in Eq.~(\ref{eq:def_estimate}) takes a role for fixing the gauge as the estimates become close to $\bm{s}^{\prime}$, which is a new way to use regularization.  
   The set $\bm{s}^{\prime}$ is a user-specified set of quantum operations.
   The choice is arbitrary and up to the user.
   We propose to use the target set as the regularization point, i.e., $\bm{s}^{\prime}=\bm{s}^{\mathrm{target}}$ in Eq. (\ref{eq:def_estimate}). 
   We discuss the choice of $\bm{s}^{\prime}$ in Sec. \ref{subsec:Choise_RegularizationFunction}. 
   
   We call the estimator defined by Eq. (\ref{eq:def_estimate}) a {\it regularized self-consistent (RSC)} estimator, and we call a quantum tomographic protocol with the RSC estimator {\it regularized self-consistent quantum tomography (RSCQT)}.      
   We have to select the value of $r_N$ carefully. 
   For example, if we select $r_N$ so large that the effect of the loss function in the minimization of Eq. (\ref{eq:def_estimate}) becomes negligible, the RSC estimate $\sest_N$ approaches $\starget$.
   Then $\bm{p}(\setIndexSequence, \sest_N)$ cannot reproduce $\bm{f}_{N}(\setIndexSequence)$ precisely for finite $N$.

   The following theorem gives a guideline to select a valid value of $r_N$.
   We use a mathematical notation, $\lsim$, in such a way that $f(N) \lsim g(N)$ indicates that, for a positive constant $a$, $f(N) \le a g(N)$ holds for any sufficiently large $N$.
   An abbreviation, $\as$, stands for {\it almost surely} in probability theory. 
   A rigorous definition of the notation is given in Appendix \ref{subsec:Def_AsymptoticNotation}.
   \begin{theorem}[Asymptotic gauge equivalence]\label{theorem:AsymptoticConvergence}
      If we select a regularization parameter satisfying
      \begin{eqnarray}
         \lim_{N\to \infty} r_{N} = 0 \ \as, \label{eq:rN_goto0}
      \end{eqnarray}
      then the sequence of the probability distributions, $\{  \bm{p}({\setIndexSequence}, \sest_{N}) \}$, converges
 to the true one $\bm{p}({\setIndexSequence}, \strue )$ almost surely, i.e., the equality, 
      \begin{eqnarray}
         \lim_{N\to \infty} \sqrt{\Loss (\bm{p}({\setIndexSequence}, \sest_{N}), \bm{p}({\setIndexSequence}, \strue ) )}  &=& 0 \ \as, \label{eq:AsymptoticConvergence_ProbDistSpace}     
       \end{eqnarray}   
       holds.
      If we select the regularization parameter satisfying
      \begin{eqnarray}
         r_N \lsim 1/N\ \as, \label{eq:rN_1overN}
      \end{eqnarray}    
      then inequalities,     
      \begin{eqnarray}
        &&\sqrt{\Loss (\bm{p}({\setIndexSequence}, \sest_{N}), \bm{p}({\setIndexSequence}, \strue ) )} \notag  \\ 
        &&\lsim \sqrt{\Loss (\bm{p}({\setIndexSequence}, \strue), \bm{f}_{N}({\setIndexSequence}) )} \label{eq:EquivalentConvergence} \\
        &&\lsim \sqrt{\frac{\ln \ln N}{N}}\ \as, \label{eq:ConvergenceRateEquivalence}
      \end{eqnarray}
      hold.
      If Eq.~(\ref{eq:rN_goto0}) is satisfied and $\setIndexSequence$ is SCIC, then the sequence of RSC estimates $\{ \sest_{N} \} $ converges to $[\strue ]$ almost surely, i.e., the equality,
      \begin{eqnarray}
         \lim_{N \to \infty}  \min \left\{ \ R(\sest_{N}, \tilde{\qoset}  ) \ \big| \ \tilde{\qoset} \in [\strue] \ \right\} &=& 0 \ \as, \label{eq:AsymptoticConvergence_ParameterSpace}
       \end{eqnarray}    
       holds.
   \end{theorem}    
   The details of the proof are given in Appendix~\ref{sec:proofOfTheorem2}. 
   Here we sketch them.
   \begin{itemize}
      \item Proof of Eqs.~(\ref{eq:AsymptoticConvergence_ProbDistSpace}), (\ref{eq:EquivalentConvergence}), and (\ref{eq:ConvergenceRateEquivalence}): We combine a property of $\sest_N$ as a minimizer with the strong law of large numbers, the central limit theorem, and the strong law of iterated logarithm \cite{Text_ProbabilityTheory_Klenke2008} in order to prove them. 
   
      \item Proof of Eq.~(\ref{eq:AsymptoticConvergence_ParameterSpace}): 
         First, we derive an inequality that any points in $\qosetsphysical$ outside $\epsilon$-neighborhood of $[\strue]$ satisfy. 
         A main mathematical tool at the derivation is the strong law of large numbers.
         Second, we prove that, for any small $\epsilon > 0$, by taking a sufficiently large $N$, $\sest_N$ does not satisfy the inequality.
         This indicates that $\sest_N$ is in the $\epsilon$-neighborhood and converges to $[\strue]$.
              
   \end{itemize}

   At the construction of the proof of Eq.~(\ref{eq:AsymptoticConvergence_ParameterSpace}), we used known results from mathematical statistics as reference.   
   If we neglect the existence of the gauge degrees of freedom in the setting of SCQT, the RSC estimator defined by Eq. (\ref{eq:def_estimate}) can be categorized into an abstract and general class of statistical estimators, called minimum contrast estimator.
   In statistical parameter estimation, some sufficient conditions for a minimum contrast estimator to asymptotically converge to the true parameter are known \cite{Text_MathematicalStatistics_Yoshida2006}.
   These results are not directly applicable to the RSC estimator in the setting of SCQT because there exist the gauge degrees of freedom.
   Nevertheless, our setting has many properties that are easy to mathematically handle, such as finite dimensional parameter space, multinomial probability distributions, and smooth parametrization of the probability distributions.
   We modified the known results to make them applicable to the setting of SCQT.
   Simultaneously, with the good properties of the setting of SCQT and the specific form of the RSC estimates $\sest_N$, we simplified the modified sufficient conditions.
   
   Suppose that $\setIndexSequence$ is SCIC and $r_N$ is chosen to satisfy Eq.~(\ref{eq:rN_goto0}). 
   Theorem \ref{theorem:AsymptoticConvergence} indicates the asymptotic gauge-equivalence of $\sest_N$, i.e., the convergence of $\sest_N$ to $[\strue]$, and this guarantees the high reliability of the RSC estimates $\sest_N$ for sufficiently large $N$.   
   The estimates are physical, because the minimization range is restricted into the physical region $\mathcal{S}$.   
   Hence, $\sest_N$ is self-consistent, stringently physical, and asymptotically gauge-equivalent, in theory. 
   Additionally, Eq. (\ref{eq:EquivalentConvergence}) guarantees that, if we choose $r_N = c /N$ where $c$ is a positive constant independent of $N$, the asymptotic convergence rate of $\bm{p}({\setIndexSequence}, \sest_{N})$ to $\bm{p}({\setIndexSequence}, \strue)$ becomes equivalent to or better than that of $\bm{f}_{N}(\setIndexSequence)$.
   This means that $\sest_N$ can reproduce $\bm{p}({\setIndexSequence}, \strue)$ at least as precise as $\bm{f}_{N}(\setIndexSequence)$ can. 
   We conjecture that the asymptotic convergence rates of $\bm{p}({\setIndexSequence}, \sest_{N})$ and $\bm{f}_{N}(\setIndexSequence)$ are equivalent because that of $\bm{f}_{N}(\setIndexSequence)$ would be optimal.
   There is still arbitrariness of selection of $c$ for tuning the value of $r_N$.
   In practice, even if $c$ is independent of $N$, too large $c$ can lead a large bias on the RSC estimates for finite $N$. 
   We can avoid to select such unreasonably large $c$ by combining the estimator with cross validation \cite{CV_Stone1974, CV_Survey_Arlot2010}, which is a standard method for selecting a regularization parameter in statistics.
   We show the performance of the combination for the case of 1-qubit system in Sec.~\ref{sec:numericalResults}.

   \subsection{Dynamics Generator Analysis}\label{subsec:DynamicsGeneratorAnalysis}
   
   Here we propose a method for extracting information of dynamics generators from an estimate of a quantum gate.
Suppose that the dynamics of a quantum state during the gate operation obeys the time-dependent version of the GKLS equation \cite{GKS_1976,Lindblad_1976,Text_BreuerPetruccione2002}.
\begin{eqnarray}
   \frac{d \rho}{dt} = \mathcal{L}_{t}(\rho) &:=& -i [H(t), \rho] + \{ J(t), \rho \} \notag\\
   &&+\sum_{\alpha , \beta=1}^{d^2 -1} K_{\alpha\beta}(t) B_{\alpha} \rho B_{\beta}^{\dagger}, \label{eq:Lindblad_timelocal}
\end{eqnarray}
where $\bm{B}:=\{ B_{\alpha} \}_{\alpha=0}^{d^2 -1}$ is an orthonormal Hermitian matrix basis satisfying $B_{0} \propto \mathbbm{1}$, $H(t) = \sum_{\alpha=1}^{d^2 -1} H_{\alpha}(t) B_{\alpha}$, $J(t) = \sum_{\alpha = 0}^{d^2 -1} J_{\alpha}(t) B_{\alpha}$, $H_{\alpha}(t) \in \mathbbm{R}$, $J_{\alpha}(t) \in \mathbbm{R}$, and $K_{\alpha \beta}(t) \in \mathbbm{C}$.
When a quantum gate $\mathcal{G}$ is implemented under Eq. (\ref{eq:Lindblad_timelocal}) from $t=0$ to $t=T$, the HS representation of the map is formally expressed as  
\begin{eqnarray}
   \mathrm{HS}(\mathcal{G}) = \mathcal{T} \exp \left[ \int_{0}^{T}\! \! \! dt\ \mathrm{HS}(\mathcal{L}_{t}) \right],
\end{eqnarray}
where $\mathcal{T}$ stands for the chronological operator.
If $H(t)$, $J(t)$, and $K(t)$ are bounded for any $t \in [0, T]$ and $T$ is finite, $\mathrm{HS}(\mathcal{G})^{-1}$ exists (see Appendix~\ref{sec:dynamicsGeneratorAnalysis}~.1 for the proof), and there exists a matrix 
$\mathrm{L}^{\mathrm{acc}}$ satisfying
\begin{eqnarray}
   \mathrm{HS}(\mathcal{G}) = \exp (\mathrm{L}^{\mathrm{acc}}).
\end{eqnarray}
We call $\mathrm{L}^{\mathrm{acc}}$ the accumulated dynamics generator of the gate $\mathcal{G}$. 
It satisfies
\begin{eqnarray}
   \mathrm{L}^{\mathrm{acc}} = \ln \mathrm{HS}(\mathcal{G}).  
\end{eqnarray}
Let us define $\mathcal{L}^{\mathrm{acc}}$ as a linear map satisfying
\begin{eqnarray}
   \mathrm{HS}(\mathcal{L}^{\mathrm{acc}}) = \mathrm{L}^{\mathrm{acc}}.
\end{eqnarray}
From the completeness of the matrix basis $\bm{B}$, the action of the map $\mathcal{L}^{\mathrm{acc}}$ can be represented in the following form.
\begin{eqnarray}
   \mathcal{L}^{\mathrm{acc}}(\rho) 
   &=& -i [H^{\mathrm{acc}}, \rho] + \{ J^{\mathrm{acc}}, \rho \} \notag \\
   && +\sum_{\alpha , \beta=1}^{d^2 -1} K_{\alpha\beta}^{\mathrm{acc}} B_{\alpha} \rho B_{\beta}^{\dagger}, \label{eq:action_map_L_acc} 
\end{eqnarray}
where
\begin{eqnarray}
   H^{\mathrm{acc}} &=& \sum_{\alpha=1}^{d^2 -1} H^{\mathrm{acc}}_{\alpha} B_{\alpha}, \\
   J^{\mathrm{acc}} &=& \sum_{\alpha=0}^{d^2 -1} J^{\mathrm{acc}}_{\alpha} B_{\alpha}.
\end{eqnarray}
The matrix $H^{\mathrm{acc}}$ represents the accumulated action of the original Hamiltonian $H(t)$ from $t=0$ to $t=T$, and $K^{\mathrm{acc}}$ represents the accumulated action of the original dissipator $K(t)$ from $t=0$ to $t=T$.
When the original generators are time-independent, i.e., $H(t) = H$, $J(t) = J$, and $K(t)=K$, the accumulated generators $\{ H^{\mathrm{acc}}, J^{\mathrm{acc}}, K^{\mathrm{acc}} \}$ are simply $H^{\mathrm{acc}} = T\cdot H$, $J^{\mathrm{acc}} = T \cdot J$, and $K^{\mathrm{acc}} = T \cdot K$. 

Let $\mathrm{L}^{\mathrm{acc}, \mathrm{cb}}$ denote the HS representation of $\mathcal{L}^{\mathrm{acc}}$ with respect to the computational basis. 
The coefficients of the accumulated generators can be calculated from $\mathcal{L}^{\mathrm{acc}}$ as follows: 
\begin{eqnarray}
   H^{\mathrm{acc}}_{\alpha} 
   &=& \frac{i}{2d} \mathrm{Tr} \left[ \mathrm{L}^{\mathrm{acc}, \mathrm{cb}} (B_{\alpha} \otimes \mathbbm{1} - \mathbbm{1} \otimes \overline{{B}_{\alpha}} )\right], \notag \\
   && \alpha = 1, \ldots, d^2 -1, \label{eq:H_acc_reconstruct}\\
   J^{\mathrm{acc}}_{\alpha} 
   &=& \frac{1}{2d (1 + \delta_{0\alpha})} \mathrm{Tr} \left[ \mathrm{L}^{\mathrm{acc}, \mathrm{cb}} (B_{\alpha} \otimes \mathbbm{1} + \mathbbm{1} \otimes \overline{{B}_{\alpha}} )\right] , \notag \\
   && \alpha = 0, \ldots , d^2 -1, \label{eq:J_acc_reconstruct}\\
   K^{\mathrm{acc}}_{\alpha\beta} &=& \mathrm{Tr} \left[ \mathrm{L}^{\mathrm{acc}, \mathrm{cb}} \left( B_{\alpha} \otimes \overline{{B}_{\beta}} \right) \right], \notag \\
   && \alpha, \beta = 1, \ldots , d^2 -1, \label{eq:K_acc_reconstruct}
\end{eqnarray}
where $\overline{B_{\alpha}}$ is the complex conjugate of the matrix basis element with respect to the computational basis representation.
The derivations of Eqs.~(\ref{eq:H_acc_reconstruct}), (\ref{eq:J_acc_reconstruct}), and (\ref{eq:K_acc_reconstruct}) are shown in Appendix~\ref{sec:dynamicsGeneratorAnalysis}.2.
After performing a self-consistent tomographic experiments and data-prosessing, we have an estimate $\mathrm{G}^{\mathrm{est}} := \mathrm{HS}(\mathcal{G}^{\mathrm{est}})$ of a gate $\mathrm{G}=\mathrm{HS}(\mathcal{G})$.
Then we can obtain an estimate of the accumulated generators in the following procedure:
\begin{enumerate}[Step 1.]
   \item Choose the computational basis as the representation basis of $\mathrm{HS}$. Then we have $\mathrm{G}^{\mathrm{est}} = \mathrm{HS}^{\mathrm{cb}}(\mathcal{G}^{\mathrm{est}})$.
   \item Calculate the matrix logarithm, $\ln \mathrm{G}^{\mathrm{est}}=:(\mathrm{L}^{\mathrm{acc}, \mathrm{cb}})^{\mathrm{est}}$.
   \item Substitute $(\mathrm{L}^{\mathrm{acc}, \mathrm{cb}})^{\mathrm{est}}$ into $\mathrm{L}^{\mathrm{acc}, \mathrm{cb}}$ in the R.H.S. of Eqs. (\ref{eq:H_acc_reconstruct}), (\ref{eq:J_acc_reconstruct}), and (\ref{eq:K_acc_reconstruct}).
\end{enumerate}

With the method explained above, we can extract information of the accumulated generators $H^{\mathrm{acc}}$, $J^{\mathrm{acc}}$, and $K^{\mathrm{acc}}$, but we cannot know the information about the original generators $H(t)$, $J(t)$, and $K(t)$ at each $t \in [0, T]$.
This is because in general quantum tomography treat a quantum gate as a black box, and a tomographic result gives the information of an input-output relation during the time period.

\section{Numerical Results}\label{sec:numericalResults}

Theorem \ref{theorem:AsymptoticConvergence} in Sec.~\ref{sec:theoreticalResults} guarantees the high reliability of the RSC estimator for asymptotically large $N$.
In practice, it is important to investigate its performances for finite $N$.
The investigation must be done by numerical experiments because calculations of an estimation error require information of the true set $\bm{s}^{\mathrm{true}}$, and the information is not available in real experiments.   
The R.H.S. of Eq.~(\ref{eq:def_estimate}) is a constraint nonlinear optimization problem, which is a main challenge at numerical implementation of the RSC estimator.
The nonlinearity is one of the major differences of SCQT from standard QT which can be formulated as a constraint linear  or convex optimization problem.
We numerically implemented the RSC estimator for 1-qubit systems with a constraint nonlinear optimization library, called IPOPT \cite{IPOPT}, and performed numerical experiments.

 We performed a Monte Carlo simulation, generated  pseudo-random data, calculated statistics like expectations.
We also performed the dynamics generator analysis proposed in Sec.~\ref{subsec:DynamicsGeneratorAnalysis}.      
We tested the performance of the RSC estimator for several settings and parameter regions of error models such as a realistic model generated by pulse dynamics with decoherence obeying a GKLS master equation and toy models of depolarizing, amplitude damping, rotation errors. 
 The observed results are both of positive and negative.
 The positive part is that they are consistent with Eqs.~(\ref{eq:AsymptoticConvergence_ProbDistSpace}), (\ref{eq:EquivalentConvergence}), and (\ref{eq:ConvergenceRateEquivalence}), and these indicate its high reliability on its prediction performance of probability distributions with finite data as well, even though there exist effects of bias originated from regularization. 
The negative part is that Hamiltonian estimated with the dynamics generator analysis includes an effect of a gauge transformation and can differ from the true value, which is not a specific feature of the RSC estimator, but is a common feature of the SCQT approach.
In this section, we briefly explain the setting and results of the numerical simulations on the realistic model.
Results on the toy models and details of the simulation are described in Appendix~\ref{sec:numericalExperiments}. 

\subsection{Setting}
The system simulated is a two-level system ($d=2$).
The target set is chosen as $\bm{s}^{\mathrm{target}}= \{ \rho^{\mathrm{target}}, \bm{\Pi}^{\mathrm{target}}, \mathcal{G}^{\mathrm{target}}_{0}, \mathcal{G}^{\mathrm{target}}_{1}, \mathcal{G}^{\mathrm{target}}_{2} \}$ such that
\begin{eqnarray}
   \rho^{\mathrm{target}} &=& |0\rangle \langle 0|, \\
   \bm{\Pi}^{\mathrm{target}} &=& \left\{ |0\rangle \langle 0|, |1\rangle \langle 1| \right\}, \\
   \mathcal{G}^{\mathrm{target}}_{0}(\rho) &=& \rho, \\
   \mathcal{G}^{\mathrm{target}}_{1}(\rho) &=& e^{-i\frac{\pi}{4}\sigma_{1}} \rho e^{i\frac{\pi}{4}\sigma_{1}}, \\
   \mathcal{G}^{\mathrm{target}}_{2}(\rho) &=& e^{-i\frac{\pi}{4}\sigma_{2}} \rho e^{i\frac{\pi}{4}\sigma_{2}},   
\end{eqnarray}
i.e., the target state is the ground state, the target POVM is the projective measurement along with the $Z$-axis, and the target gates are the identity, $\frac{\pi}{2}$-rotation along with the $X$-axis, and $\frac{\pi}{2}$-rotation along with the $Y$-axis.
The true set $\bm{s}^{\mathrm{true}}$ is chosen as a set that is close to the target set but it includes coherent errors and decoherence.
In the realistic model, the state and POVM are affected by a depolarizing error, and the gates are generated by a rectangular pulse with decoherence obeying a GKLS master equation \cite{BriegelEnglert_PRA47_3311_1993}.
We chose an experimental schedule $\setIndexSequence$ consisting of 45 operation sequences, which is SCIC. 
Details of the model is described in Appendix~\ref{subsec:setting_numerical_experiments}.

\begin{figure*}[tb]
   \begin{center}
      \includegraphics[width=\linewidth]{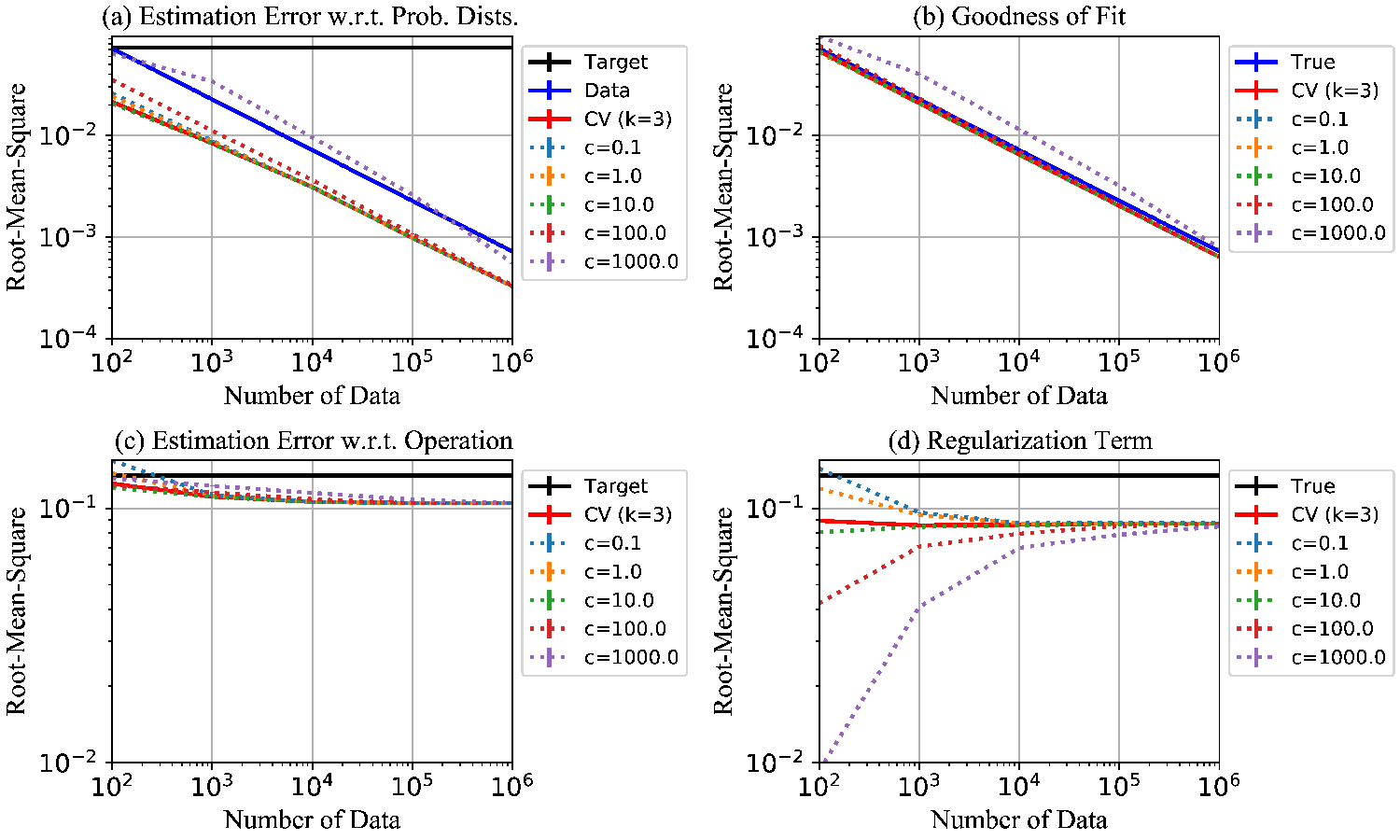}
      \caption{Plots of Root-Mean-Squares (RMS) of loss and regularization functions for the RSC estimator against number of data $N$ when the error is generated by the GKLS master equation. Panel (a) is for the estimation error in the space of the probability distributions. Panel (b) is for the goodness of fit to data. Panel (c) is for the estimation error in the space of quantum operations. Panel (d) is for the regularization term to the target set. All of horizontal and vertical axes are log-scale. See main texts for the details.}
      \label{fig:type0_main_lindblad01}
   \end{center}
\end{figure*}

We select the regularization parameter as $r_N = c / N$, where $c$ is a constant positive value.
The selection of $c$ is up to the user of the estimator.
In order to check the effect of the selection on the performance, we set $c$ in a wide range, $10^{-1}, 1, 10, 10^2, 10^3$, and we combined the RSC estimator with a $k$-fold cross validation, which selects a reasonable value of $c$ from the set of candidate values.
The procedure of the data-processing at the $k$-fold cross validation is explained in Appendix~\ref{sec:crossValidation}.
The computational cost for $k$-fold cross validation procedures becomes larger as $k$ becomes larger. 
In order to keep the computational cost as small as possible, we set $k=3$.
We performed a Monte Carlo simulation with $N=10^2$ to $10^6$.
Statistics like expectations, variances, and standard deviations are calculated with $500$ iterations.

\subsection{Numerical Result 1: Loss and Regularization}

At the first analysis, we investigate behaviors of quantities related to the loss function $\Loss$ in Eq.~(\ref{def_loss_squared}) and regularization function $\Regu$ in Eq.~(\ref{def_reg_squared}) in order to test how the performance of the RSC estimator for finite $N$ differs from the result of Theorem \ref{theorem:AsymptoticConvergence} that holds for asymptotically large $N$.
Figure \ref{fig:type0_main_lindblad01} visualizes the results.
There are four panels in Figure \ref{fig:type0_main_lindblad01}, and horizontal axes of the panels are the amount of data $N$. 

\begin{figure*}[tb]
   \begin{center}
      \includegraphics[width=\linewidth]{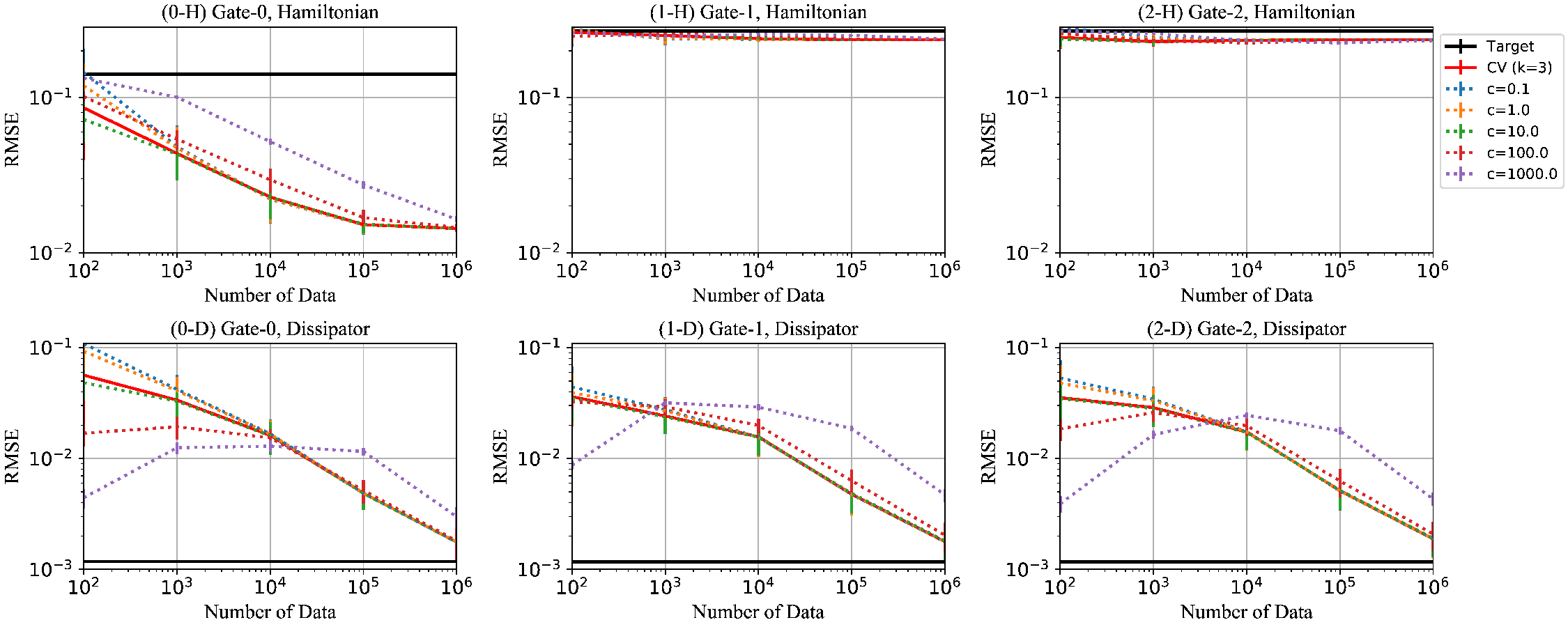}
      \caption{Root-Mean-Squared Error (RMSE) of the estimated Lindbladian against the number of data $N$ when the error is generated by the GKLS master equation. The number and letter at each panel label corresponds to the gate number (0, 1, 2) and part of Lindbladian (H for Hamiltonian, and D for Dissipator). All of horizontal and vertical axes are log-scale. See the main texts for the details.}
      \label{fig:type1_main_lindblad01}
   \end{center}
\end{figure*}

Panel (a) of Figure  \ref{fig:type0_main_lindblad01} is for the root-mean square of estimation error from the true set $\strue$ w.r.t. the probability distributions, i.e., $\sqrt{\mathbbm{E} [  \Loss (\bm{p}({\setIndexSequence}, \bf{s}), \bm{p}({\setIndexSequence}, \strue ) )} ]$.
The black solid line is for $\bm{s} = \starget$, and it quantifies the discrepancy between $\starget$ and $\strue$ in the space of probability distributions, which is independent of $N$.
The blue solid line is for empirical distributions, $\sqrt{\mathbbm{E} [  \Loss (\bm{f}_{N}({\setIndexSequence}), \bm{p}({\setIndexSequence}, \strue ) )} ]$, which scales as $1/\sqrt{N}$.
The red solid line is for ${\bf s} = \sest_N$ with the cross validation, and the other dotted lines are for ${\bm s} = \sest_N$ with fixed $c$.  
Line style and color of $\sest_N$ are common in all panels of  Figures \ref{fig:type0_main_lindblad01}, \ref{fig:type1_main_lindblad01}, and \ref{fig:type2_main_lindblad01}. 
Lines of $\sest_N$ converges to zero as $N$ increases, and they are, except for the line of the largest $c$, almost parallel with and below the line of $\bm{f}_N$. 
The convergence in the space of probability distributions means the convergence to the gauge-equivalence class $[\strue]$ because of the SCIC of the experimental schedule and Theorem \ref{theorem:GaugeEquivalence-SCIC}.
This is consistent with Eqs.~(\ref{eq:AsymptoticConvergence_ProbDistSpace}) and (\ref{eq:ConvergenceRateEquivalence}). 
The panel also shows that the cross validation selects the best value of $c$ in the candidates on average.

It is interesting that the estimation errors of $\sest_N$ are smaller than that of $\bm{f}_N$ (except for $c=10^3$), and the gap remains up to asymptotically large amount of data, at least up tp $N=10^6$.
We observed the same tendency at the other error models as well.
This means that, if we choose a reasonable value of the regularization parameter, which is possible by using the cross validation, the RSC estimator has good prediction ability of the true probability distributions that is higher than experimental data itself.
There are four possible origins of the gap: (i) inequality constraints of physicality, (ii) equality constraints of physicality, (iii) regularization, and (iv) gauge degrees of freedom.
It is known that, in the standard quantum tomography, the inequality constraints contribute to reduction of estimation error \cite{QST_Sugiyama2012}.
Such effect exists as well in the SCQT, but it would not be the main origin of the gap because the effect is expected to decrease as $N$ increases.
The equality constraints must be one of the main origin, because it reduces the degrees of freedom of $\bm{s}$ and $\bm{p}(\setIndexSequence, \bm{s})$ while there are no restrictions on $\bm{f}_N$.
Regularization can also be the main origin, because it tends to reduce the variance of the estimator, although it introduce a bias.
Whatever the origin, the figure indicates the RSC estimator's high predictability of probability distributions.
 
In general, a regularization can cause a large bias that makes the estimation error of an estimator with the regularization worse than an estimator without regularizations.  
However, Panel (a) indicates that the RSC estimators seem to avoid such a bias. 
Indeed, our theoretical result (Theorem \ref{theorem:AsymptoticConvergence}) guarantees that, if we appropreately select the regularization parameter, the effect of the bias asymptotically converges to zero. 
For finite amount of data, the numerical result (Panel (a) of Figure \ref{fig:type0_main_lindblad01}) shows that the bias does not make the estimator's performance worse, and on the contrary the performance is better than those of the empirical distributions, although the main origin might not be the regularization. 
Thus, our regularization does not cause trouble both for finite and infinite amount of data.
 
Panel (b) of Figure  \ref{fig:type0_main_lindblad01} is for the goodness of fit to data, i.e., $\sqrt{\mathbbm{E} [  \Loss (\bm{p}({\setIndexSequence}, \bm{s}), \bm{f}_N )} ]$.
The blue line is for $\bm{s}=\strue$, and note that it's equivalent to the blue line at Panel (a) due to the symmetry of $\sqrt{\Loss}$ w.r.t. the first and second variables. 
For $\bm{s}=\sest_N$, the lines correspond to the first term of R.H.S. of Eq.~(\ref{eq:def_estimate}) after its minimization process. 
The lines are almost similar except for the largest $c$.
This means that the selection of the regularization parameter $c$ does not affect on the goodness of fit of the RSC estimator to data if $c$ is not too large.

Panel (c) of Figure  \ref{fig:type0_main_lindblad01} is for the estimation error in the space of quantum operations.
The vertical axis is the root-mean-squared error from $\strue$, $\sqrt{ \mathbbm{E} [ \Regu ( \bm{s} , \strue)]}$, for $\bm{s}=\starget$ and $\bm{s} = \sest_N$.
In Panels (c) and (d), we have performed a gauge transformation on $\sest_N$ that diagonalizes the POVM in order to adjust a reference frame for comparing to $\strue$.
The lines of $\sest_N$ are below that of $\starget$, but they do not convergent to zero.
Panels (a) and (c) indicate that the RSC estimates converge to a point in the gauge-equivalence class $[\strue]$ and the point is different from $\strue$.
This is as expected because there is an arbitrary choice of gauge-fixing method and the gauge-fixing with the squared 2-norm distance to the target set, $\Regu (\bm{s}, \starget)$, does not lead the estimates to the true set.
We investigate the discrepancy of $\sest_N$ and $\strue$ later at the explanation of Figures \ref{fig:type1_main_lindblad01} and \ref{fig:type2_main_lindblad01}.

Panel (d) of Figure \ref{fig:type0_main_lindblad01} is for the root-mean of the regularization term without the regularization parameter, $\sqrt{ \mathbbm{E} [ \Regu ( \bm{s} , \starget)]}$, for $\bm{s}=\strue$ and $\bm{s}=\sest_N$.
Note that the black lines at Panels (c) and (d) are equivalent due to the symmetry of $\Regu$.
For regions of small $N$, $\sest_N$ tends to be closer to $\starget$ as $c$ becomes larger.
This is as expected because larger $c$ causes an effect that makes its estimate closer to $\starget$.

\subsection{Numerical Result 2: Dynamics Generator Analysis}

As shown in the previous subsection, the RSC estimator gives estimates converge to the gauge-equivalence set $[\strue]$, but the convergence point is different from $\strue$. 
This is due to the existence of the gauge degrees of freedom.
We investigate the discrepancy with the dynamics generator analysis.

Figure \ref{fig:type1_main_lindblad01} shows the root-mean-squared errors (RMSE) of the RSC estimator to the true state.
We have performed a gauge transformation on $\sest_N$ that diagonalizes their POVM in order to adjust a reference frame to $\strue$.
There are six panels.
Three panels at the upper row are for the RMSE of the hamiltonian part of the Lindbladian of gate-$0$, -$1$, and -$2$, respectively.
Other panels at the lower row are for the RMSE of the dissipator part of Lindbladian of each gate.
At the all panels, the black solid line is for $\starget$, red solid lines are for $\sest_N$ with the cross validation, and the other dashed lines are for $\sest_N$ with a fixed $c$.
These six panels indicate that the main sources of the non-convergence to $\strue$ are the hamiltonian parts of gate-1 and gate-2 (Panels (1-H) and (2-H)).

\begin{figure*}[bt]
   \begin{center}
      \includegraphics[width=\linewidth]{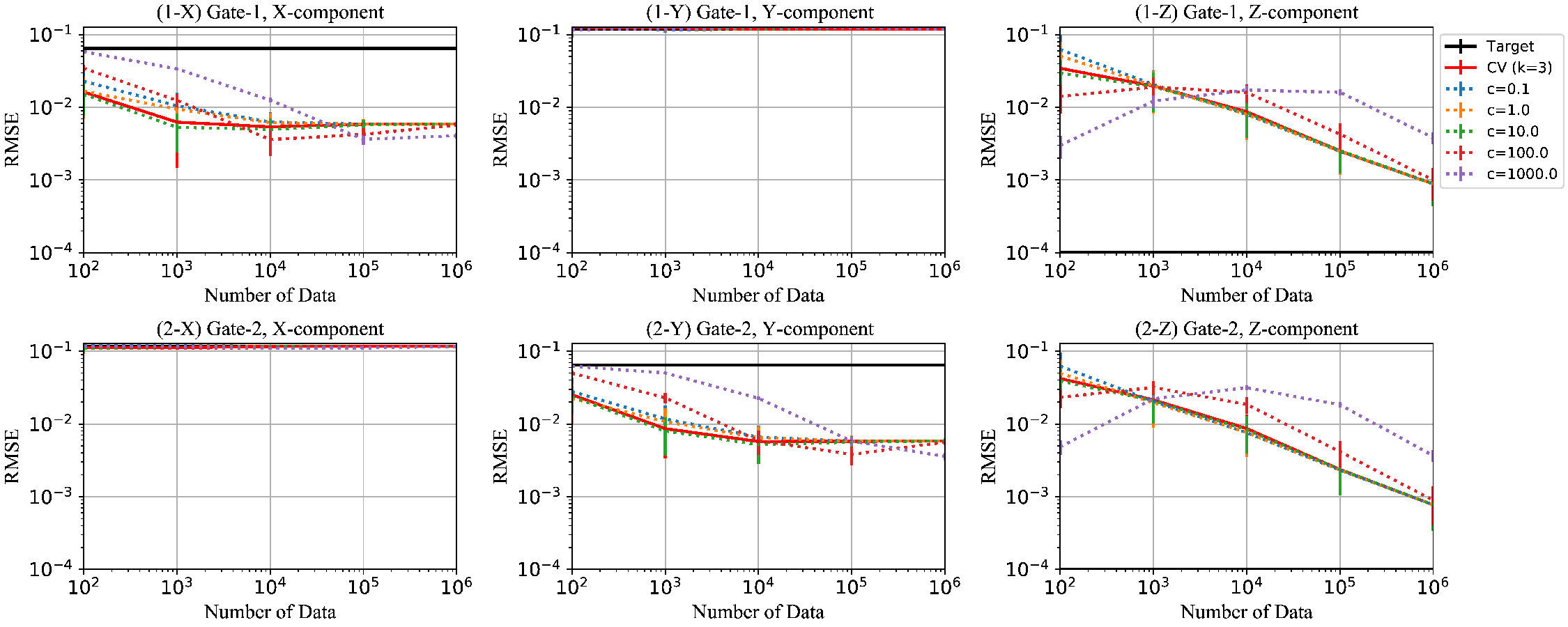}
      \caption{Root-Mean-Squared Error (RMSE) of the estimated Hamiltonian components against the number of data $N$ when the error is generated by the GKLS master equation. The number and letter at each panel label corresponds to the gate number (0, 1, 2) and component. All of horizontal and vertical axes are log-scale. See the main texts for the details.}
      \label{fig:type2_main_lindblad01}
   \end{center}
\end{figure*}

More details of the estimation errors of Hamiltonian are shown in Figure \ref{fig:type2_main_lindblad01}, which is for the RMSE of the $X$-, $Y$-, and $Z$-components of the estimated Hamiltonian.
There are six panels.
Three panels at the upper row are for the RMSE of $X$-, $Y$-, and $Z$-components of gate-1, respectively.
Other panels at the lower row are for the RMSE of them of gate-2.
Line style and color are same as in Figure \ref{fig:type1_main_lindblad01}. 
Behaviors of the lines for $\sest_N$ in the six panels can be classified into the following three types:
(i) they decrease almost monotonically (Panels (1-Z) and (2-Z)), 
(ii) they are below the black line but converge to a finite value (Panels (1-X) and (2-Y)),
and (iii) they are the same order of the black line (Panels (1-Y) and (2-X)).
This classification has a symmetry on $X$ and $Y$ for gate-1 ($\pi$/2-rotation along with $X$-axis) and gate-2 ($\pi$/2-rotation along with $Y$-axis).
These behavior can be explained by the commutation relation between the ideal Hamiltonian of each gate and the generator of the gauge transformation as follows.
Let $\mathrm{G}_i = \exp (\mathrm{L}_i)$ denote the matrix representation of the $i$-th gate of the convergence point of the RSC estimates, where $\mathrm{L}_i$ denote their Lindbladian.
The convergence point is gauge-equivalent to $\strue$, there exists a gauge transformation between them.
Let $A$ denote the matrix representation of the gauge transformation.
Because of the invertibility of $A$, there exists a matrix $a$ satisfying $A = e^a$.
Let $\mathrm{L}_i^{\mathrm{target}}$ denote the ideal Lindbladian of the $i$-th gate, which consists of the Hamiltonian part only and $\Delta_i $ denote the discrepancy between $\mathrm{L}_i$ and $\mathrm{L}_{i}^{\mathrm{target}}$, i.e., $\Delta_i = \mathrm{L}_i - \mathrm{L}_{i}^{\mathrm{target}}$.
Then 
\begin{eqnarray}
   \mathrm{G}_i 
   &=& A \mathrm{G}_{i}^{\mathrm{true}} A^{-1} \\
   &=& \exp\left( e^a \mathrm{L}_i e^{-a} \right) \\   
   &\approx& \exp \left(\mathrm{L}_i + [a, \mathrm{L}_{i}^{\mathrm{target}}] \right), \label{eq:gauge_transformed_lindbladian}
\end{eqnarray}
where we assumed $\|a\| \ll 1$, used Tailer expansion, and neglected higher order terms. 
Eq.~(\ref{eq:gauge_transformed_lindbladian}) indicates that the gauge transformation changes the Lindbladian from $\mathrm{L}_i$ to $\mathrm{L}_i + [a, \mathrm{L}_{i}^{\mathrm{target}} ]$ (approximately).
The Lindbladians contain parts of Hamiltonian and Dissipators.
For simplicity, let us focus on the Hamiltonian part.
In the dynamics generator analysis, we have performed the gauge transformation that makes POVM diagonal.
At the reference frame of diagonal POVM, the remaining gauge degrees of freedom in the Hamiltonian part is the rotation along with $Z$-axis, because rotations along with other axes makes POVM non-diagonal.
Then the hamiltonian part of $a$ contains $Z$-component only.
On the other hand, the target Lindbladian contains only $X$-component for gate-1 and only $Y$-component for gate-2.
Hence, the commutator $[a, \mathrm{L}_{i}^{\mathrm{target}}]$ leads to $Y$-component for gate-1 and to $X$-component for gate-2.
 
 At the discussion above, we ignored the higher order terms of the Taylor expansion.
 When $\| a \|$ is not so small, such higher order terms become non-negligible. 
 Actually we observed such cases in our numerical simulations.
 A result of the cases are shown in Appendix~\ref{sec:numericalExperiments}.
 The results of the dynamics generator analysis shown here indicate that the RSC estimates is affected by uncontrollable gauge transformation and the estimated Hamiltonian can be different from the true value in some non-negligible amount.
 The discrepancy makes the RSC estimates not useful for experimentalists to perform further improvement of their gate operations.
 Although we investigated the performance of the RSC estimator only, we believe that this defect is common for all characterization methods in the self-consistent approach, because the gauge-degrees of freedom remain in any way.

\section{Discussion}\label{sec:discussions}

   In this section, we discuss suitability of our approach for the accuracy validation and improvement steps (see Appendix~\ref{sec:roleInExperiment} for their details).
   In Sec.~\ref{subsec:UseInAccuracyValidationStep}, we discuss the suitability of the self-consistent approach for its use at the accuracy validation step in general.
   In Sec.~\ref{subsec:Choise_RegularizationFunction}, we explain the suitability of the use of a regularization for the gauge-fixing in the self-consistent approach.   
   In Sec.~\ref{subsec:improvementAndValidationForQEC}, we discuss the importance and suitability of the tomographic approach for its use at the accuracy validation and improvement steps for quantum error correction codes.
   We discuss implementation costs in Sec.~\ref{subsec:ComputationCosts}.
   A method for pre-evaluating a sufficient amount of data is explained in Sec.~\ref{subsec:pre-evaluationOfTheAmountOfData}.

   \subsection{Use in Accuracy Validation Step}\label{subsec:UseInAccuracyValidationStep}
   
      Here we discuss suitability of the SCQT approach with the accuracy validation step.
      At the step, accuracies of elemental quantum operations used in a QIP protocol are evaluated.
      If they are sufficiently high, the device is considered to be ready for performing the QIP protocol.
      If not, we need to step back and repeat improvement trial until they pass the accuracy validation test.
      The accuracy validation test is based on a layered structure of circuit-model quantum computation.
      
      Quantum computing is probabilistic, and the performance of a quantum computing device, which must be large-scale in practice, is evaluated by a closeness of an implemented (noisy) probability distribution, $\bm{p}^{\mathrm{true}}$, from the target (ideal and noiseless) probability distribution, $\bm{p}^{\mathrm{target}}$.
      The closeness is typically evaluated by the 1-norm, $\| \bm{p}^{\mathrm{true}} -  \bm{p}^{\mathrm{target}} \|_1$, but other figures of merits can be chosen.
      Let $f^{\mathrm{total} }$, a function of two probability distributions, denote a figure of merit of the performance of the total system.
      If $f^{\mathrm{total}}(\bm{p}^{\mathrm{true}}, \bm{p}^{\mathrm{target}}) $is sufficiently small (highly accurate), the total device pass the accuracy validation test.
      If not, a developer needs to improve at least one component of the device and then needs to evaluate component-wise accuracy with another figure of merit, say $f^{\mathrm{comp}}$.
      There are two approaches to connect $f^{\mathrm{total} }$ and $f^{\mathrm{comp}}$.
      One approximates the accuracy of the total system by the summation of each component's accuracy, and the other rigorously bounds the total accuracy by it. 
      \begin{eqnarray}
         f^{\mathrm{total}}(\bm{p}^{\mathrm{true}}, \bm{p}^{\mathrm{target}}) 
         &\approx& \sum_{i} f^{\mathrm{comp}}(c_i^{\mathrm{true}}, c_{i}^{\mathrm{target}}), \label{eq:approximationReduction}\\
         f^{\mathrm{total}}(\bm{p}^{\mathrm{true}}, \bm{p}^{\mathrm{target}}) 
         &\le& \sum_{i} f^{\mathrm{comp}}(c_i^{\mathrm{true}}, c_{i}^{\mathrm{target}}), \label{eq:overevaluationReduction}
      \end{eqnarray}
      where $i$ is an index of components in a quantum circuit for the QIP protocol, $c_{i}^{\mathrm{true}}$ and $c_{i}^{\mathrm{target}}$ are $i$-th implemented and target components, respectively.
      The infidelity function of states, the average gate infidelity of gates, and depolarizing error rates are categorized into the approximation reduction (Eq.~(\ref{eq:approximationReduction})) \cite{Text_NielsenChuang2000}.   
      The diamond norm distance of gates is categorized into the over-evaluation reduction (Eq.~(\ref{eq:overevaluationReduction})) \cite{DiamondNorm_Kitaev97}.

      In the SCQT approach, a suitable SCQT estimator is expected to give an estimate that is almost gauge equivalent to the true set of elemental quantum operations for sufficiently large amount of data.
      Then the predicted probability distribution from the SCQT estimate, $\bm{p}(\bm{s}^{\mathrm{scqt}}_N )$ is expected to be sufficiently close to the true probability distribution $\bm{p}^{\mathrm{true}} = \bm{p}(\bm{s}^{\mathrm{true}})$, i.e., $\bm{p}(\bm{s}^{\mathrm{scqt}}_N ) \approx \bm{p}(\bm{s}^{\mathrm{true}})$.
      Then the total accuracy is approximated and reduced to component accuracies as
      \begin{eqnarray}
         &f^{\mathrm{total}}& \left( \bm{p}(\bm{s}^{\mathrm{true}}), \bm{p}(\bm{s}^{\mathrm{target}}) \right) \notag \\
         &\approx& f^{\mathrm{total}} \left( \bm{p}(\bm{s}^{\mathrm{scqt}}_N ), \bm{p}(\bm{s}^{\mathrm{target}}) \right) \label{eq:1stApproximation}\\
         & \approx& (\le) \sum_{i} f^{\mathrm{comp}}(c_{i, N}^{\mathrm{scqt}} , c_{i}^{\mathrm{target}}). \label{eq:2ndApproximation}
      \end{eqnarray}  
      Some of reductions above assume physicality of the components, and the physicality of the estimation results is required to a SCQT method for combining that with the reduction.
      As proved in Sec.~\ref{sec:theoreticalResults}, the RSC estimator proposed in Eq.~(\ref{eq:def_estimate}) gives a sequence of physical estimates that converge to the gauge equivalent class $[\bm{s}^{\mathrm{true}}]$.
      Hence, the approximation at Eq.~(\ref{eq:1stApproximation}) becomes more precise as $N$ becomes larger, and it becomes the equation at the limit of $N$ going to infinity.
      The tightness of the approximation or over-evaluation at Eq.~(\ref{eq:2ndApproximation}) depends on that of the reduction method, which is independent of the choice of the SCQT estimator.
      Therefore, as long as the amount of data is sufficiently large and the reduction method is valid, the use of an estimate of an asymptotically gauge-equivalent estimator at the accuracy test of components is valid, and such an estimator including the RSC estimator is suitable for the accuracy validation test.
      The performance of the SCQT estimator for finite amount of data depends on the choice of a gauge fixing method.
      
      We consider that at least the method must be compatible with the physicality constraints of estimates, because it is unclear how to use unphysical estimates at the accuracy validation and improvement steps. 
      Additionally a suitable method is expected to fix the gauge in such a way that each component accuracy $f^{\mathrm{comp}}(c_{i, N}^{\mathrm{scqt}} , c_{i}^{\mathrm{target}})$ becomes as small as possible, which leads to a tight evaluation of the total accuracy $f^{\mathrm{total}}$.
      The regularization used in Eq.~(\ref{eq:def_estimate}) has been designed for meeting the expectation.
      We discuss this point in Sec.~\ref{subsec:Choise_RegularizationFunction}.

   \subsection{Choice of Regularization}\label{subsec:Choise_RegularizationFunction}
   
       The main purpose of this paper is to propose a reliable tomographic estimator. 
     We require the estimator to return a physical estimate that can reproduce experimental data and predict results of a QIP experiment in the future precisely.
     A physical argument that minimizes the loss function, i.e., $\argmin_{\qoset \in \qosetsphysical} \Loss (\bm{p}({\setIndexSequence}, \qoset ), \bm{f}_{N}({\setIndexSequence})  )$ might look suitable for the request.
     However, since there exist gauge degrees of freedom, the argument is not unique.
     In order to obtain an estimate from multiple candidates, we have to fix the gauge.
     It is desirable to choose a gauge-fixing method suitable for validation and improvement after characterization.
     A typical task at the validation and improvement step is to estimate a difference between $\strue$ and $\starget$, say $D (\strue, \starget)$ by evaluating the difference between $\sest_N$ and $\starget$, $D (\sest_N , \starget)$.
     Suppose that there are two gauge-fixing methods $\mathrm{A}$ and $\mathrm{B}$.
     Their respective estimates, obtained from experimental data, are denoted as $\sest_{\mathrm{A}, N}$ and $\sest_{\mathrm{B}, N}$.
     If $\bm{p}({\setIndexSequence}, \sest_{\mathrm{A}, N})$ is as close to $\bm{f}_{N}(\setIndexSequence)$ as $\bm{p}({\setIndexSequence}, \sest_{\mathrm{B},N})$ is and $D (\sest_{\mathrm{A}, N}, \starget) < D (\sest_{\mathrm{B}, N}, \starget)$, we consider method $\mathrm{A}$ better because the difference $D (\sest_{\mathrm{B},N}, \starget) - D (\sest_{\mathrm{A},N}, \starget)$ is mainly caused by the difference of gauge degrees of freedom that is experimentally indistinguishable.
     In order to reduce such fake effect on estimates, we fix the gauge such that estimates are as close to the target $\starget$ as they can describe experimental data precisely.
      
      In Eq. (\ref{eq:def_estimate}), we choose the squared $2$-norm as the regularization.
      This is for the simplicity of mathematical and numerical treatments.
      We can replace the $2$-norms in the loss function and in the regularization with any other norms. 
      The estimator with other norms is also asymptotically gauge equivalent because any norms can be upper-bounded by the $2$-norm in finite dimensional complex spaces \cite{Text_MatrixAnalysis_HornJohnson2013}.  
      In quantum information theory, some norms like the trace-norm and diamond-norm have operational meanings \cite{Text_NielsenChuang2000,DiamondNorm_Kitaev97}.
      A regularization using such norms might be more suitable from the perspective of validation after characterization, but numerical treatments of such norms become hard and computational costs at the minimization increase.
      
      As numerically shown in Sec.~\ref{sec:numericalResults}, the existence of the gauge-degrees of freedom can cause discrepancy between true and estimated Hamiltonians of the RSC estimator, which is a common feature of the SCQT approach.
      The possible discrepancy makes the performance of the current form of the RSC estimator at the use for further improvement of accuracy low. 
      One possible direction toward improving the performance is to exploit prior information on the experiment with the regularization term.

    \subsection{Accuracy Validation and Improvement Steps for Quantum Error Correction}\label{subsec:improvementAndValidationForQEC}
  
    Validation of accuracies of elementary quantum operation implemented for quantum error correction is one of the most important application of estimation results in quantum characterization.
    A depolarizing error model, which is uniquely characterized only with the average gate fidelity, is a popular error model used in theoretical or numerical analysis of quantum error correction codes' performance.
    However, recent numerical studies \cite{QEC_Coherence_Suzuki2017} indicate that the average gate fidelity is not enough for predicting the performance of a quantum error correction in some realistic settings.
    Hence, it may not be appropriate to compare the average gate fidelities estimated with a characterization method to a threshold value calculated with depolarizing error models even if the method is highly reliable. 
    Results of the comparison using the average gate fidelity become more unreliable if RB is used because of its possible low reliability \cite{RB_Reliability_Epstein2014,RB_Reliability_Proctor2017,RB_Reliability_Qi2018}.
    RSCQT can reliably provide more detailed information of errors in real experiments (with discrepancy originated from the gauge degrees of freedom).
    Such information must be contributory to figuring out more realistic error models to be analyzed in theoretical and numerical studies of quantum error correction as well as further improving accuracies of quantum operations implemented in a lab.    
        
  \subsection{Related Work on Dynamics Generator Analysis}        
  Here we discuss relation of the dynamics generator analysis proposed in Sec.~\ref{subsec:DynamicsGeneratorAnalysis} to known methods.
  
  In recent experiments on superconducting quantum circuits \cite{Sheldon2016PRA93, Patterson2019arXiv}, experimentalists try to estimate the accumulated Hamiltonian $H^{\mathrm{acc}}$ in which experiments and data-processing procedures are different from the methods proposed here.
  They report that calibration methods for gates using the estimated information worked well. 
  The data-processing procedures depend on specific models of accumulated Hamiltonians and do not take into account the effects of decoherence during the gate operations. 
  On the other hand, the method we propose is very general, there are no assumptions on the accumulated generators (the dimension of the system is assumed to be known), and in the data-processing both effects of Hamiltonian and decoherence are taken into account.
  Therefore, our method can give us more accurate information of the accumulated generators, which would be useful for calibration.

  An error generator, defined as $\ln \left\{ (\mathrm{G}^{\mathrm{target}})^{-1} \mathrm{G} \right\}$, is estimated with results of gate-set tomography in \cite{SCQT_GST_BlumeKohout2017}.
When a target gate is unitary, any gate can be decomposed into the form of $\mathcal{G} = \mathcal{G}^{\mathrm{target}} \circ \mathcal{E}$, where $\mathcal{E}:= (\mathcal{G}^{\mathrm{target}})^{-1} \circ \mathcal{G}$. 
This leads to 
\begin{eqnarray}
   \mathrm{E}:= \mathrm{HS}(\mathcal{E})
   &=& \mathrm{HS}( (\mathcal{G}^{\mathrm{target}})^{-1} \circ \mathcal{G}) \notag \\
   &=&  \mathrm{HS}( \mathcal{G}^{\mathrm{target}})^{-1} \mathrm{HS}(\mathcal{G}) \notag \\
   &=& (\mathrm{G}^{\mathrm{target}})^{-1} \mathrm{G}.
\end{eqnarray}
The error generator, $\ln \mathrm{E}$, can be considered as a representation of errors on the accumulated generators.
However, in general $\ln \mathrm{E}$ and $\ln \mathrm{G}^{\mathrm{target}}=:(\mathrm{L}^{\mathrm{acc}})^{\mathrm{target}}$ are not commutable, and $\mathrm{L}^{\mathrm{acc}} \neq (\mathrm{L}^{\mathrm{acc}})^{\mathrm{target}} + \ln \mathrm{E}$ because
\begin{eqnarray}
   \mathrm{G} = \exp ( \mathrm{L}^{\mathrm{acc}}) 
   &=& \mathrm{G}^{\mathrm{target}} \mathrm{E} \notag \\
   &=& \exp( ( \mathrm{L}^{\mathrm{acc}})^{\mathrm{target}}) \exp(\ln \mathrm{E}) \notag \\
   &\neq& \exp ( ( \mathrm{L}^{\mathrm{acc}})^{\mathrm{target}} + \ln \mathrm{E} )
\end{eqnarray}
Therefore the error generator $\ln E$ does not represent the direct discrepancy of the accumulated generators.
On the other hand, if we define $\Delta \mathrm{L} :=\ln \mathrm{G} - \ln \mathrm{G}^{\mathrm{target}} = \mathrm{L}^{\mathrm{acc}} - (\mathrm{L}^{\mathrm{acc}})^{\mathrm{target}}$, $\mathrm{L}^{\mathrm{acc}} = (\mathrm{L}^{\mathrm{acc}})^{\mathrm{target}} + \Delta \mathrm{L}$ holds by definition.
We consider $\Delta \mathrm{L}$ or $\left\{ H_{\alpha}^{\mathrm{acc}} - (H_{\alpha}^{\mathrm{acc}})^{\mathrm{target}} \right\}_{\alpha=1}^{d^2 -1}$ more suitable for the use in a calibration process.
In theory of quantum information, especially in quantum error correction, an error model on a quantum gate is typically introduced as $\mathcal{G} = \mathcal{E}^{\prime}\circ \mathcal{G}^{\mathrm{target}}$. 
Note that the timing of the error's action is different from $\mathrm{E}$.
If the purpose of analysis is to know information of $\mathcal{E}^{\prime}$, for comparison to numerical simulation of a quantum error correction code, for example, $\mathrm{E}^{\prime}:= \mathrm{G} (\mathrm{G}^{\mathrm{target}})^{-1}$ would be an appropriate quantity to analyze.

   \subsection{Implementation Costs}\label{subsec:ComputationCosts}

      The RSC estimator proposed here has superior properties such as asymptotic gauge equivalence and probably optimal convergence rate, but one disadvantage is the high cost of experiments and data processing.
      For multi-qubit systems, the number of gate index sequences for a SCIC experiment, $|\setIndexSequence|$,  increases exponentially with respect to the number of qubits.
      The cost of data processing also increases exponentially.  
      The exponential scaling is common in tomographic methods, where the experimental cost of SCQT is about the same as that of GST and is higher than that of standard QT.
      The numerical cost of RSC estimator depends on the choice of the loss and regularization functions and optimization algorithm, but in general it is higher than that of standard QT because the number of parameters to be estimated is much higher.
      Comparison to GST or pyGSTi is a bit obscure, because they use an approximated likelihood function as a loss function, which is different from our choice, and the physicality constraints are not fully taken into account at the optimization \cite{SCQT_GST_BlumeKohout2017, pyGSTi, pyGSTi_FAQ}, whereas the constraints are taken into account in the RSC estimator.
      Suppose that, for GST and RSCQT, we have chosen the same experimental setting, the same loss function, and the same optimization algorithm with fully taking into the physicality constraints.
      Then only difference between them is how to fix the gauge.
      GST perform the gauge fixing data-processing separately after the optimization of the loss function, and in the approach we need to perform an additional optimization over the gauge degrees of freedom.
      On the other hand, in the RSCQT approach, the gauge-fixing is simultaneously performed during the optimization of the objective function.
      The objective function consists of loss and regularization functions.
      If we choose the squared 2-norm as the regularization, the numerical cost of the objective function is dominated by that of the loss function because the nonlinearity of the loss function is much higher than a quadratic function in general.
      The regularization function is a function of the set of quantum operations $\bm{s}$ and the gauge degrees of freedom does not appear explicitly in the regularization.
      In the RSCQT approach, we can avoid the optimization over the gauge degrees of freedom.
      The gauge optimization problem contains the matrix inverse (see Eq.~(\ref{eq:gaugeTransformation})) with the physicality constraints.
      The highly nonlinear constraint optimization can become numerically unstable and hard to be solve.
      This might be a reason that the current version of pyGSTi cannot take into account the full physicality constraints.  
      Therefore, because of the difference of the gauge fixing methods, for a fixed regularization parameter, the computational cost of the RSC estimator would be lower than that of GST.
      From the perspective of numerical stability, the RSCQT approach would be superior than GST as well.
      When we combine a $k$-fold cross validation with the RSC estimator, we have to perform the optimization of the objective function many times.
      In that case, it is unclear which computational cost is lower, which depends on how hard the optimization of the gauge degrees of freedom in GST with full physicality constraints is.       
      
      In quantum computation based on the circuit model, a computational process is constructed with combinations of 1-qubit state preparations, 1-qubit measurements, 1-qubit gates, and 2-qubit gates \cite{Text_NielsenChuang2000}.
      If we restrict the use of the RSC estimator to such small subsystems, the exponential increase of the implementation costs mentioned above poses no problem.
      Let $n_Q$ denote the number of qubits in a device. 
      In cases, where qubits are aligned at each node on a 2-dimensional square-grid lattice, the total number of possible locations of 1-qubit and nearest neighbor 2-qubit operations increase linearly with respect to $n_{Q}$.
      Even if there is concern about crosstalk errors and we need to evaluate nearest $k$-qubit subsystems, the scaling of  the cost of characterization with the RSC estimator still remains linear with respect to $n_Q$, where $k$ is assumed to be small and independent of $n_Q$.
      Therefore if we focus on reliable characterization of elementary quantum operations on the physical layer, the high implementation cost of RSC estimator would not fatal disadvantage.
      Naturally, lower computational cost is better and therefore it is important to develop more stable, more accurate, and faster numerical algorithms for solving the minimization in Eq. (\ref{eq:def_estimate}). 
      A issue in the nearest future is a numerical implementation of the RSC estimator for a 2-qubit system.

   \subsection{Pre-evaluation of the amount of data}\label{subsec:pre-evaluationOfTheAmountOfData}   
      
      As theoretically proven and numerically indicated, the estimation error of probability distributions of the RSC estimator is smaller than that of empirical distributions.
      Hence, we can over-evaluate the estimation error of the RSC estimator by evaluating that of empirical distributions of which analytical form of the variance for any finite $N$ is known.
      The data processing of the RSC estimator is complicated and it is hard to rigorously evaluate its estimation error for finite $N$, but an upper bound of the error is readily calculated from the variance.
      Then we can evaluate the amount of data $N$ for achieving a given value of precision or estimation error on the RSC estimator.
      At the calculation of the variance, we need the information of the unknown true probability distribution, and the rigorous calculation is impossible.
      However, over-evaluation by an error model with a bit larger error amount is possible.
      From such rough pre-evaluation, we can estimate a sufficient amount of data for the precision before an experiment.
      We can update the amount after obtaining experimental data by calculating the variance from their empirical distributions.
      If the updated amount is larger than the pre-evaluated amount, it is better to perform additional experiments until gathering data up to the updated amount.

\section{Conclusion}\label{sec:conclusion}
   
   In this paper, we considered a quantum characterization problem whose purpose is to reliably characterize accurate elemental quantum operations implemented in quantum information processing.
   We derived a sufficient condition on experimental designs which enables us to access all information of the set of unknown state preparations, measurements, and gates except for the gauge degrees of freedom. 
   We also proposed a self-consistent estimator with regularization and physicality constraints.
   We theoretically proved that the sequence of estimates converges to the gauge-equivalence class of the prepared true set of operations at the limit of the data size going to infinity.
   This guarantees the high reliability of the estimation results for sufficiently large amount of data.
   We also theoretically derived the rate of the asymptotic convergence, which is expected to be optimal.
   These are the first mathematically rigorous proofs of asymptotic behaviors of a self-consistent quantum tomography method.
   Additionally, we also proposed how to use the estimation results for the accuracy improvement step.
   In addition, we performed numerical experiments for a 1-qubit system, and numerically analyzed performances of the proposed estimator.
   The numerical results compatible with the theoretical results, and it is numerically shown that the proposed estimator has predictability of the true probability distributions higher than that of empirical distributions, even though there exists a bias originated from the regularization. 
   The numerical results also showed that the existence of the gauge degrees of freedom makes it difficult to directly use the estimation result for the accuracy improvement step, which would be a common feature of the self-consistent approach.
   In order to make quantum technologies, e.g., quantum computation, quantum communication, and quantum sensing, more practical, it is indispensable to develop more reliable characterization method for accuracy validation and accuracy improvement of elemental quantum operations.
   Theoretical and numerical results indicate that the method is suitable for the reliable accuracy validation and needs additional ingenuity for contributing to the accuracy improvement.

\section*{Acknowledgements}
TS would like to thank Yasunobu Nakamura, Yutaka Tabuchi, and Shuhei Tamate for their helpful discussions on realistic error models in superconducting quantum circuits, Yasunari Suzuki, Haruhisa Nagata, and Kazuki Fukui for their useful advice on numerical simulation, and Rekishu Yamazaki, Dany Lachance-Quirion, and Florian R. Schlederer for their helpful comments on the manuscript.
This work was supported by JSPS KAKENHI Grant Numbers JP24700273, JP16K13775, JST ERATO Grant Number JPMJER1601, and ``Funds for the Development of Human Resources in Science and Technology" under MEXT, through the ``Home for Innovative Researchers and Academic Knowledge Users (HIRAKU)" consortium.

\newpage
\appendix

\section{Role in Experiment}\label{sec:roleInExperiment}

\begin{figure}[b]
   \begin{center}
      \includegraphics[width=0.8\linewidth]{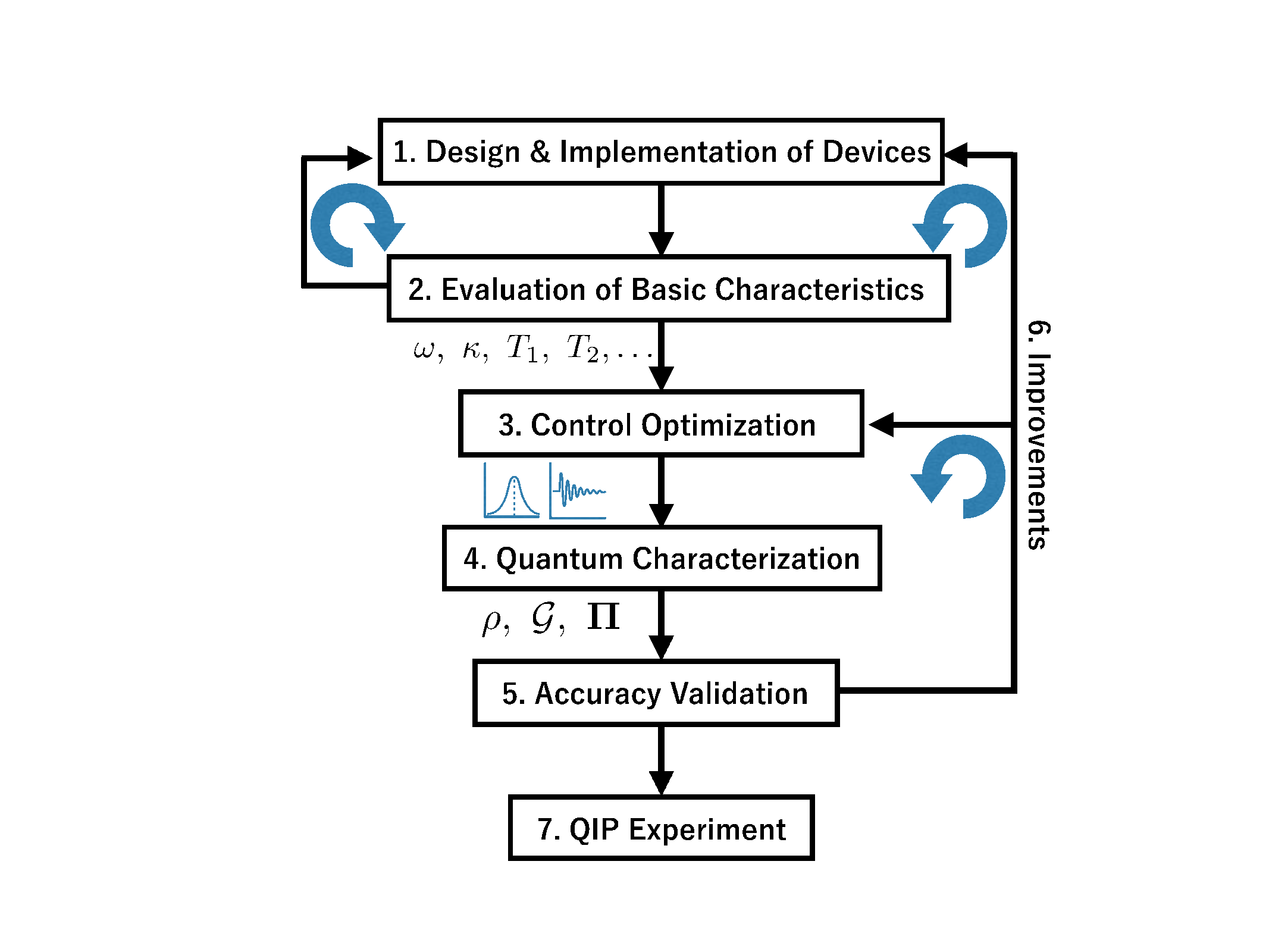}
      \caption{Flow chart of a pre-experiment toward realizing a QIP experiment. Numbers in the chart correspond to the step numbers. Details are explained in the main text at Appendix~\ref{sec:roleInExperiment}.}
      \label{fig:RoleInQIP}
   \end{center}
\end{figure}

In a QIP experiment, quantum characterization takes roles to obtain information of quantum operations implemented and to provide the information to accuracy validation and improvement steps.
Therefore reliability of a characterization method used is important because reliabilities of the validation and improvement are based on it.
In order to clarify these roles and importance, we explain the procedures of a pre-experiment for a QIP experiment (see the flow chart in Fig.~\ref{fig:RoleInQIP}).
Based on the explanation of the roles, we discuss requirements on a characterization method in Appendix~\ref{sec:requiredConditions},

\begin{enumerate}[{\bf Step 1.}]
\item {\bf Design and Implementation of Devices}\\
In this step, experimentalists make designs of experimental devices and implement them.
First, they choose a target QIP experiment to be realized.
Typical examples of the target are traditional quantum algorithms like Shor's factorization \cite{Shor1994}, quantum/classical hybrid algorithms like variational quantum eigensolver \cite{VQE}, quantum supremacy protocols like random unitary sampling \cite{RUS}, and quantum error correction \cite{QEC}.
Next, they pre-evaluate performances and functionalities required for experimental devices.
Based on the results of the pre-evaluation, they make designs of devices toward realizing the target QIP experiment with sufficient accuracy and implement prototypes.
The devices include multi-qubit systems as well as control systems, readout systems, and packages for isolating multi-qubit systems from environmental systems.
\item {\bf Evaluation of Basic Characteristics}\\
In this step, experimentalists evaluate basic characteristics of the multi-qubit system.
They perform experiments with the implemented devices for obtaining basic characteristics like resonant frequencies of qubits and resonators, coherence times $T_1$ and $T_2$, and coupling parameters.
Rough calibration processes for such experiments are also included in this step.
If the basic characteristics obtained are out of their expectation, they go back to the 1st step.
\item {\bf Control Optimization}\\
They perform experiments for optimizing control parameters in order to implement highly accurate elemental quantum operations.
For example, most solid qubit platforms use electromagnetic pulses for implementing quantum operations, and in that case their pulse shapes are tunable to some extent.
Pulse shapes themselves or a few characteristics of pulse shape such as width, amplitude, and relative phase are control parameters in the case. 
\item {\bf Quantum Characterization}\\
They perform experiments and data-processing on quantum characterization for obtaining mathematical representation of implemented quantum operations such as density matrix, POVM, HS matrix, or for obtaining a few characteristic parameters related to the implementation accuracies such as fidelity.
Quantum tomographic protocols and randomized benchmarking protocols are used for the former and latter purposes, respectively.   
\item {\bf Accuracy Validation}\\
In this step, they judge whether the accuracies of the operations implemented are valid or not.
We intentionally separate this step from Step 4, because the judgement process using results of quantum characterization is different from the characterization itself.
First, they calculate a figure of merit on implementation accuracy from the results of quantum characterization, if necessary.
The figure of merit should be selected such that it is suitable for the target QIP experiment chosen at Step 1.
If the value has been already obtained at Step 4, they skip this part.  
Next, they compare the value to a reference value, which can be provided by theory like a threshold of a quantum error correction code or by numerical simulations of the target experiment with an error model, or can be experimental values reported by their past experiments or by other experimental groups.   
This is the accuracy validation test.
If the result of the test is ``Invalid", they proceed to Step 6.
If it is ``Valid", they proceed to Step 7. 
\item {\bf Improvement}\\
In most cases, the comparison in Step 5 would result in ``Invalid", i.e., the implementation accuracies are not enough, and experimentalists go back to the 3rd step and perform control optimization again.
This is the sixth step, and we call it the improvement step.
If they repeat this cycle of re-calibration (Steps~3, 6), quantum characterization (Step~4), and validation test (Step~5) several times and judge it hard to achieve the valid accuracy only by such software optimization, they would return to the 1st step and repeat the steps mentioned above again.
\item {\bf QIP experiment}\\
When the value calculated is sufficiently high compared to the reference value, they conclude that the accuracies of the implemented elemental quantum operations are valid for performing the target experiment with the accuracy they expect and proceed to the main QIP experiment.
\end{enumerate}
As explained above, results of a quantum characterization is used at the validation step (Step~5) and the improvement step (Step~6).
Quantum characterization take a role to provide information of elemental quantum operation implemented to these steps.

\section{Requirements for Quantum Characterization}\label{sec:requiredConditions}

There are several possible requirements on quantum characterization used for a QIP experiment. 
Performances of a characterization protocol must be evaluated by figures of merits appropriate with the requirements. As far as the authors recognize, discussions on the requirements have not been active in the community despite of its importance in the development of a quantum computer. In this section, we explain and discuss five important requirements, taking into consideration the roles explained in Appendix \ref{sec:roleInExperiment}. We also discuss how much the proposed protocol (RSCQT) and known representative ones such as RB and GST satisfy each requirement. 

\begin{figure}[t]
   \begin{center}
      \includegraphics[width=0.8\linewidth]{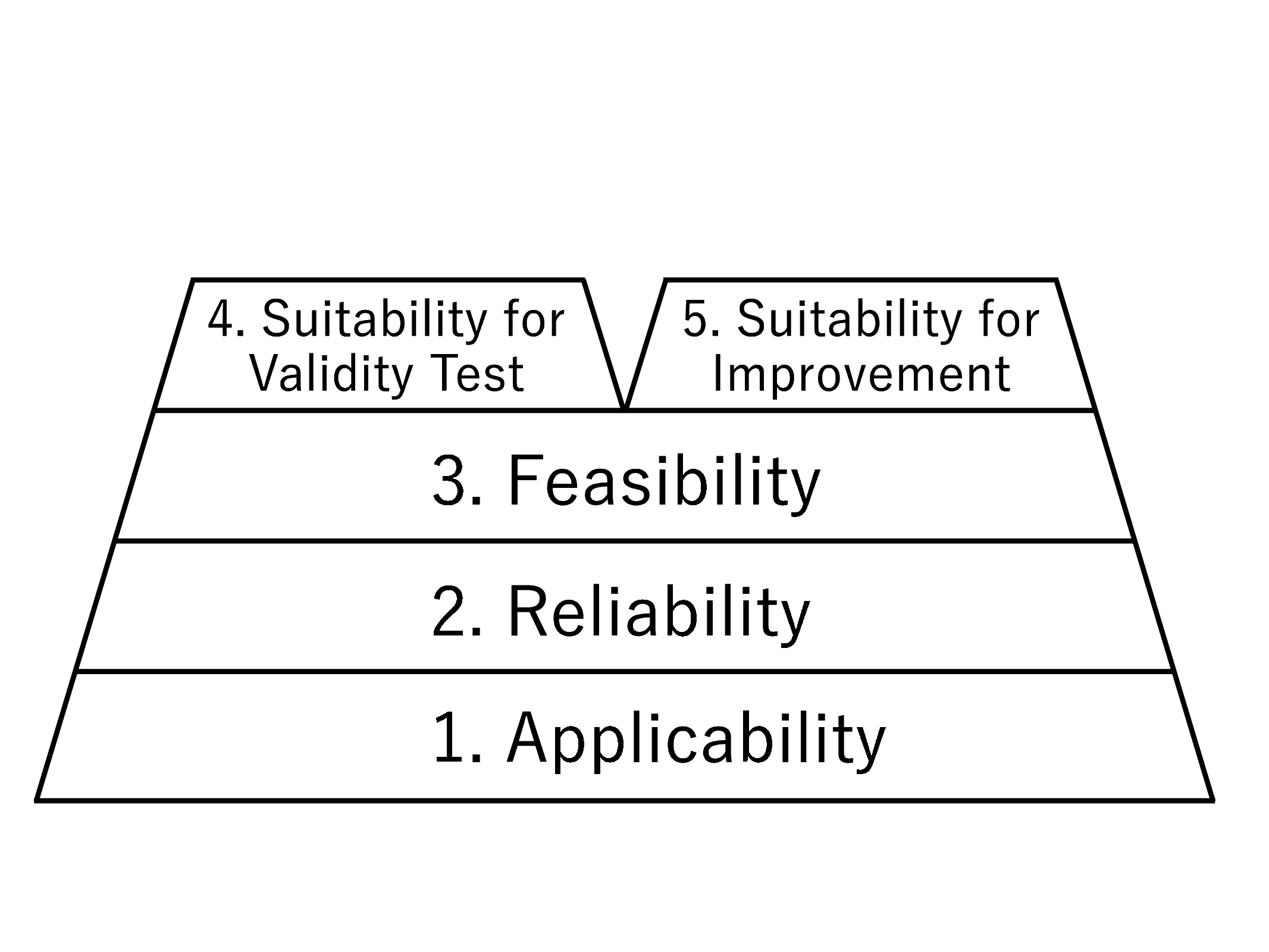}
      \caption{Layered structure of requirements for a quantum characterization protocol. Numbers in the figure correspond to the requirement numbers. Lower requirement is more fundamental. Details are explained in the main text at Appendix~\ref{sec:requiredConditions}.}
      \label{fig:requiredConditions}
   \end{center}
\end{figure}

The whole set of five requirements has a layered structure as depicted in Fig. ~\ref{fig:requiredConditions}.
Those requirements in the lower layer is more fundamental than others in the upper layer.
Discussion on how much one requirement is satisfied, i.e., high, middle, low, etc., makes sense only when lower requirements are satisfied.

As discussed below, a single characterization protocol does not need to satisfy all of the requirements in practice. 
We need to use several protocols so that they compensate for each other's shortcomings. 
The following discussion would be helpful to understand significance of our results along with the requirements.

\begin{enumerate}[{\bf R1.}]
   \item {\bf Applicability}\\
	   A protocol needs to be applicable to all elemental quantum operations used in the target experiment.
	   A standard QT protocol is applicable to one type of quantum operations.
	   Some protocols in self-consistent quantum tomography like GST and RSCQT satisfy the applicability to all types of operations.
	   RB protocols are not applicable to state preparation and measurement, but are applicable to a specific class of quantum gates.

   \item {\bf Reliability}\\
   
    We require a characterization protocol highly reliable.
    The output of a highly reliable protocol is guaranteed to be sufficiently close to actually implemented objects. 
    In other words, reliability is the concept on how close they are. 
    In particular, high reliability is desired at the accuracy validation and improvement steps of highly accurate quantum operations. 
    The closeness is quantified by how small an estimation error is.
    There are many figures of merits for estimation errors in Statistics.
    In the main text, we chose the root-mean-squared error in probability space, $\sqrt{\mathbbm{E} [  \Loss (\bm{p}({\setIndexSequence}, {\bf s}^{\mathrm{est}}_N), \bm{p}({\setIndexSequence}, \strue ) )} ]$, and that in the parameter space, $\sqrt{ \mathbbm{E} [ \Regu ( \bm{s}^{\mathrm{est}}_N , \strue)]}$ as the estimation errors.

    When considering the reliability, we need to take into account statistical errors, systematic errors such as pre-knowledge errors, violation of assumptions, and numerical accuracy, which are explained below.
    An estimation error of a protocol includes effects from all of them.

   	\begin{enumerate}[(i)]
	   \item Statistical errors\\
             Outcomes of a quantum measurement are obtained probabilistically, which is a principle of quantum theory.
             Then there exists a statistical fluctuation on finite amount of data.
             An amount of data available in any real experiments is always, and the finiteness of the data leads to statistical errors on results of a protocol.
             There are many different concepts and figures of merits for quantifying statistical errors.
             
             The first one is the convergence of an estimation result sequence of a protocol to the true value of the estimation object at the limit of the amount of data going to infinity.
             In statistics, this property of an estimator is called {\it asymptotic consistency} or {\it consistency} in short, but in this paper we use {\it asymptotic convergence} as the same meaning in order to avoid confusion with {\it self-consistency} which appears frequently in this paper and is a concept on a treatment of estimation object in quantum tomographic protocols.
             When a protocol has the asymptotic convergence, it is guaranteed that we can obtain an estimate much closer to the true value as the amount of data becomes larger.
             This is an expected and necessary condition for a quantum characterization protocol.

             When the asymptotic convergence of a protocol is proved, it is also important to analyze a statistical error for finite data, because the asymptotic convergence is just an asymptotic property.
             Derivation of the analytical form of a statistical error is possible only for simple estimators such as linear estimators in standard QT.
             Evaluation of statistical errors for more complicated estimators like the RSC estimator needs numerical experiments.
                          
             As proven in Sec.~\ref{subsec:AsymptoricallyGauge-EquivalentEstimator}, the RSC estimator has the asymptotic convergence to the gauge equivalent class for arbitrary finite dimensional systems under the SCIC condition and invertibility of gates. 
             Furthermore, the dominant rate of the convergence scaling is also derived.
             A behavior of its statistical errors has been investigated numerically for 1-qubit cases.

             We mainly choose the probability distance between the estimate and true set of quantum operations or its mean squared error in Secs.~\ref{sec:theoreticalResults} and \ref{sec:numericalResults}.
             The goodness of fit (GOF) to data obtained is another popular figure of merit related to statistical fluctuation.
             We need to be careful about interpretation of GOF because it is different from a discrepancy from the true set.
             Suppose that we consider a loss minimization estimator without a regularization.
             Its estimates are not unique, and assume that we choose a gauge fixing method in some way.
             Then the GOF of the unregularized estimator is better than that of the RSC estimator by definition.
             However, statistical errors of the unregularized estimator can be worse than those of the RSC estimator since over-fitting to data may occur at the unregularized estimator.
             As shown in Sec.~\ref{sec:numericalResults}, an estimation error of empirical distributions can be worse than that of the RSC estimator.
             The RSC estimator has worse GOF than the unregularized one, but a statistical error of the regularized estimator can be smaller than that of data which are bases of the data-fitting of the unregularized one.
             Data has statistical fluctuations, and unregularized estimator can be over-fitten to the fluctuations.
             This example indicate that GOF is not an appropriate figure of merit for statistical error.      
             
             \item Systematic errors\\
             An estimation error includes effects of statistical errors and systematic errors.
             When a systematic error exists, it remains even at the limit of the amount of data going to infinity.
             In other words, we cannot reduce the systematic error by increasing the amount of data.
             Systematic errors originate from the estimator itself or our pre-knowledge error.
             Here we focus on systematic errors from pre-knowledge error.
             
             All protocols require some pre-knowledge on settings of the characterization experiments, or equivalently there are assumptions on any protocols, such as measurement outcomes obey identically and independently distributed (iid) probability distributions, the dimension of the system is known, mathematical representations of a part of quantum operations are known, physical errors occurring on an estimation object can be described by a model, etc.
             For example, mathematical representations of state preparations and measurements are assumed to be known in the standard quantum process tomography.
             We need to model their actions in order to proceed data-processing in the standard quantum process tomography.
             Let use call the part of quantum operations to be known a tester.
             There are unknown imperfections and uncontrollable degrees of freedom in experimental devices, and these leads to a discrepancy between our pre-knowledge (or model) on the action of the tester and the actual action.
             The pre-knowledge error in the standard QPT is called SPAM error.
             In general, assumptions can be invalid in real experiments, and then pre-knowledge error occurs.
             In the SCQT approach, effects of SPAM error is reduced to be zero, which is a superior property compared to the standard QT.
             Known dimension and iid are assumed in GST and RSCQT.
             Then there is a pre-knowledge error on them, systematic errors occurs on GST and RSCQT.
             Error models are not assumed in the SCQT approach. 
             RB protocols also assume known dimension and IID condition.
             A systematic error on the standard RB (SRB) originated from assumption on distributions and types of physical errors is recently investigated numerically, which reports it can be non-negligible \cite{RB_Reliability_Proctor2017}.
             The interleaved RB, which is the most popular RB protocol, is based on SRB and introduce an approximation of physical errors by depolarizing error models at a conversion process of decay rates \cite{RB_interleaved_Magesan2012}.
             From these facts, there is a possibility that RB protocols have non-negligible systematic errors.
             Additional numerical investigation on the possibility is important to guarantee the reliability of RB protocols.
             A method for detecting possible non-negligible violation of the iid condition using GST data is proposed \cite{SCQT_GST_BlumeKohout2017}.
             It becomes more important to develop such methods for violation of assumptions as the implemented accuracies becomes higher.
             At the moment, it is unclear to the authors whether any modification of the method from GST to RSCQT is possible or not.    

             \item Numerical accuracy\\
                In general, data-processing for quantum characterization is involved with constraint optimization at data-fitting process.
                Accuracy and stability of its optimization results depend on the choice of optimization algorithm, option settings such as selections of tuning parameters and starting point, and numerical implementation.
                Data-processing with different choices of them can give us different results of characterization even from same data.
                In particular, the optimization problem in the SCQT approach includes highly nonlinear terms and its solution tends to become unstable. 
                It would be desirable that there is a (de facto) standard library for each characterization protocol.
                A library pyGSTi is provided for GST.
                
                There are many possible requirements for a standard library such as transparency and user-friendliness.
                From a perspective of reliability, it is important that numerical evaluations of the library's performance have been done for many different settings covering examples typical in experiments.
                Here a setting of the performance evaluation includes dimensions and numbers of quantum systems, target set of quantum operations, physical error models on quantum operations, experimental schedules, amounts of data, optimization algorithm and information of its option, pre-knowledge error models, and so on.                
                It would be important for the community to select a standard set of settings to be analyzed for the performance evaluation of a characterization protocol. 
          \end{enumerate} 
          
     Basically, a reliability of a protocol is guaranteed or proven by theory and numerical experiments.
     Physical experiments cannot prove the reliability because complete information of the true estimation object is not available in physical experiments and we cannot calculate estimation errors of the protocol.
     When a protocol is proven to be reliable by theory and numerical experiments, a result of the protocol in a physical experiment is believed to be reliable as long as assumptions that the proof of the reliability bases on are valid in the physical experiment.
     An error bar, which includes (i), (ii), and (iii), attached to a result of the protocol is a useful information to quantitatively represent reliability of the results.
     Reliability of a protocol for calculating the error bar also needs to be proven theoretically or numerically.
     If a characterization protocol and a protocol for calculating error bars are reliable, a result with a small error bar is interpreted as much reliable than that with a large error bar. 
     There are two popular protocols for calculating an error bar.
     One is based on normal distribution approximation, and the other is bootstrapping.
     The reliability of these protocols are based on asymptotic theory of random variables, in which a bias is considered to be negligible since the amount of data is sufficiently large. 
     In the tomographic approach including the standard QT and SCQT, however, there can be a non-negligible bias of a characterization protocol originated from the physicality constraints, even for experimentally large amount of data.
     When the quantum operations implemented become closer to ideal ones such as pure states, projective measurements, and unitary operations, which is the case of highly accurate quantum operations, the effect of the bias remains until larger amount of data \cite{QST_Sugiyama2012}.
     Investigation of the bias effect on error bar for experimentally typical amount of data is important for judging the reliability of a protocol for calculating an error bar.      
     
     Error amplification, or long gate sequence, is an approach to increase reliability by repeating gates, which is expected to reduce effects of statistical and/or systematic errors.
     RB experiments contain randomized error amplification and GST experiments contain structured error amplification.
     Theoretical results of RSCQT in the paper hold for experiments with structured error amplification as well, but its numerical implementation has not been achieved yet.

   \item {\bf Feasibility}\\
            Experimental costs and data-processing costs of a protocol need to be a feasible amount.  
            Here experimental costs are a number of experimental schedules, numbers of repetitions of each schedule, and time periods of each schedule.
            Data-processing costs are memory size and computational time. 
            At the evaluation of costs, not only their scaling but also their absolute values are important.
            
   	   Experimentalists may need to repeat the cycle of Steps 3, 4, 5, 6 many times for improvement. 
	   Then smaller cost per single run of a protocol is better. 
            At the current numerical implementation of the RSC estimator with 3-fold cross validation for a 1-qubit, a typical computational time period for obtaining a single estimate is in the order of a few seconds on a single CPU (2.4 GHz).
            There is room for further speeding up on coding.
            Its numerical implementation for a 2-qubit system that can run within a feasible time period is a challenging task to be tackled.
   
   \item {\bf Suitability for Validation Test}\\
            Results of a protocol are used at the accuracy validation test in Step. 5, and the protocol needs to be suitable for the validation test.
            The comparison at the test is done along with a figure of merit appropriate for the target QIP experiment.
            
            Quantum error correction is one of the most important target experiments at present and in the near future.
            As discussed in Sec.~\ref{subsec:improvementAndValidationForQEC}, recent numerical results indicate that the average gate fidelity is not enough for predicting a performance of a quantum error correction code \cite{QEC_Coherence_Suzuki2017}.
            RB protocols may lack suitability for validation test toward quantum error correction.
            Additionally, there can be a bias on results of SRB whose size dramatically differs for experimental settings \cite{RB_Reliability_Proctor2017} and there can be a bias on results of IRB with factor of 2 at least \cite{RB_Reliability_Epstein2014}. 
            This indicates that a straightforward comparison of RB infidelity to a theoretical value or an experimental value on other platform can be invalid.            
            
            Derivation of an appropriate figure of merit is an important issue in the theory of quantum error correction.
            Although the explicit form of the appropriate figure of merit has not been known, it must be a function of an elemental quantum operation.
            In the SCQT approach, full information of each operation is available, and we can calculate the value of the figure of merit in any functional form if it is given.

   \item {\bf Suitability for Improvement}\\ 
            Results of a protocol is used at the accuracy improvement step (Step. 6), and the protocol needs to be suitable for the improvement.   
            At the improvement step, experimentalists need some hints for improving the insufficient accuracy up to a sufficient level.
            It is desirable to extract information useful for further improvements. 
            Even if the amount of information is not enough for specifying types of errors and origins of noises, such partial information would be helpful for experimentalists to narrow possible candidates of types of errors.
            After the selected candidates of errors, experimentalists may design additional experiments specified for the candidates and perform the experiments for figuring out the details of errors.
            When there are possibilities of existence of physical errors such as crosstalk and leakage errors, experiments should be modified for detecting their effects and we need to spread out the dimension of the system of interest used in data-processing. 
            
            The authors consider that the suitability of these protocols for the improvement need further theoretical or numerical investigations, even though there are experimental results that report usefulness of RB protocols and GST for accuracy improvement.     
            First, we explain the reasons for RB protocols.
            There are two concerns on RB protocols. 
            The first concern of RB protocols is that there can be a bias on results of SRB and IRB as discussed in R4.  
            Comparison of RB results of an experiment to those of other platform can be invalid.
            However, comparison between RB results obtained from the same device may be somewhat valid if the change of control parameters during the calibration is small because types and degrees of violation of assumptions in a RB protocol can be almost equivalent.
            In this case, smaller RB infidelity would be better.
            The second concern on RB protocols is that quantum gates implemented by different accumulated Hamiltonian errors can have the same value of the average gate fidelity. 
            Standard and interleaved RB protocols returns estimates of the average gate fidelity, with possible bias.
            The information of the average gate fidelity only is not enough for determining values of control parameters. 
            Unitarity or purity RB protocols returns partial information on Hamiltonian, but it is still not enough for determining all parameters of Hamiltonian.  
            Next, we explain the reasons for GST.
            Our concern on GST is the possible unphysicality of state preparation and measurement in the current version of pyGSTi.
            A gauge fixing method that gives unphysical operations includes artificial effects on estimates which are expected to make them far from the true operations because all true operations are physical.
            It would be valuable for developing better improvement protocol to investigate reasons that usefulness of RB and GST protocols are reported in many experiments even though there are several concerns mentioned above. 
\end{enumerate} 

We summarize the evaluation of requirements for RB, GST, and RSCQT discussed above in Table~\ref{table:Comparison_RB_GST_RSCQT}. 

    \begin{table*}[bt]
       \centering
       \begin{tabular}{|l||l|l|l|}
       \hline
        \multicolumn{1}{|c||}{Requirements} & \multicolumn{1}{|c|}{RB} & \multicolumn{1}{|c|}{GST} & \multicolumn{1}{|c|}{RSCQT} \\
        \hline
        \hline
        R1. Applicability & Middle & High & High \\
                          & \ - Limited to a class & & \\
                          & \ \ \ of gates & & \\
                          & & & \\
        \hline    
        R2. Reliability & Low & High (theoretically) & High (theoretically) \\
                        & \ - Bias & Low (numerically) & Low (numerically) \\
                        &  & \ - Nonlinearity of optimization & \ -  Nonlinearity of optimization \\ 
                        &  & \ - Possible unphysicality & \   - No error amplification \\ 
                        & & & \\
        \hline
        R3. Feasibility & Middle & Low & Low \\
                        & \ - Sampling of gate sequences & \ - High costs of experiments  & \ - High costs of experiments \\     
                        & \ \ \ with different length & \ \ \ and data-processing & \ \ \ and data-processing \\
                        & & & \\
        \hline
        R4. Suitability for  & Low & High (theoretically) & High (theoretically) \\
        \phantom{R4.} Validity Test & \ - Low reliability & Low (numerically) & Low (numerically) \\
            & \ - Partial information of errors & \ - Low reliability & \ - Low reliability \\
            &  & & \\
        \hline
        R5. Suitability for & Low? & Low? & Low \\
        \phantom{R5.} improvement & \ - Low reliability & \ - Low reliability & \ - Low reliability \\
                       & \ - Partial information of errors & \ - Low feasibility & \ - Low feasibility \\
                       & \ - Gauge degrees of freedom? & \ - Gauge degrees of freedom? & \ - Gauge degrees of freedom \\
                      & & & \\ 
        \hline
       \end{tabular}
       \caption{Summary of requirements for RB, GST, and RSCQT discussed in Appendix~\ref{sec:requiredConditions}. Each requirement is evaluated by `Low', `Middle', or `High'. Reasons of `Low' and `Middle' are attached below each degree. A question mark, `?', attached to an evaluation or reason means that information for the evaluation is not enough and further theoretical or numerical investigations are necessary.}
       \label{table:Comparison_RB_GST_RSCQT}
   \end{table*}

\section{Parametrization of Quantum Operations}\label{sec:parametrization}

   In this section, we explain real vector parametrization of state preparation, measurement (POVM), and gate, which were used in numerical experiments explained in Sec. \ref{sec:numericalResults}.
   Let $\mathcal{H}$ denote a quantum system of interest.
   The dimension of $\mathcal{H}$ is finite, denoted by $d$.
   
   Let $\rho$ denote a density matrix on $\mathcal{H}$, which is a $d \times d$ complex matrix that is trace-one and positive semidefinite, i.e., 
   \begin{eqnarray}
      \Trace [\rho] = 1 \ \& \ \rho \succeq 0.  \label{eq:physicality_state}
   \end{eqnarray}
   Let ${\bm{B}} = \{ B_{\alpha} \}_{\alpha=0}^{d^2 -1}$ denote a $d \times d$ Hermitian orthonormal matrix basis with $B_0 = I/\sqrt{d}$.
   From the completeness and orthogonality of the basis, we can uniquely expand any density matrix by $\bm{B}$ as
   \begin{eqnarray}
      \rho = \sum_{\alpha=0}^{d^2 -1} \rho_{\alpha} B_{\alpha}.
   \end{eqnarray} 
   From the Hermiticity of $\bm{B}$ and $\rho$, the expansion coefficients $\rho_{\alpha}$ are real.
   The trace-one condition leads to $\rho_0 = 1/\sqrt{d}$.
   The other $(d^2 -1)$ real numbers parametrize a density matrix.
   The positive-semidefiniteness condition restricts the possible range of $(\rho_1 , \ldots , \rho_{d^2 -1})$ into a compact convex region in $\mathbbm{R}^{d^2 -1}$.  
   
   There are two objects of description of a quantum measurement as a quantum operation, a probability distribution of measurement outcome and state transformations with respect to the obtained outcome.
   A positive operator-valued measure (POVM) can treat a probability distribution of measurement outcome only.
   A quantum measurement process can treat both of them.
   In the main text, we do not consider a quantum state after measurement, and here we only explain POVM.
   Let us assume that the set of possible outcomes of a measurement is discrete and finite.
   Then a POVM $\bm{\Pi} = \{ \Pi_{x} \}_{x=0}^{m-1}$ is a discrete and finite set of $d \times d$ Hermitian matrices that is sum-identity and positive-semidefinite each, i.e.,
   \begin{eqnarray}
      \sum_{x=0}^{m-1} \Pi_{x} = I \ \& \ \Pi_{x} \succeq 0. \label{eq:physicality_povm}
   \end{eqnarray}  
   From the sum-identity condition, one of $m$ elements of the POVM is fixed, e.g., as $\Pi_{m-1} = I - \sum_{x=0}^{m-2} \Pi_{x}$.
   We expand the $(m-1)$ matrices as a density matrix.
   \begin{eqnarray}
      \Pi_x = \sum_{\alpha=0}^{d^2 -1} \Pi_{x, \alpha} B_{\alpha}, \ x= 0, \ldots , m-2.
   \end{eqnarray}
   From the Hermiticity of $\bm{\Pi}$ and each $\Pi_x$, the $(m-1) \times d^2$ expansion coefficients $\Pi_{x, \alpha}$ are real and parametrize a POVM.
   The positive-semidefiniteness condition restricts the possible range of the parameters into a compact convex region in $\mathbbm{R}^{(m-1) d^2}$.
   
   A quantum gate transform a state $\rho$ to another state $\rho^{\prime}$, and the action is described by a linear map $\mathcal{G}: \rho \mapsto \rho^{\prime} = \mathcal{G}(\rho)$ that is completely positive (CP) and trace-preserving (TP).
   An action of a linear map can be represented by a matrix.
   Let us choose a matrix representation along with the real vector representation of a state $|\rho \drangle := (\rho_0 , \ldots , \rho_{d^2 -1})^T \in \mathbbm{R}^{d^2}$ with respect to the basis $\bm{B}$.
   Let $\mathrm{G}$ denote the matrix representation of $\mathcal{G}$.
   Then it is a $d^2 \times d^2$ real matrix.
   The TP condition leads to equations,
   \begin{eqnarray}
      \mathrm{G}_{0\beta} = \delta_{0\beta}\ (\beta = 0, \ldots, d^2 -1 ). \label{eq:TP_HS}
   \end{eqnarray}
   The other $(d^2 -1) \times d^2$ real numbers $G_{\alpha\beta}\ (\alpha = 1, \ldots , d^2 -1 , \beta = 0, \ldots , d^2 -1)$ and parametrize a gate.
   The CP condition restricts the possible range of the parameters into a compact convex region in $\mathbbm{R}^{(d^2 -1) d^2}$.   
   We introduce another matrix representation of a linear map, called Choi-Jamilkowski matrix, $\mathrm{CJ}(\mathcal{G}) \in \mathbbm{C}^{d^2 \times d^2}$, in order to treat the CP condition mathmatically.
   Let us define a vector $|I\drangle := \sum_{i=0}^{d}|i\rangle |i\rangle \in \mathbbm{C}^{d^2}$.
   With the vector, the CJ matrix is defined as
   \begin{eqnarray}
      \mathrm{CJ}(\mathcal{G}) := (\mathcal{G} \otimes \mathcal{I}) |I\drangle \dlangle I|,
   \end{eqnarray}
   where $\mathcal{I}$ is the identity map on $\mathbbm{C}^{d \times d}$.
   The CP condition on $\mathcal{G}$ is equivalent to the matrix inequality,
   \begin{eqnarray}
      \mathrm{CJ}(\mathcal{G}) \succeq 0. \label{eq:CP_CJ}
   \end{eqnarray}
   An explicit form of the relation between CJ and HS matrices is given at Eq.~(\ref{eq:HS_CJ_relation}).

   For a given set of quantum operations consisting of $n_{\mathrm{s}}$ states, $n_{\mathrm{p}}$ POVMs, and $n_{\mathrm{g}}$ gates, i.e., $\{ \rho_{0}, \ldots , \rho_{n_{\mathrm{s}}-1},\ \bm{\Pi}_{0}, \ldots , \bm{\Pi}_{n_{\mathrm{p}}},\  \mathcal{G}_{0}, \ldots , \mathcal{G}_{n_{\mathrm{g}}-1} \}$, the real vector $\bm{s}$ for the set is the set of parameters for each operation, i.e., 
   \begin{eqnarray}
      \bm{s} &=& ( \rho_{0, 1}, \ldots , \rho_{0, d^2 -1},\ \ldots , \rho_{n_{\mathrm{s}}-1, 1}, \ldots , \rho_{n_{\mathrm{s}}-1, d^2 -1}, \notag \\
       && \ \Pi_{0,0}, \ldots , \Pi_{0, d^2 -1}, \ldots , \Pi_{m-2, 0}, \ldots , \Pi_{m-2, d^2 -1},  \notag \\
       && \ G_{0,10}, \ldots , G_{0,d^{2}-1 d^2 -1}, \ldots , G_{n_{\mathrm{g}}-1,10}, \ldots , \notag \\ 
       && \ G_{n_{\mathrm{g}}-1,d^{2}-1 d^2 -1} )^T . \label{eq:parametrization_s}
   \end{eqnarray}
   Theoretical results in Sec. \ref{sec:theoreticalResults} hold for any $n_{\mathrm{s}}$, $n_{\mathrm{p}}$, and $n_{\mathrm{g}}$.
   The parametrization in Eq. (\ref{eq:parametrization_s}) is used in numerical experiments for a 1-qubit system explained in Sec. \ref{sec:numericalResults}, which were performed for the case of $n_{\mathrm{s}}=1$, $n_{\mathrm{p}}=1$, and $n_{\mathrm{g}}=3$.

\section{Gauge degrees of freedom}\label{sec:gaugeDegreesOfFreedom}

   In this section, we explain a mathematical treatment of the gauge degrees of freedom.
   For a given set of quantum operations $\bm{s}$, a gauge transformation $\mathcal{A}$ is an invertible map from a set of quantum operations to another set that satisfies $\mathcal{A}(\bm{s}) \in [ \bm{s} ]$, in which a set of quantum operations and a parameter characterizing the set are identified as explained in Sec. \ref{sec:settingAndNotation}.
   Let $\mathcal{A}_{\mathrm{s}}$, $\mathcal{A}_{\mathrm{p}}$, and $\mathcal{A}_{\mathrm{g}}$ denote the maps corresponding to the action of $\mathcal{A}$ on a state, POVM, and gate, respectively, e.g., for $\bm{s} = \{ \rho , \bm{\Pi}, \mathcal{G} \}$, $\mathcal{A}(\bm{s}) = \{ \mathcal{A}_{\mathrm{s}}(\rho), \mathcal{A}_{\mathrm{p}}(\bm{\Pi}), \mathcal{A}_{\mathrm{g}}(\mathcal{G}) \}$.  
   By definition, gauge equivalent sets give an equivalent probability distribution for a given experimental setting.
   The functionalities of $\mathcal{A}_{\mathrm{s}}$, $\mathcal{A}_{\mathrm{p}}$, and $\mathcal{A}_{\mathrm{g}}$ are limited into linear because of the linearity of the Born's rule.
      
   Let $X$ denote a $d \times d$ complex matrix with the form $X = \sum_{\alpha=0}^{d^2 -1} X_{\alpha} B_{\alpha}$, where $\bm{B}= \{ B_{\alpha} \}$ is an orthonormal matrix basis.
   We introduce a vectorization of $X$ with respect to the basis, $| X \drangle := \sum_{\alpha=0}^{d^2 -1} X_{\alpha} \bm{e}_{\alpha}$, where $\{ \mathbbm{e}_\alpha \}_{\alpha=0}^{d^2 -1}$ is an orthonormal basis on $\mathbbm{C}^{d^2}$.
   The vectorization keep{\dgreen s} the value of the Hilbert-Schmidt inner product of two matrices, i.e.,   
   \begin{eqnarray}
      \Trace [Y^{\dagger} X] = \dlangle Y | X \drangle .
   \end{eqnarray}
   An action of a linear map on $\rho$ can be represented by a matrix on the vectorization $|\rho\drangle$, as in the case of a gate explained in Sec. \ref{sec:parametrization}.
   The matrix representation of a linear map, say $\mathcal{F}$, is called the Hilbert-Schmidt (HS) representation, and we use a notation, $\mathrm{HS}(\mathcal{F})$ for the representation. 
   Let $\mathrm{A}$ denote the matrix representation of $\mathcal{A}_{\mathrm{s}}$, i.e.,
   \begin{eqnarray}
      |\mathcal{A}_{\mathrm{s}}(\rho) \drangle = A |\rho\drangle .
   \end{eqnarray} 
   Then the Born's rule enacts the action of $\mathcal{A}_{\mathrm{p}}$, and $\mathcal{A}_{\mathrm{g}}$ as follows.
   \begin{eqnarray}
      \dlangle \mathcal{A}_{\mathrm{p}}\Pi_{x}| &=& \dlangle \Pi_{x}| A^{-1}, \\
      \mathrm{HS}(\mathcal{A}_{\mathrm{g}}(\mathcal{G})) &=& A\, \mathrm{HS}(\mathcal{G}) A^{-1}.
   \end{eqnarray}   
   Therefore, the action of a gauge transformation $\mathcal{A}$ on a state, POVM, and gate is characterized by a $d^2 \times d^2$ matrix as
   \begin{eqnarray}
      \begin{array}{cccc}
                           & |\rho\drangle & \mapsto & A |\rho\drangle \\
       \mathcal{A} :& \dlangle \Pi_{x} | & \mapsto & \dlangle \Pi_{x} | A^{-1} \\
                           & G & \mapsto & A G A^{-1}  
      \end{array}\label{eq:gaugeTransformation}
   \end{eqnarray}

      In general, a gauge transformation $\mathcal{A}$ is an invertible linear map, and a transformed set $\mathcal{A}(\bm{s})$ is not guaranteed to be physical even if the original set $\bm{s}$ is physical.
      If we require Hermiticity of the transformed state $\mathcal{A}_{\mathrm{s}}(\rho)$ as well, $\mathcal{A}_{\mathrm{s}}$ is an Hermiticity-preserving map.
   When we choose $\bm{B}$ as each $B_{\alpha}$ is Hermitian, the HS representation of an Hermiticity-preserving (HP) map is a real matrix.
   If we require the trace-oneness of the transformed state, the $\mathcal{A}_{\mathrm{s}}$ is a trace-preserving map. 
   If we choose $\bm{B}$ such that $B_{0} = I / \sqrt{d}$, the HS representation of a trace-preserving map satisfies  $A_{0\beta} = \delta_{0\beta}$ for $\beta = 0, \ldots , d^2 -1$.
   In that case, the representation can be written as
   \begin{eqnarray}
      A = \left[
         \begin{array}{cc}
            1 & \mathbf{0}^{T} \\
            \mathbf{b} & C
         \end{array}
      \right],
   \end{eqnarray}
   where $\bm{b}$ is a $d^2 -1$ dimensional vector and $C$ is a $d^2 - 1 \times d^2 -1$ matrix. 
   $T$ denotes the transposition with respect to the indexing of the HS representation.
   $\bm{0}^{T}$ denotes the $d^2 -1$ dimensional zero vector transposed.
   The inverse $A^{-1}$ has the following form
      \begin{eqnarray}
      A^{-1} = \left[
         \begin{array}{cc}
            1 & \mathbf{0}^{T} \\
            - C^{-1} \mathbf{b} & C^{-1}
         \end{array}
      \right],
   \end{eqnarray}
   Let us choose an Hermitian orthonormal matrix basis $\bm{B}$ satisfying $B_{0}= I / \sqrt{d}$.
   Then the vectorization of a density matrix $\rho $ is represented as
   \begin{eqnarray}
            |\rho \drangle = \frac{1}{\sqrt{d}}\left[
         \begin{array}{c}
            1 \\
            \bm{v}
         \end{array}
      \right]. 
   \end{eqnarray}
   The parameter vector $\bm{v} \in \mathbbm{R}^{d^2 -1}$ is a generalized Bloch vector. 
   The transformed vectorized density matrix via a TPHP gauge transformation is 
   \begin{eqnarray}
      A |\rho\drangle = \frac{1}{\sqrt{d}}
      \left[
      \begin{array}{cc}
         1 \\
         C\bm{v} + \bm{b} 
      \end{array}   
      \right].
   \end{eqnarray}
   From the singular value decomposition of $C$, the matrix $C$ contains actions of rotation and rescaling of $\bm{v}$.
   The vector $\bm{b}$ acts as the origin shift.
   Therefore the actions of a gauge transformation are categorized into rotation, rescaling, and origin shift.
   The rescaling and origin shift can cause an unphysical $\mathcal{A}(\bm{s})$.
   The number of degrees of freedom characterizing a TPCP gauge transformation is $d^4 - d^2$.

\section{Regularization}\label{sec:regularization}
   
In this section, we briefly explain conventional purpose of using a regularization in applied mathematics.
After that, we describe differences between regularization in RSCQT and that in conventional settings, from the viewpoints of motivation and three mathematical properties.

Regularization is an attractive way to stably solve inverse problems \cite{Tikhonov1943, Tikhonov1963}, i.e., the solution is not much changed when the observed data are slightly fluctuated. 
The regularization has been much attention in many other fields of mathematical sciences including integral equation \cite{IntegralEquation_Phillips1961}, signal processing \cite{Filtering_Foster1961}, statistics and machine learning \cite{Ridge_Hoerl1962, Ridge_HoerlKennard1970, Lasso_Tibshirani1996}. 
It is also used in variants of standard QT \cite{BlumeKohout_HedgedMLE2010, Gross_CompressedSensing2010, Flammia_CompressedSensing2012}. 
A purpose to use regularization depends on each research field. 
For example, it is used to avoid over-fitting to observed data in machine learning, to make a solution stable or to solve ill-posed problems in inverse problems, and among others. 
Moreover, in order to improve interpretability of models in regression problems by increasing sparsity of estimates or to obtain a smooth function in non-parametric estimation, the regularization also plays an important role in statistics. 
From the viewpoint of Bayesian analysis, a regularization can be regarded as exploiting prior information with respect to model parameters under some conditions. 

Our purpose of introducing a regularization into the setting of SCQT is to fix the gauge degrees of freedom, suitable for the improvement and validation steps.
This is originated from the role of quantum characterization in quantum information processing and our choice of self-consistent approach. 
This is new and quite different from the the conventional purposes of regularization mentioned above, although the use of $\starget$ in the regularization function can be regarded as a use of prior information of the target set of quantum operations that we aim to implement. 
Additionally, a mathematical framework of RSCQT has at least the following three attributes: (i) non-uniqueness of the true solution of the original (unregularized) problem, (ii) restricted parameter space, (iii) non-linear parameterization, of which difficulties make our problem more complicated than the previous studies. 
Hereafter, we briefly explain these three attributes. 

First, there exist the gauge degrees of freedom, which is originated from the self-consistent approach and Born's rule, a fundamental principle of quantum theory.
We cannot determine parameters of interests only from experimental data, and such estimation problem can be categorized into an ill-posed problem in the inverse problem. 
A conventional approach to the ill-posed problem in applied mathematics is to neglect or remove such unaccessible degrees of freedom. 
On the other hand, we cannot neglect or remove the gauge degrees of freedom because at the validation and improvement steps after characterization each mathematical representation of state, measurement, and gate is necessary.
In order to separate the representations from each other, we need to fix the gauge. 

Second, the region of possible parameters is constraint, which is originated from the requirement of physicality on estimates of quantum operations. 
When an accuracy of quantum operations is high, the true set lies close to the boundary of the physical region. 
If we require physicality on estimates, we have to take the boundary into account. 
In standard QT, the boundary affects on the performance of estimators \cite{QST_Sugiyama2012} for finite data. 
In RSCQT, the dimension of the parameter space is much larger than that of standard QT, and the analysis of the boundary effect becomes much harder. 

Third, the parametrization of probability distributions is non-linear, which is originated from the self-consistent approach.
A non-linear function in a loss function is often analyzed in the inverse problems \cite{Regularization_Review_Benning2018}. 
The asymptotic convergence for non-linear Tikhonov regularization was derived in \cite{NonlinearTikhonov_Seidman1989} under an assumption that the true solution of the original (unregularized) problem is unique, and \cite{NonlinearTikhonov_Engl1989} showed its rate of convergence under a similar assumption. 
Since their proofs are shown by exploiting the uniqueness of the original solution, it is non-trivial to extend their results to our framework with the gauge degrees of freedom. 

A previous study, which takes three attributes, (i), (ii) and (iii), exists \cite{NonlinearTikhonov_Johansen1997}. 
However, the study only considers a case that the regularization parameter is fixed, and does not show the asymptotic convergence to equivalence class of the true parameter nor the convergence rate of estimator. 
As a matter of fact, when we fix the regularization parameter, a bias caused by a regularization remains even at the limit of data size going to infinity, and the asymptotic convergence does not hold.

\section{Cross Validation}\label{sec:crossValidation}
Cross validation is a standard method for selecting a regularization parameter in statistics and machine learning \cite{CV_Stone1974,CV_Survey_Arlot2010}.
In the numerical experiments reported in Sec.~\ref{sec:numericalResults}, we combined the RSC estimator with $k$-fold cross validation ($k=3$).
Roughly speaking, the $k$-fold cross validation selects a regularization parameter from the perspective of prediction.
If we calculate both of the goodness of fit and the estimate from common data, an over-fitting to the data occurs, and the performance of predicting the true probability distributions or the goodness of fit to different data can become worse, as explained at sentences of R2 (i) in Sec.~\ref{sec:requiredConditions}.
The over-fitting problem is caused by the statistical dependence of the data for calculating the estimate and goodness of fit. 
In order to avoid the over-fitting problem, the cross validation divides the data into two parts.
One is for calculating an estimate, which is called {\it learning data}.
The other is for calculating the goodness of fit, which is called {\it test data}.
This division makes learning data and test data statistically independent.
In order to reduce an effect of the way of division, divisions are differently performed $k$ times.
A goodness of a regularization parameter is evaluated by an average value of the goodness of fit over $k$ divisions. 
We explain the details of the procedure of $k$-fold cross validation below.

Suppose that we performed experiments with a SCIC set of experimental schedules $\setIndexSequence$ and obtained experimental data $D_{N}$ with an amount of data $N$.
For a given coefficient $c$, a regularization parameter in the RSC estimator is calculated from the coefficient and amount of data as $r_{N} = c / N$. 
Let $\bm{c} = \{ c_{1}, \ldots , c_{n_{\mathrm{r}}}\}$ denote a set of candidates of regularization parameter coefficients.
Let $k$ denote a positive integer larger than or equal to 2.
The $k$-fold cross validation selects a value from $\bm{c}$ for the RSC estimator along with the following procedure.
\begin{enumerate}[{\bf Step 1.}]
\item {\bf Data Division}\\
We randomly divide the data into $k$ distinct parts as equally as possible.
Let $D_{N, 1}, \ldots , D_{N,k}$ denote the $k$ parts of $D_{N}$ ($D_{N} = \cup_{j=1}^{k} D_{N, j}$).
We introduce a notation for complement sets $\overline{D_{N, j}} := D_N \backslash D_{N, j}$, $j=1, \ldots , k$.
At the $j$-th division $D_N = D_{N,j} \cup \overline{D_{N,j}}$, $D_{N,j}$ is the test data and $\overline{D_{N,j}}$ is the learning data. 
Let $N_j$ and $\overline{N_{j}}$ denote the amounts of data for $D_{N,j}$ and $\overline{D_{N, j}}$, respectively.
Roughly speaking, $N_j \approx N/k$ and $\overline{N_{j}} \approx N - N/k$ hold.
\item {\bf Calculation of Empirical Distributions}\\
We calculate empirical distributions from each $D_{N,j}$ and $\overline{D_{N, j}}$.
Let $\bm{f}_{N_j}(\setIndexSequence)$ and $\bm{f}_{\overline{N_{j}}}(\setIndexSequence)$ denote the set of empirical distributions calculated from $D_{N,j}$ and $\overline{D_{N, j}}$, respectively.
For simplicity of notation, we omit $\setIndexSequence$ from the notation of the set of empirical distributions below in this section.  
\item {\bf Calculation of Cross Validation Losses}\\
First, we calculate multiple RSC estimates from complement data and coefficient candidate.
Next, we calculate values of loss functions for cross validation.
Let $\ell$ denote an index for the candidates of regularization parameter. 

For $\ell = 1, \ldots , n_{\mathrm{r}}$, we repeat the following procedures:
   \begin{enumerate}[{\bf 3.1}]
      \item {\bf Calculation of Estimates}\\
      We calculate each RSC estimate from complement empirical distribution $\bm{f}_{\overline{N_{j}}}$ and a regularization parameter $c_{\ell}/\overline{N_{j}}$ for $j=1, \ldots , k$ along with Eq.~(\ref{eq:def_estimate}). 
       Let $\qoset^{\est}_{\overline{N_{j}}}(c_{\ell})$  denote the estimates, in which their dependency on $c_{\ell}$ is explicitly shown in the notation for clarifying the dependence.
      \item {\bf Calculation of Cross Validation Losses}\\
      We calculate values of the loss function in Eq.~(\ref{def_loss_squared}) between the probability distributions predicted by the estimates calculated in the previous sub-step from the leaning data and the empirical distributions calculated from the test data for $j=1, \ldots , k$.
      We calculate the arithmetic mean, which is the definition of the cross validation (cv) loss for a candidate $c_{\ell}$.
      Let $\Loss^{\mathrm{cv}}(c_{\ell})$ denote the cv loss of $c_{\ell}$.
      An explicit mathematical form of the cv loss is as follows: 
      \begin{eqnarray}
         \Loss^{\mathrm{cv}}(c_{\ell}) := \frac{1}{k} \sum_{j=1}^{k} \Loss \left( \bm{p}(\setIndexSequence , \qoset^{\est}_{\overline{N_{j}}}(c_{\ell})), \bm{f}_{N_{j}} \right). \label{def:loss_cv}
      \end{eqnarray}
   \end{enumerate}
   At the end of Step 3, we have a set of values of cv loss, $\left\{ \Loss^{\mathrm{cv}}(c_{\ell}) \right\}_{\ell=1}^{n_{\mathrm{r}}}$.
\item {\bf Selection of Regularization Parameter}\\
   From $\bm{c}$, we choose the coefficient candidate $c_{\ell}$ that has the minimal value of the cv loss.
   Let $c^{\mathrm{cv}}$ denote the selected coefficient.  
   It is defined as
   \begin{eqnarray}
      c^{\mathrm{cv}} := \argmin_{c \in \bm{c}} \Loss^{\mathrm{cv}}( c ).     
   \end{eqnarray}
\end{enumerate}

In the procedure of $k$-fold cross validation explained above, we need to perform the optimization for calculating an RSC estimate $k n_{\mathrm{r}}$ times.
After the procedure, we obtain the selected coefficient $c^{\mathrm{cv}}$.
Finally, we calculate the RSC estimate $\qoset^{\mathrm{est}}_{N}(c^{\mathrm{cv}})$ with the total data $D_N$ and the selected regularization parameter $r_N^{\mathrm{cv}}:= c^{\mathrm{cv}}/N$.
The estimate $\qoset^{\mathrm{est}}_{N}(c^{\mathrm{cv}})$ is the result of the RSC estimator with $k$-fold cross validation.
In total, we need to perform the optimization $(k n_{\mathrm{r}} + 1)$ times for the combination. 
The $k n_{\mathrm{r}}$ times optimizations are additional cost for using $k$-fold cross validation.

\section{Proof of Theorem 1}\label{sec:proofOfTheorem1}

    In this section, we give a proof of Theorem \ref{theorem:GaugeEquivalence-SCIC}. 
    First, we mention two lemmas on vector bases as a preparation for the proof.
    Let $\dim$ denote a finite positive integer.
    We will set $\dim = d^2$ in the proof.
    \begin{lemma}\label{lemma:UniqueExistenceOfMatrixBetweenTwoBases}
      Let $\{ \bm{a}_{i} \}_{i=1}^{\dim}$ and $\{ \bm{b}_{i} \}_{i=1}^{\dim}$ denote bases of a $\dim$-dimensional complex vector space $\mathbbm{C}^{\dim}$. Then there exists a unique invertible $\dim \times \dim$ matrix $C$ satisfying
      \begin{eqnarray}
         \bm{b}_{i} = C \bm{a}_{i}, \ i = 1, \ldots , \dim.
      \end{eqnarray}  
   \end{lemma}
   \textbf{Proof} (Lemma \ref{lemma:UniqueExistenceOfMatrixBetweenTwoBases}):\ 
   Let $\{ \mathbbm{e}_{i}\}_{i=1}^{d}$ denote an orthonormal basis of $\mathbbm{C}^{\dim}$.
   There exist unique invertible matrices $A$ and $B$ satisfying
   \begin{eqnarray}
      \bm{a}_{i} &=& A \mathbbm{e}_{i}, \ \forall i = 1, \ldots , \dim, \\
      \bm{b}_{i} &=& B \mathbbm{e}_{i}, \ \forall i = 1, \ldots , \dim. 
   \end{eqnarray} 
   Then
   \begin{eqnarray}
      \bm{b}_{i} = B A^{-1} \bm{a}_{i}
   \end{eqnarray}
   holds, and $C= B A^{-1}$. From the uniqueness and invertibility of $A$ and $B$, $C$ is also unique and invertible.
   $\square$
   
   \begin{lemma}\label{lemma:MatrixEquivalenceConditionWithTwoBases}
      Let $\{ \bm{a}_{i} \}_{i=1}^{\dim}$ and $\{ \bm{b}_{i} \}_{i=1}^{\dim}$ denote bases of $\mathbbm{C}^{dim}$. 
      If matrices $X$, $Y \in \mathbbm{C}^{\dim\times \dim}$ satisfy
      \begin{eqnarray}
         \bm{b}_{j} \cdot X \bm{a}_{i}= \bm{b}_{j}\cdot Y \bm{a}_{i}, \forall  i, j = 1, \ldots , \dim,
      \end{eqnarray}  
      then $X=Y$ holds.
   \end{lemma}
   \textbf{Proof} (Lemma \ref{lemma:MatrixEquivalenceConditionWithTwoBases}):\
   As introduced in the proof of Lemma \ref{lemma:UniqueExistenceOfMatrixBetweenTwoBases}, there exist unique invertible matrices $A$ and $B$ satisfying 
   \begin{eqnarray}
      \bm{a}_{i} &=& A \mathbbm{e}_{i}, \ \forall i = 1, \ldots , \dim, \\
      \bm{b}_{i} &=& B \mathbbm{e}_{i}, \ \forall i = 1, \ldots , \dim. 
   \end{eqnarray}    
   Then
   \begin{eqnarray}
       &&\bm{b}_{j} \cdot X \bm{a}_{i} = \mathbbm{e}_{j} \cdot B^{\dagger} X A \mathbbm{e}_{i} \\
       &=& \bm{b}_{j} \cdot Y \bm{a}_{i} = \mathbbm{e}_{j} \cdot B^{\dagger} Y A \mathbbm{e}_{i}.
   \end{eqnarray}
   Therefore we have $B^{\dagger} X A = B^{\dagger} Y A$.
   From the invertibility of $A$ and $B$, $X = Y$ holds. 
   $\square$ \\

   Second, we introduce a lemma on informationally complete sets of states and POVMs with gauge-equivalence. 
   In the following, we set $\dim = d^2$.
   Let $|\rho \drangle$, $|\bm{\Pi}\drangle = \{ |\Pi_{x}\drangle \}_{x \in \mathcal{X}}$, and $\mathrm{G}$ denote a vectorized representation of a density matrix $\rho$, the same representation of a POVM $\bm{\Pi}$, and a HS representation of a TPCP map $\mathcal{G}$ \cite{Text_QStateEstimation_Paris2004,TN_Wood2015}.
   The vectors $|\rho \drangle$ and $|\Pi_{\omega}\drangle$ are in $\mathbbm{C}^{\dim}$ and the matrix $G$ is in $\mathbbm{C}^{\dim \times \dim}$.
   Then generalized Born's rule can be rewritten with the vector representation as  
   \begin{eqnarray}
      p(x | \rho, \mathcal{G}, \bm{\Pi}) 
      &=& \Trace \left[ \Pi_{x} \mathcal{G} (\rho) \right] \\
      &=& \dlangle \Pi_{x}| G | \rho \drangle.
   \end{eqnarray}
   Note that 
   \begin{eqnarray}
      \dlangle \mathcal{G}^{\mapAdjoint}(\Pi_{x}) | = \dlangle \Pi_{x} | G 
   \end{eqnarray}
   holds.

   \begin{lemma}\label{lemma:UniqueLinearGaugeTransformation}
      Suppose that $\{ |\rho^{i}\drangle \}_{i=1}^{N_{s}}$ and $\{ |\tilde{\rho}^{i}\drangle \}_{i=1}^{N_{s}} $ are state-informationally complete and $\{ | \bm{\Pi}^{j} \drangle \}_{j=1}^{N_{p}}$ and $\{ | \tilde{\bm{\Pi}}^{j}\drangle \}_{j=1}^{N_{p}}$ are POVM-informationally complete. 
      If 
      \begin{eqnarray}
         \dlangle \Pi^{j}_{x} | \rho^{i}\drangle = \dlangle \tilde{\Pi}^{j}_{x} | \tilde{\rho}^{i}\drangle, \label{eq:EqualProbabilities}   
      \end{eqnarray} 
      holds for any $i$, $j$, and $x$, then there exists a unique invertible matrix $A$ satisfying
      \begin{eqnarray}
         |\tilde{\rho}^{i} \drangle &=& A |\rho^{i}\drangle , \label{eq:GaugeTransformation_State_InLemma}\\
         \dlangle \tilde{\Pi}^{j}_{x}| &=& \dlangle \Pi^{j}_{x}| A^{-1}, \label{eq:GaugeTransformation_POVM_InLemma}     
      \end{eqnarray}
      for any $i$, $j$, and $x$.  
   \end{lemma}   
  \textbf{Proof} (Lemma \ref{lemma:UniqueLinearGaugeTransformation}):
      We divide each set into a linear independent subset subscripted with $1$ and the residual subset subscripted with $2$.   
      \begin{eqnarray}
         \{ |\rho^{i}\drangle \}_{i=1}^{N_{s}} 
         &=& \mathcal{S}_{1} \cap \mathcal{S}_{2}, \\
         \mathcal{S}_{1} &:=& \{ |\rho^{i}\drangle \}_{i=1}^{\dim}, \\
         \mathcal{S}_{2} &:=& \{ |\rho^{i}\drangle \}_{i=\dim + 1}^{N_{s}},\\
         \{ |\tilde{\rho}^{i}\drangle \}_{i=1}^{N_{s}} 
         &=& \tilde{\mathcal{S}}_{1} \cap \tilde{\mathcal{S}}_{2}, \\
         \tilde{\mathcal{S}}_{1} &:=& \{ |\tilde{\rho}^{i}\drangle \}_{i=1}^{\dim}, \\
         \tilde{\mathcal{S}}_{2} &:=& \{ |\tilde{\rho}^{i}\drangle \}_{i=\dim + 1}^{N_{s}},\\  
         \{ | \bm{\Pi}^{j}\drangle \}_{j=1}^{N_{p}} 
         &=& \mathcal{P}_{1} \cap \mathcal{P}_{2}, \\
         \mathcal{P}_{1} &:=& \{ | \Pi_{x_{j}}^{j}\drangle \}_{j=1}^{\dim}, \\
         \mathcal{P}_{2} &:=& \{ | \Pi_{x_{j}}^{j}\drangle \}_{j=\dim+1}^{|\mathcal{X}|\cdot N_{p}}, \\ 
         \{ | \tilde{\bm{\Pi}}^{j}\drangle \}_{j=1}^{N_{p}} 
         &=& \tilde{\mathcal{P}}_{1} \cap \tilde{\mathcal{P}}_{2}, \\
         \tilde{\mathcal{P}}_{1} &:=& \{ | \tilde{\Pi}_{x_{j}}^{j}\drangle \}_{j=1}^{\dim}, \\
         \tilde{\mathcal{P}}_{2} &:=& \{ | \tilde{\Pi}_{x_{j}}^{j}\drangle \}_{j=\dim+1}^{|\mathcal{X}| \cdot N_{p}}. 
      \end{eqnarray} 
       During the devision process, if necessary, we relabel the indices $i$, $j$, and $\omega$ so that $\mathcal{S}_{1}$ and $\mathcal{P}_{1}$ are bases of $\mathbbm{C}^{\dim}$.      
       From Lemma \ref{lemma:UniqueExistenceOfMatrixBetweenTwoBases}, there exist unique invertible matrices $A$ and $B$ satisfying
      \begin{eqnarray}
         |\tilde{\rho}^{i}\drangle &=& A |\rho^{i}\drangle, \forall i = 1, \ldots , d , \\
         \dlangle \tilde{\Pi}_{x_{j}}^{j}| &=& \dlangle \tilde{\Pi}_{x_{j}}^{j} | B, \forall j = 1, \ldots , \dim.
      \end{eqnarray}
      Then
      \begin{eqnarray}
         \dlangle \Pi^{j}_{x_{j}} | \rho^{i}\drangle 
         = \dlangle \tilde{\Pi}^{j}_{x} | \tilde{\rho}^{i}\drangle   
         = \dlangle \Pi^{j}_{x_{j}} |B A | \rho^{i}\drangle    
      \end{eqnarray}
      holds for $i, j = 1, \ldots , d$.
      From Lemma \ref{lemma:MatrixEquivalenceConditionWithTwoBases}, we have $B=A^{-1}$.
      Therefore it is proven that there exists a unique matrix $A$ satisfying Eqs. (\ref{eq:GaugeTransformation_State_InLemma}) and (\ref{eq:GaugeTransformation_POVM_InLemma}) for the linear independent subsets $\mathcal{S}_{1}$, $\tilde{\mathcal{S}}_{1}$, $\mathcal{P}_{1}$, and $\tilde{\mathcal{P}}_{1}$.
      
      Suppose that $\dim < k \le N_{\mathrm{s}}$ in the case of $\dim < N_{\mathrm{s}}$. 
      We can span any residual vectors in $\mathcal{S}_{2}$ and $\tilde{\mathcal{S}}_{2}$ by $\mathcal{S}_{1}$ and $\tilde{\mathcal{S}}_{1}$ as 
      \begin{eqnarray}
         |\rho^{k}\drangle 
         &=& \sum_{i=1}^{\dim}c_{ki} |\rho^{i}\drangle, \\  
         |\tilde{\rho}^{k}\drangle 
         &=& \sum_{i=1}^{\dim} \tilde{c}_{ki} |\tilde{\rho}^{i}\drangle
         = A\left( \sum_{i=1}^{\dim} \tilde{c}_{ki} |\rho^{i}\drangle  \right).
      \end{eqnarray} 
      Then from Eq. (\ref{eq:EqualProbabilities}),  
      \begin{eqnarray}
         &&\dlangle \Pi_{x_{j}}^{j} |\rho^{k} \drangle 
         = \dlangle \Pi_{x_{j}}^{j}| \left( \sum_{i=1}^{\dim}c_{ki} |\rho^{i}\drangle \right)  \\
         &=& \dlangle \tilde{\Pi}_{x_{j}}^{j} | \tilde{\rho}^{k} \drangle  
         = \dlangle \Pi_{x_{j}}^{j}|B A \left( \sum_{i=1}^{\dim}\tilde{c}_{ki} |\rho^{i}\drangle \right)  \\ 
         && \hspace{16mm} = \dlangle \Pi_{x_{j}}^{j}| \left( \sum_{i=1}^{\dim}\tilde{c}_{ki} |\rho^{i}\drangle \right)  
      \end{eqnarray}
      holds for $j=1, \ldots , \dim$, and we have
      \begin{eqnarray}
         \sum_{i=1}^{\dim}c_{ki} |\rho^{i}\drangle
         = \sum_{i=1}^{\dim}\tilde{c}_{ki} |\rho^{i}\drangle.   
      \end{eqnarray}
      Then
      \begin{eqnarray}
         |\tilde{\rho}^{k}\drangle = A |\rho^{k}\drangle 
      \end{eqnarray}
      holds for $k = \dim + 1, \ldots , N_{\mathrm{s}}$. 
      
      In the same way as the state vector, we can prove 
      \begin{eqnarray}
         \hspace{-4mm}
         \dlangle \tilde{\Pi}_{x_{j}}^{j} | = \dlangle \Pi_{x_{j}}^{j} | A^{-1},\ \forall j = \dim +1 , \ldots , |\mathcal{X} | \cdot N_{\mathrm{p}}.
      \end{eqnarray}
      $\square$
      In the proof above, we assumed that numbers of possible outcomes are common for all $\bm{\Pi}^j$ for simplicity.
      A generalization to cases that each POVM has different number of elements is straightforward.
      
      Now we are ready for proving Theorem \ref{theorem:GaugeEquivalence-SCIC}.\\
      \textbf{Proof} (Theorem \ref{theorem:GaugeEquivalence-SCIC}): \
      When $\tilde{\qoset} \in [\qoset]$, $\bm{p}({\setIndexSequence}, \tilde{\bm{s}}) = \bm{p}({\setIndexSequence}, \tilde{\bm{s}})$ holds by definition of the gauge-equivalence. 
      Here we prove the opposite direction, i.e., when $\setIndexSequence$ is SCIC and each gate in $\qoset$ has the inverse, which can be unphysical, then $\bm{p}({\setIndexSequence}, \tilde{\qoset}) = \bm{p}({\setIndexSequence}, \qoset )$ implies $\tilde{\qoset} \in [\qoset]$.
      When $\setIndexSequence$ is SCIC, it includes a set of index sequences satisfying Eq. (\ref{eq:SCIC_state_povm}).
      $\setIndexSequence_{\mathrm{s}}$ and $\setIndexSequence_{\mathrm{p}}$ included in $\setIndexSequence$ are state- and POVM-informationally complete, respectively.
      From Lemma \ref{lemma:UniqueLinearGaugeTransformation}, equations 
      \begin{eqnarray}
         \dlangle \tilde{\Pi}^{\indexSequence_{\mathrm{p}}}_{x} | \tilde{\rho}^{\indexSequence_{\mathrm{s}}} \drangle 
         = \dlangle \Pi^{\indexSequence_{\mathrm{p}}}_{x} | \rho^{\indexSequence_{\mathrm{s}}} \drangle
      \end{eqnarray}
      imply that there exists a unique matrix $A$ such that
      \begin{eqnarray}
         |\tilde{\rho}^{\indexSequence_{\mathrm{s}}}\drangle 
         &=& A |\rho^{\indexSequence_{\mathrm{s}}}\drangle, \ \forall \indexSequence_{\mathrm{s}} \in \setIndexSequence_{\mathrm{s}}, \label{eq:GaugeTransformation_State}\\
         \dlangle \tilde{\Pi}^{\indexSequence_{\mathrm{p}}}_{x}| 
         &=& \dlangle \Pi^{\indexSequence_{\mathrm{p}}}_{x}| A^{-1}, \ \forall \indexSequence_{\mathrm{p}} \in \setIndexSequence_{\mathrm{p}}, \ x \in \mathcal{X}. \label{eq:GaugeTransformation_POVM}    
      \end{eqnarray} 
      The SCIC $\setIndexSequence$ also includes a set of index sequences satisfying Eq.~(\ref{eq:SCIC_state_gate_povm}).   
      Then
      \begin{eqnarray}
         \dlangle \Pi^{\indexSequence_{\mathrm{p}}}_{x} | \mathrm{G}_{i_{\mathrm{g}}} | \rho^{\indexSequence_{\mathrm{s}}} \drangle
         &=& \dlangle \tilde{\Pi}^{\indexSequence_{\mathrm{p}}}_{x}| \tilde{\mathrm{G}}_{i_{\mathrm{g}}} | \tilde{\rho}^{\indexSequence_{\mathrm{s}}} \drangle \\
         &=& \dlangle \Pi^{\indexSequence_{\mathrm{p}}}_{x} | A^{-1} \tilde{\mathrm{G}}_{i_{\mathrm{g}}} A | \rho^{\indexSequence_{\mathrm{s}}} \drangle  
      \end{eqnarray}   
      hold for $\indexSequence_{\mathrm{s}} \in \setIndexSequence_{\mathrm{s}}$, $i_{\mathrm{g}} = 1, \ldots , \numGates$, $\indexSequence_{\mathrm{p}} \in \setIndexSequence_{\mathrm{p}}$, and $x \in \mathcal{X}$.
      From Lemma \ref{lemma:MatrixEquivalenceConditionWithTwoBases}, we have
      \begin{eqnarray}
         \mathrm{G}_{i_{\mathrm{g}}} = A^{-1} \tilde{\mathrm{G}}_{i_{\mathrm{g}}} A 
         \Leftrightarrow  \tilde{\mathrm{G}}_{i_{\mathrm{g}}} = A \mathrm{G}_{i_{\mathrm{g}}} A^{-1},  \label{eq:GaugeTransformation_Gate}  
      \end{eqnarray} 
      for $i_{\mathrm{g}} = 1, \ldots , \numGates$.
      With Eq.~(\ref{eq:GaugeTransformation_Gate}), Eqs.~(\ref{eq:GaugeTransformation_State}) and (\ref{eq:GaugeTransformation_POVM}) can be rewritten as
      \begin{eqnarray}
         A \mathrm{G}^{\indexSequence_\mathrm{s}} A^{-1}|\tilde{\rho} \drangle 
         &=& A \mathrm{G}^{\indexSequence_\mathrm{s}} |\rho\drangle, \\
         \dlangle \tilde{\Pi}_{x} | A \mathrm{G}^{\indexSequence_\mathrm{p}} A^{-1}
         &=& \dlangle \Pi_{x} | \mathrm{G}^{\indexSequence_{\mathrm{p}}} A^{-1},
      \end{eqnarray}
      where $\mathrm{G}^{\indexSequence_\mathrm{s}}$ and $\mathrm{G}^{\indexSequence_{\mathrm{p}}}$ are HS representations of gates constructed by applying $\mathcal{G}_{i_{\mathrm{g}}}$ along with $\indexSequence_\mathrm{s}$ and $\indexSequence_\mathrm{p}$, respectively.
      From the invertibility of $A$ and $\mathrm{G}_{i_{\mathrm{g}}}$, we obtain
      \begin{eqnarray}
         |\tilde{\rho} \drangle &=& A |\rho \drangle , \label{eq:state_A} \\
         \dlangle \tilde{\Pi}_{x} | &=& \dlangle \Pi_{x}| A^{-1}. \label{eq:povm_A}
      \end{eqnarray}
      Let $\indexSequence$ denote an arbitrary gate index sequences whose each element is in $\{ 1, \ldots , \numGates \}$. 
      The length of $\indexSequence$ is arbitrary, and $\indexSequence$ itself is not necessarily in $\setIndexSequence$.
      Eqs. (\ref{eq:state_A}), (\ref{eq:povm_A}), and (\ref{eq:GaugeTransformation_Gate}) lead
      \begin{eqnarray}
         \bm{p}^{\indexSequence}(\tilde{\qoset}) = \bm{p}^{\indexSequence}(\qoset).  
      \end{eqnarray}
      Therefore $\tilde{\qoset} \in [\qoset]$.
      $\square$
      
      In the proof above, the numbers of states and POVMs are assumed to be 1.
      A generalization of the proof to cases of multiple states and POVMs is straightforward.

\section{Proof of Theorem 2}\label{sec:proofOfTheorem2}
   
   In this section, we give a proof of Theorem \ref{theorem:AsymptoticConvergence}.
   In order to clarify the roles of each properties of the RSC estimator in the setting of SCQT and each condition mentioned in Theorem \ref{theorem:AsymptoticConvergence}, we split the proof into two parts.
   First, we prove theorems for a general setting of statistical parameter estimation with the following conditions:\\
   
   \vspace{-2mm}
   \noindent {\bf Conditions}\\
   \vspace{-5mm}
   \begin{enumerate}
      \item[C.1] The parameter space $\qosetsphysical$ is a compact subset of a Euclidean space, and the parametrization of $\bm{p}(\setIndexSequence, \bm{s})$ is continuous.
      \item[C.2] The regularization function $\Regu (\bm{s}, \bm{s}^{\prime})$ is positive and bounded.
      \item[C.3] The regularization parameter $r_N$ is positive and satisfies $\displaystyle \lim_{N\to\infty}r_N = 0 \ \as$.
      \item[C.4] The regularization parameter $r_N$ is positive and satisfies $r_N \lsim \Loss ( \bm{p}({\setIndexSequence}, \strue), \bm{f}_{N}(\setIndexSequence)) $.
      \item[C.5] For a given $\strue \in \mathcal{S}$, a point $\bm{s} \in \mathcal{S}$ satisfying $\bm{p}^{\indexSequence}(\qoset) = \bm{p}^{\indexSequence}(\strue) $ is uniquely determined up to the gauge equivalence.
      	In other words, $\qoset \notin [\strue ]  \Leftrightarrow \exists \indexSequence \in \setIndexSequence$ such that $\norm{ \bm{p}^{\indexSequence}(\qoset) - \bm{p}^{\indexSequence}(\strue )}_{2} > 0$. 
   \end{enumerate}        
   Second, we show that the theorems for the general setting are applicable to the RSC estimator in the setting of SCQT.
   In Appendix~\ref{subsec:Def_AsymptoticNotation}, we give a rigorous definition of the asymptotic notation, $\lsim$.
   The definition is used in proofs in this section.    
   In Appendix~\ref{subsec:Rate_EmpiricalDistribution}, we introduce a lemma that is about bounds of the asymptotic convergence rate of the empirical distributions to the true probability distributions. 
   The lemma is used in Appendix \ref{subsec:Proof_EquivalentConvergence_ConvergenceRateEquivalence}. 
   In Sec.~\ref{subsec:Proof_AsymptoticConvergence_ProbDistSpace}, we prove Eq.~(\ref{eq:AsymptoticConvergence_ProbDistSpace}).
   In Appendix~\ref{subsec:Proof_EquivalentConvergence_ConvergenceRateEquivalence}, we prove Eqs.~(\ref{eq:EquivalentConvergence}) and (\ref{eq:ConvergenceRateEquivalence}).
   In Appendix~\ref{subsec:Proof_AsymptoticConvergence_ParameterSpace}, we prove Eq.~(\ref{eq:AsymptoticConvergence_ParameterSpace}).
   Main tools used in the proofs are the property of $\sest_N$ as the minimizer of the objective function, the triangle inequality of norms, the strong law of large numbers, the central limit theorem, and the strong law of iterated logarithm.

   \subsection{Definition of the Asymptotic Notation}\label{subsec:Def_AsymptoticNotation}

      We give a rigorous definition of the asymptotic notation, "$\lsim$", introduced in the main text and used in Theorem \ref{theorem:AsymptoticConvergence}.   
      Suppose that $f(N)$ and $g(N)$ are positive functions of the data size $N > 0$. 
      Then the notation is defined as 
      \begin{eqnarray}
         f (N) \lsim g(N) \overset{\mathrm{def}}{\Longleftrightarrow} \limsup_{N\to \infty} \frac{f (N)}{g(N)} < \infty.
      \end{eqnarray}
      Equivalently, $f (N) \lsim g(N)$ holds if and only if there exists a positive real numbers $a$ and $N_0$ such that 
      \begin{eqnarray}
         f (N) \le a g (N), \ \forall \ N \ge N_0.
      \end{eqnarray}
      This is equivalent to the big O notation, $f(N) \in O\large( g(N) \large)$, in computer science.

    \subsection{Asymptotic convergence rate of empirical distributions}\label{subsec:Rate_EmpiricalDistribution}
    
       We introduce a lemma for proving Eqs. (\ref{eq:EquivalentConvergence}) and (\ref{eq:ConvergenceRateEquivalence}).
       \begin{lemma}\label{lemma:AsymptoticScaling_empiricalDistribution}
       The asymptotic convergence rate of $\bm{f}_{N}(\setIndexSequence)$ to $\bm{p}(\setIndexSequence, \strue)$ is bounded as 
       \begin{eqnarray}
          \hspace{-9mm}
          \frac{1}{\sqrt{N}}
          \lsim \sqrt{ \Loss \big( \bm{p}({\setIndexSequence}, \strue), \bm{f}_{N}(\setIndexSequence) \big) }
          \lsim \sqrt{\frac{\ln\ln N}{N}}\ \as .  \label{eq:ConvergenceRate_EmpiricalDistributions}
       \end{eqnarray}
       \end{lemma}
       \noindent {\bf Proof} (Lemma \ref{lemma:AsymptoticScaling_empiricalDistribution}) : 
       First, we prove the left inequality of Eq.~(\ref{eq:ConvergenceRate_EmpiricalDistributions}) by contradiction to the central limit theorem.
       We assume
       \begin{eqnarray}
          \Loss \left( \bm{p}({\setIndexSequence}, \strue), \bm{f}_{N}({\setIndexSequence}) \right) 
          < \frac{C}{N} \ \as, 
       \end{eqnarray}
       for arbitrary positive constant $C$ and sufficiently large $N$.
       Then, due to the dominant convergence theorem, we obtain
       \begin{eqnarray}
          \Expectation \left[ \Loss (\bm{p}({\setIndexSequence}, \strue), \bm{f}_{N}({\setIndexSequence}) ) \right] 
          < \frac{C}{N}, \label{eq:TobeContradiction_1overNscaling} 
       \end{eqnarray}
       for arbitrary positive constant $C$ and sufficiently large $N$, where $\Expectation$ denote the expectation with respect to the observed measurement outcomes. 
       On the other hand, the central limit theorem \cite{Text_ProbabilityTheory_Klenke2008} leads to 
       \begin{eqnarray}
          \Expectation \left[ \Loss \big(\bm{p}({\setIndexSequence}, \strue), \bm{f}_{N}({\setIndexSequence}) \big) \right] \propto \frac{1}{N}. \label{eq:CentralLimitTheorem_Expectation}   
       \end{eqnarray}
       Eq. (\ref{eq:TobeContradiction_1overNscaling}) contradicts Eq.~(\ref{eq:CentralLimitTheorem_Expectation}).
       Therefore there exists a positive number $a$ such that 
       \begin{eqnarray}
          \frac{a}{N} \le \Loss \big( \bm{p}({\setIndexSequence}, \strue), \bm{f}_{N}({\setIndexSequence}) \big)  \ \as , \label{eq:LowerBound_EmpiricalDistributionScaling}
       \end{eqnarray}
       for any sufficiently large $N$, and we obtain the first inequality to be proved.
       
       The right inequality of Eq.~(\ref{eq:ConvergenceRate_EmpiricalDistributions}) is the strong law of iterated logarithm \cite{Text_ProbabilityTheory_Klenke2008} itself.
    $\square$

    \subsection{Proof of Eq. (\ref{eq:AsymptoticConvergence_ProbDistSpace})}\label{subsec:Proof_AsymptoticConvergence_ProbDistSpace}
    
    We prove Eq.~(\ref{eq:AsymptoticConvergence_ProbDistSpace}) in Theorem \ref{theorem:AsymptoticConvergence}.    
    First, we prove the equivalent statement under Conditions C.1, C.2, and C.3.
    
       \begin{theorem}\label{theorem:AsymptoticConvergence_ProbDistSpace}
	If C.1, C.2, and C.3 are satisfied, then
	\begin{eqnarray}
   	\lim_{N\to\infty}\sqrt{\Loss ( \bm{p}({\setIndexSequence}, \sest_N), \bm{p}(\setIndexSequence, \strue))} = 0 \ \as
	\end{eqnarray}
	holds.
       \end{theorem}	
	\noindent {\bf Proof} (Theorem \ref{theorem:AsymptoticConvergence_ProbDistSpace}):\\
	Under Condition C.1, there exists an argument minimizing the objective function $F_N (\bm{s})$ over $\mathcal{S}$. 
	Then, we have
	\begin{eqnarray}
	\hspace{-5mm}
	  &&\Loss ( \bm{p}({\setIndexSequence}, \sest_N), \bm{f}_{N}(\setIndexSequence)) \notag \\
	  \hspace{-5mm}
      	  &&\le \Loss ( \bm{p}({\setIndexSequence}, \sest_N), \bm{f}_{N}(\setIndexSequence))
            + r_{N} R(\bm{s}^{\mathrm{est}}_{N}, \bm{s}^{\mathrm{target}}) \\
            \hspace{-5mm}
            &&= \min_{\bm{s}\in\mathcal{S}} F_{N}(\bm{s}) \\
            \hspace{-5mm}
            &&\le F_{N}(\strue) \\
            \hspace{-5mm}
            &&= \Loss ( \bm{p}({\setIndexSequence}, \strue), \bm{f}_{N}(\setIndexSequence)) 
            + r_{N} R( \strue, \bm{s}^{\mathrm{target}}) \label{eq:Lestemp_smaller_than_Ltrueemp_plus_regularization} \\
            \hspace{-5mm}
            &&\to 0 \ \mbox{as}\ N\to \infty \ \as. 
         \end{eqnarray}
         Here we used the strong law of large numbers \cite{Text_ProbabilityTheory_Klenke2008} and Conditions C.2 and C.3.
   
         By using the triangle inequality of $\sqrt{\Loss}$, we have
         \begin{eqnarray}
            &&\sqrt{\Loss ( \bm{p}({\setIndexSequence}, \sest_N), \bm{p}(\setIndexSequence, \strue))} \notag \\
            &&\le \sqrt{\Loss ( \bm{p}({\setIndexSequence}, \sest_N), \bm{f}_{N}(\setIndexSequence))} \notag \\
            &&\phantom{=}+ \sqrt{\Loss ( \bm{p}({\setIndexSequence}, \strue), \bm{f}_{N}(\setIndexSequence))}  \\
            &&\to 0 \ \mbox{as} \ N \to \infty \ a.s..
         \end{eqnarray}
         $\square$   
    
    Let us move on to the proof of Eq.~(\ref{eq:AsymptoticConvergence_ProbDistSpace}).
    In the setting of SCQT, we can choose a continuous parametrization of probability distributions, and the continuous parameter space can be compact subset of a Euclidean space. 
    Hence, Condition C.1 is satisfied.
    Condition C.2, the positivity and boundedness of $\Regu$, is satisfied in the RSC estimator.
    Condition C.3 is Eq.~(\ref{eq:rN_goto0}) itself.  
    Therefore, Theorem \ref{theorem:AsymptoticConvergence_ProbDistSpace} is applicable to the RSC estimator in the setting of SCQT, and Eq.~(\ref{eq:AsymptoticConvergence_ProbDistSpace}) holds. 
    $\square$

    \subsection{Proof of Eqs. (\ref{eq:EquivalentConvergence}) and (\ref{eq:ConvergenceRateEquivalence})}\label{subsec:Proof_EquivalentConvergence_ConvergenceRateEquivalence}
    
    We prove Eqs.~(\ref{eq:EquivalentConvergence}) and (\ref{eq:ConvergenceRateEquivalence}) in Theorem \ref{theorem:AsymptoticConvergence}.    
    First, we prove an equivalent statement under condition C.1, C.2, and C.4.      
    
      \begin{theorem}\label{theorem:EquivalentConvergence}
	If C.1, C.2, and C.4 are satisfied, then
	\begin{eqnarray}
   	&&\sqrt{\Loss ( \bm{p}({\setIndexSequence}, \sest_N), \bm{p}(\setIndexSequence, \strue))} \notag \\
   	&&\lsim \sqrt{\Loss ( \bm{p}({\setIndexSequence}, \strue), \bm{f}_{N}(\setIndexSequence))} \ \as
	\end{eqnarray}
	holds.
      \end{theorem}	
	\noindent {\bf Proof} (Theorem \ref{theorem:EquivalentConvergence}):\\
	As shown in the proof of Theorem \ref{theorem:AsymptoticConvergence_ProbDistSpace}, Eq. (\ref{eq:Lestemp_smaller_than_Ltrueemp_plus_regularization}) holds under Conditions C.1 and C.2.
	By combining Condition C.2 and C.4 with Eq. (\ref{eq:Lestemp_smaller_than_Ltrueemp_plus_regularization}), we obtain
	\begin{eqnarray}
	\hspace{-5mm}
      	   &&\Loss ( \bm{p}({\setIndexSequence}, \sest_N), \bm{f}_{N}(\setIndexSequence)) \notag \\
	   \hspace{-5mm}
      	   &&\lsim \left\{ 1 + R( \strue, \bm{s}^{\mathrm{target}}) \right\} \Loss ( \bm{p}({\setIndexSequence}, \strue), \bm{f}_{N}(\setIndexSequence)) \\
	   \hspace{-5mm}
      	   &&\lsim \Loss ( \bm{p}({\setIndexSequence}, \strue), \bm{f}_{N}(\setIndexSequence)) \ \as. \label{eq:1}  
   	\end{eqnarray}
     
         By using the triangle inequality of $\sqrt{\Loss}$ and Eq. (\ref{eq:1}), we have
         \begin{eqnarray}
            &&\sqrt{\Loss ( \bm{p}({\setIndexSequence}, \sest_N), \bm{p}(\setIndexSequence, \strue))} \notag \\
            &&\le \sqrt{\Loss ( \bm{p}({\setIndexSequence}, \sest_N), \bm{f}_{N}(\setIndexSequence))} \notag \\
            && \phantom{\le}  + \sqrt{\Loss ( \bm{p}({\setIndexSequence}, \strue), \bm{f}_{N}(\setIndexSequence))}  \\
            &&\lsim \sqrt{\Loss ( \bm{p}({\setIndexSequence}, \strue), \bm{f}_{N}(\setIndexSequence))} \notag \\
            && \phantom{\lsim} + \sqrt{\Loss ( \bm{p}({\setIndexSequence}, \strue), \bm{f}_{N}(\setIndexSequence))}   \\
            &&= 2 \sqrt{\Loss ( \bm{p}({\setIndexSequence}, \strue), \bm{f}_{N}(\setIndexSequence))} \\
            &&\lsim \sqrt{\Loss ( \bm{p}({\setIndexSequence}, \strue), \bm{f}_{N}(\setIndexSequence))} \ \as.  
         \end{eqnarray}
         $\square$    
    
    Let us move on to the prove Eqs.~(\ref{eq:EquivalentConvergence}) and (\ref{eq:ConvergenceRateEquivalence}) in Theorem \ref{theorem:AsymptoticConvergence}. 
    Conditions C.1 and C.2 are satisfied in the setting of SCQT as explained in the end of Appendix~\ref{subsec:Proof_AsymptoticConvergence_ProbDistSpace}.
    If we select a regularization parameter satisfying $r_N \lsim 1/N \ \as$ (Eq. (\ref{eq:rN_1overN})), inequalities
    \begin{eqnarray}
       r_N \lsim \frac{1}{N} \lsim \Loss \big( \bm{p}({\setIndexSequence}, \strue), \bm{f}_{N}(\setIndexSequence) \big) \ \as \label{eq:rN_scaling_emp}
    \end{eqnarray}
    hold because of the left inequality in Lemma \ref{lemma:AsymptoticScaling_empiricalDistribution}, and Condition C.4 is satisfied.
    Therefore Theorem \ref{theorem:EquivalentConvergence} is applicable to the RSC estimator in the setting of SCQT, and we obtain Eq.~(\ref{eq:EquivalentConvergence}).
    Eq.~(\ref{eq:ConvergenceRateEquivalence}) is given by combining the right inequality in Lemma \ref{lemma:AsymptoticScaling_empiricalDistribution} and Theorem \ref{theorem:EquivalentConvergence}.
    $\square$

      \subsection{Proof of Eq. (\ref{eq:AsymptoticConvergence_ParameterSpace})}\label{subsec:Proof_AsymptoticConvergence_ParameterSpace}
      
   We prove that a sequence of estimates $\{ \sest_{N} \} $ converges to the gauge equivalence class $[ \strue ]$ almost surely at the limit of $N$ going to infinity.
   To prove that, we modify the proof of Theorem 4.4 in \cite{Text_MathematicalStatistics_Yoshida2006} to make it applicable to the RSC estimator in the setting of SCQT, which is an ill-posed problem caused by the existence of the gauge degrees of freedom.
   We define 
   \begin{eqnarray}
     \Regu (\qoset, [ \strue ]):= \min \left\{\  \Regu (\qoset , \qoset^{\prime}) \ \big|\  \qoset^{\prime} \in [ \strue ]\,  \right\}
   \end{eqnarray}
   as a (squared) distance between $\qoset$ and the gauge equivalence class $[\strue]$.

   \begin{theorem}\label{theorem:StrongConsistency}
      If Conditions C.1, C.2, C.3, and C.5 are satisfied,   
      the sequence of RSC estimates $\{ \sest_{N} \} $ converges to $[\strue ]$ almost surely, i.e., 
      \begin{eqnarray}
         \lim_{N \to \infty}  \Regu ( \sest_{N}, [ \strue ])= 0 \ \as .
      \end{eqnarray}
   \end{theorem}
   \noindent 
   \textbf{Proof} (Theorem \ref{theorem:StrongConsistency}): 
   First, we derive an inequality that any points in $\qosetsphysical$ outside the $\epsilon$-neighborhood of $[\strue]$ satisfy.
   For a given $\epsilon > 0$, we define 
   \begin{eqnarray}
      \eta_{\epsilon} 
      &:=& \min_{\qoset \in \qosetsphysical} \left\{  
         \sqrt{\Loss (\bm{p}(\setIndexSequence, \strue), \bm{p}(\setIndexSequence, \qoset)}
      \ ; \right. \notag \\
      &&\left. \hspace{24mm} \Regu ( \qoset, [ \strue ]) \ge \epsilon  \right\}.
   \end{eqnarray}   
    Since $\bm{p}^{\indexSequence}(\qoset)$ are continuous functions over the compact set $\qosetsphysical$ (Condition C.1), the minimal value $\eta_{\epsilon}$ exists. 
    From Condition C.5, $\eta_{\epsilon } > 0$.
   
   The following arguments hold almost surely.
   From the strong law of large numbers \cite{Text_ProbabilityTheory_Klenke2008}, for any $\indexSequence \in \setIndexSequence$
   \begin{eqnarray}
      \lim_{N \to \infty} \bm{f}^{\indexSequence}_{N} = \bm{p}^{\indexSequence} (\strue )\ \as .\label{eq:StrongLawOfLargeNumbers}
   \end{eqnarray}
    From Eq.~(\ref{eq:StrongLawOfLargeNumbers}) and Condition C.3, for every $\epsilon > 0 $, there exists a constant $N(\epsilon)$ such that 
   \begin{eqnarray}
      \hspace{-7mm}
      N \ge N(\epsilon) \Rightarrow 
     \left\{
      \begin{array}{l}
        \sqrt{\Loss (\bm{f}_{N}(\setIndexSequence), \bm{p}(\setIndexSequence, \strue))} < \frac{ \eta_{\epsilon}}{4} \\
        \sqrt{ r_{N} \Regu \left( \strue , \starget \right) }  < \frac{ \eta_{\epsilon}}{4}
      \end{array} \right. . 
      \label{eq:TwoInequalities_Nepsilon}
    \end{eqnarray}
    Then, for every $\qoset$ satisfying $ \Regu ( \qoset, [ \strue ]) \ge \epsilon$, we have
    \begin{eqnarray}
     & & \sqrt{\Loss \bigl( \bm{f}_{N}(\setIndexSequence), \bm{p}(\setIndexSequence, \qoset) \bigr)} \notag \\
     & & \ge \sqrt{\Loss (\bm{p}(\setIndexSequence, \qoset), \bm{p}(\setIndexSequence, \strue))}  \notag \\
     & &  \hspace{4mm} - \sqrt{\Loss (\bm{f}_{N}(\setIndexSequence), \bm{p}(\setIndexSequence, \strue))} \\
     & &> \sqrt{\Loss (\bm{p}(\setIndexSequence, \qoset), \bm{p}(\setIndexSequence, \strue))}
            - \frac{ \eta_{\epsilon}}{4} \\
     & &\ge \eta_{\epsilon}  -   \frac{ \eta_{\epsilon}}{4}  \\
     & &= \frac{3}{4}\eta_{\epsilon },
    \end{eqnarray}    
    where we used the triangle inequality for the $2$-norm ($\sqrt{\Loss}$).
    Therefore, an inequality
    \begin{eqnarray}
       \min_{\qoset \in \qosetsphysical} \left\{ \left. 
       \Loss (\bm{p}(\setIndexSequence, \qoset), \bm{f}_{N}(\setIndexSequence)) \
        \right|\   \Regu ( \qoset, [ \strue ]) \ge \epsilon \  \right\} \notag \\
          > \frac{9}{16}\eta_{\epsilon }^{2}  
    \end{eqnarray}
    holds.
    Then we obtain
    \begin{eqnarray}
       & &
       \hspace{-4mm}
       \min_{\qoset \in \qosetsphysical} \left\{ F_{N}(\qoset) \ \big| \    \Regu ( \qoset, [ \strue ]) \ge \epsilon   \right\} \\
       & &
       \hspace{-4mm}
       \ge  \min_{\qoset \in \qosetsphysical} \left\{ \left. 
         \Loss \bigl( \bm{p}(\setIndexSequence, \qoset), \bm{f}_{N}(\setIndexSequence) \bigr) \ 
         \right|\   \Regu \left( \qoset, [ \strue ] \right) \ge \epsilon   \right\} \\
       & &
       \hspace{-4mm}
       > \frac{9}{16}\eta_{\epsilon }^{2}. \label{eq:Inequality_outsideShouldSatisfy}
    \end{eqnarray}

    Next, we show that $\sest_N$ does not satisfy Eq.~(\ref{eq:Inequality_outsideShouldSatisfy}). 
    Since $\sest_N$ is the argument minimizing $F_{N}(\qoset)$ over $\qoset \in \qosetsphysical$, $F_{N}(\sest_{N}) \le F_{N}(\qoset)$ holds for any $\qoset \in \qosetsphysical$. 
    Then, from Eq. (\ref{eq:TwoInequalities_Nepsilon}), we have
    \begin{eqnarray}
        F_{N}(\sest_{N}) 
        &\le& F_{N}(\strue) \\  
        &=& \Loss \bigl( \bm{p}(\setIndexSequence, \strue), \bm{f}_{N}(\setIndexSequence) \bigr) \notag \\
        &&  + r_{N} \Regu (\strue , \starget) \\
        &<& \left( \frac{\eta_{\epsilon}}{4} \right)^2 + \left( \frac{\eta_{\epsilon}}{4} \right)^2 = \frac{2}{16}\eta_{\epsilon }^{2}\\
        &<& \frac{9}{16}\eta_{\epsilon }^{2}.  
    \end{eqnarray}
    Hence, $\sest_N$ does not satisfy Eq.~(\ref{eq:Inequality_outsideShouldSatisfy}), and it means that $\sest_N$ is in the $\epsilon$-neighborhood of $[\strue]$.  
    Thus, we obtain
    \begin{eqnarray}
        N \ge N(\epsilon )  \Rightarrow    \Regu ( \sest_{N}, [ \strue ]) <  \epsilon. 
    \end{eqnarray}
    Since $\epsilon$ is an arbitrary positive number, we obtain 
    \begin{eqnarray}
        \lim_{N \to \infty}  \Regu( \sest_{N}, [ \strue ])=0 \ \as .
    \end{eqnarray}
    $\square$
    
    Let us move on to the proof of Eq.~(\ref{eq:AsymptoticConvergence_ParameterSpace}).
    Condition C.1 and C.2 are satisfied in the setting of SCQT as explained in the ends of Appendix~\ref{subsec:Proof_AsymptoticConvergence_ProbDistSpace} and \ref{subsec:Proof_EquivalentConvergence_ConvergenceRateEquivalence}. 
    Condition C.3 is Eq.~(\ref{eq:rN_goto0}) itself.
    When $\setIndexSequence$ is SCIC, Condition C.5 is satisfied (Theorem \ref{theorem:GaugeEquivalence-SCIC}).
    Therefore Theorem \ref{theorem:StrongConsistency} is applicable to the RSC estimator in the setting of SCQT, and it leads to Eq.~(\ref{eq:AsymptoticConvergence_ParameterSpace}) in Theorem \ref{theorem:AsymptoticConvergence}. 
    $\square$

\section{Proofs on Dynamics Generator Analysis}\label{sec:dynamicsGeneratorAnalysis}

In this section, we prove the existence of the matrix logarithm of the HS representation of a gate and Eqs.~(\ref{eq:H_acc_reconstruct}), (\ref{eq:J_acc_reconstruct}), and (\ref{eq:K_acc_reconstruct}) in Sec. \ref{subsec:DynamicsGeneratorAnalysis}, under conditions of finite energy and finite time period.

\subsection{Proof of the Existence of the Matrix Logarithm}

   We give the proof of the existence of the matrix logarithm, which is assumed in the dynamics generator analysis proposed in Sec.~\ref{subsec:DynamicsGeneratorAnalysis}.
   In the vectorized state representation, the time-dependent version of the GKLS equation is rewritten as
   \begin{eqnarray}
      \frac{d}{dt} |\rho(t) \drangle &=& \mathrm{HS}(\mathcal{L}_{t}) | \rho (t) \drangle, \\
      \Leftrightarrow \frac{d}{dt} \mathrm{HS}(\mathcal{G}_{t})|\rho(0) \drangle &=& \mathrm{HS}(\mathcal{L}_{t}) \mathrm{HS}(\mathcal{G}_{t}) | \rho (0) \drangle,\label{eq:Lindblad_vec_2}
   \end{eqnarray}
   where we used $|\rho(t)\drangle = \mathrm{HS}(\mathcal{G}_{t}) |\rho(0) \drangle$, and $\mathcal{G}_{t}$ is defined as a map corresponding to the gate implemented with the dynamics during the time period $[0, t]$, and $\mathcal{G}_{T}$ corresponds to $\mathcal{G}$ in Sec. \ref{subsec:DynamicsGeneratorAnalysis}.
   Eq.~(\ref{eq:Lindblad_vec_2}) holds for arbitrary $\rho(0)$, and it implies 
   \begin{eqnarray}
      \frac{d}{dt} \mathrm{HS}(\mathcal{G}_{t}) &=& \mathrm{HS}(\mathcal{L}_{t}) \mathrm{HS}(\mathcal{G}_{t}). 
   \end{eqnarray}
   Therefore $\mathrm{HS}(\mathcal{G}_{t})$ is a solution of the homogeneous first-order linear differential equation.
   The general theory of differential equations guarantees the unique existence of the solution, and the following equality holds (Problems 4a in Sec. 6.5,  pp.507-508 in \cite{HornJohnson_textbook_1991}),
   \begin{eqnarray}
      \det \mathrm{HS}(\mathcal{G}_t ) = \exp \left[ \int_{0}^{t}\!\! dt^{\prime} \Trace \left\{ \mathrm{HS}(\mathcal{L}_{t^{\prime}}) \right\} \right] \label{eq:det_HS_G}.
   \end{eqnarray}
   When $H(t^{\prime})$, $J(t^{\prime})$, and $K(t^{\prime})$ are bounded for any $t^{\prime} \in [0, t]$ and $t$ is finite, $\int_{0}^{t} dt^{\prime}\! \Trace \left\{ \mathrm{HS}(\mathcal{L}_{t^{\prime}}) \right\} > - \infty$ and $\det \mathrm{HS}(\mathcal{G}_t ) > 0$ hold. 
   This implies that $\mathrm{HS}(\mathcal{G}_t )$ is invertible.  
   Every invertible matrix can be written as the exponential of a complex matrix (Exercises 2.9 and 2.10 in \cite{Hall_textbook_2015}), and for every $\mathrm{HS}(\mathcal{G}_{t})$, there exists a matrix $X(t)$ that satisfies 
   \begin{eqnarray}
      \mathrm{HS}(\mathcal{G}_t ) = \exp \left[ X(t) \right] .
   \end{eqnarray}
   $\square$
   
   Note that the trace in the R.H.S. of Eq.~(\ref{eq:det_HS_G}) can be rewritten as
   \begin{eqnarray}
      \Trace \left\{ \mathrm{HS}(\mathcal{L}_{t}) \right\} 
      = \Trace \left\{ \mathrm{HS}^{\mathrm{cb}}(\mathcal{L}_{t}) \right\}     
      = 2d J_{0}(t),
   \end{eqnarray}
   and this means that the Hamiltonian part does not affect on the invertibility.
   When the dynamics is trace-preserving, $J$ and $K$ are related as
   \begin{eqnarray}
      J(t) = - \frac{1}{2} \sum_{\alpha, \beta = 1}^{d^2 -1} K_{\alpha\beta}(t) B_{\beta}^{\dagger} B_{\alpha},
   \end{eqnarray}
   and 
   \begin{eqnarray}
      J_{0}(t) 
      &=& \Trace \left\{ B_{0}^{\dagger} J(t) \right\} \\
      &=& - \frac{1}{2\sqrt{d}} \Trace \left\{ \sum_{\alpha, \beta = 1}^{d^2 -1} K_{\alpha\beta}(t) B_{\beta}^{\dagger} B_{\alpha} \right\} \\
      &=& -\frac{1}{2\sqrt{d}} \Trace \left\{ K(t) \right\}.
   \end{eqnarray}
   Therefore $K(t)$ affects on the invertibility through $J_0(t)$.
   When $K(t)$ is positive semidefinite, the dynamics becomes completely positive, and $\Trace \left\{ K(t) \right\} \ge 0$ and $J_{0}(t) \le 0$ hold.
   Then $\Trace \left\{ \mathrm{HS}(\mathcal{L}_{t}) \right\} 
 \ge 0 > -\infty$ holds and the inverse extists.

\subsection{Proof of Eqs.~(\ref{eq:H_acc_reconstruct}), (\ref{eq:J_acc_reconstruct}), and (\ref{eq:K_acc_reconstruct})}

For simplicity of notation, we omit the superscript, $\mathrm{acc}$, below.
Eq. (\ref{eq:action_map_L_acc}) can be rewritten as
\begin{eqnarray}
   \mathcal{L}(\rho) 
   &=& -i [H, \rho] + \{ J, \rho \} +\sum_{\alpha=1}^{d^2 -1} K_{\alpha\beta} B_{\alpha} \rho B_{\beta}^{\dagger} \\  
   &=& -i \sum_{\alpha=1}^{d^2 -1} H_{\alpha} \left( B_{\alpha} \rho - \rho B_{\alpha} \right)
           + \sum_{\alpha=0}^{d^2 -1} J_{\alpha} \left( B_{\alpha} \rho + \rho B_{\alpha} \right) \notag \\
    &&       + \sum_{\alpha, \beta = 1}^{d^2 -1} K_{\alpha\beta} B_{\alpha} \rho B_{\beta}^{\dagger}.
\end{eqnarray}
In the matrix vectorization, or the HS representation w.r.t. the computational basis, $|X\drangle := \sum_{i,j} X_{ij}|i\rangle|j\rangle$, an equality, $|ABC\drangle = A \otimes C^{T} |B\drangle$, holds.
Then
\begin{eqnarray}
   &&\hspace{-10mm}\mathrm{HS}^{\mathrm{cb}}(\mathcal{L}) |\rho \drangle  \notag \\
   &=& | \mathcal{L}(\rho) \drangle \\
   &=& -i \sum_{\alpha=1}^{d^2 -1} H_{\alpha} \left( | B_{\alpha} \rho \drangle - |\rho B_{\alpha}\drangle \right) 
           + \sum_{\alpha=0}^{d^2 -1} J_{\alpha} \left( | B_{\alpha} \rho \drangle + |\rho B_{\alpha}\drangle \right) \notag \\ 
    &&       + \sum_{\alpha, \beta = 1}^{d^2 -1} K_{\alpha\beta} |B_{\alpha} \rho B_{\beta}^{\dagger} \drangle  \\
    &=&  \left\{ -i \sum_{\alpha=1}^{d^2 -1} H_{\alpha} \left( B_{\alpha} \otimes \mathbbm{1} - \mathbbm{1} \otimes B_{\alpha}^{T} \right) \notag \right. \\
     &&   + \sum_{\alpha=0}^{d^2 -1} J_{\alpha} \left( B_{\alpha} \otimes \mathbbm{1} + \mathbbm{1} \otimes B_{\alpha}^{T} \right) \notag \\           
     && \left.  + \sum_{\alpha, \beta = 1}^{d^2 -1} K_{\alpha\beta} B_{\alpha} \otimes \overline{B_{\beta}}
           \right\} |\rho \drangle \\
    &=&  \left\{ -i \sum_{\alpha=1}^{d^2 -1} H_{\alpha} \left( B_{\alpha} \otimes \mathbbm{1} - \mathbbm{1} \otimes \overline{B_{\alpha}} \right) \right. \notag \\
     && + \sum_{\alpha=0}^{d^2 -1} J_{\alpha} \left( B_{\alpha} \otimes \mathbbm{1} + \mathbbm{1} \otimes \overline{B_{\alpha}} \right)  \notag \\
     && \left. + \sum_{\alpha, \beta = 1}^{d^2 -1} K_{\alpha\beta} B_{\alpha} \otimes \overline{B_{\beta}}
           \right\} |\rho \drangle                
\end{eqnarray} 
hold, and we have
\begin{eqnarray}
    \mathrm{L}^{\mathrm{cb}}
    := \mathrm{HS}^{\mathrm{cb}}(\mathcal{L})
    &=& -i \sum_{\alpha=1}^{d^2 -1} H_{\alpha} \left( B_{\alpha} \otimes \mathbbm{1} - \mathbbm{1} \otimes \overline{B_{\alpha}} \right) \notag \\
    & &      + \sum_{\alpha=0}^{d^2 -1} J_{\alpha} \left( B_{\alpha} \otimes \mathbbm{1} + \mathbbm{1} \otimes \overline{B_{\alpha}} \right) \notag \\
    & &      + \sum_{\alpha, \beta = 1}^{d^2 -1} K_{\alpha\beta} B_{\alpha} \otimes \overline{B_{\beta}}. \label{eq:HScb_L}   
\end{eqnarray}  
By combining Eq. (\ref{eq:HScb_L}) with the orthonormality and Hermiticity of $\bm{B}$, we obtain Eqs. (\ref{eq:H_acc_reconstruct}), (\ref{eq:J_acc_reconstruct}), and (\ref{eq:K_acc_reconstruct}). \\
$\square$

\section{Numerical Experiments}\label{sec:numericalExperiments}

   We describe details of numerical experiments for 1-qubit system explained in Sec. \ref{sec:numericalResults}.
   
   \subsection{Setting}\label{subsec:setting_numerical_experiments}
   
      Three quantum gates are implemented with a Hamiltonian model \cite{Krantz2019}.
\begin{eqnarray}
   H(t) = - \frac{\Delta \omega}{2} \sigma_{3} + \frac{f(t)}{2} \left\{ \cos (\phi ) \sigma_{1} + \sin(\phi) \sigma_2 \right\},
\end{eqnarray}
where $\Delta \omega$ is the frequency detuning, $f(t)$ is the pulse shape, and $\phi$ is the relative phase.
For simplicity, we choose a rectangular pulse,
\begin{eqnarray}
 f(t) = \left\{
\begin{array}{cc}
A & (0 \le t \le W) \\
0 & \mathrm{otherwise}
\end{array}
\right. . 
\end{eqnarray}
The ideal case corresponds to the combination of $\Delta \omega = 0$, $A \cdot W = 0~(\mathcal{G}_{0}^{\mathrm{target}}),\ \pi /2 ~(\mathcal{G}_{1}^{\mathrm{target}},~ \mathcal{G}_{2}^{\mathrm{target}})$, and $\phi = 0 ~(\mathcal{G}_{0}^{\mathrm{target}}, ~\mathcal{G}_{1}^{\mathrm{target}}), ~\pi/2 ~(\mathcal{G}_{2}^{\mathrm{target}})$.
We choose the gate time as $15~\mathrm{ns}$ and $W=10~\mathrm{ns}$ with coherent errors shown in Table \ref{table:CoherentErrorAGF}. 
Decoherence is modeled by the following three dissipation operators \cite{BriegelEnglert_PRA47_3311_1993} in the GKLS master equation.
\begin{eqnarray}
   \sqrt{\Gamma_+} |1\rangle \langle 0|,\ 
   \sqrt{\Gamma_-} |0\rangle \langle 1|, \ 
   \sqrt{\Gamma_{\phi}} \frac{\sigma_{3}}{\sqrt{2}}.  
\end{eqnarray}
Relations between the dissipation ratios, $\Gamma_{+}$, $\Gamma_{-}$, $\Gamma_{\phi}$ and coherence times $T_1$, $T_2$, $T_{\phi}$, and the thermal population $p_{\mathrm{th}}$ are given as  
\begin{eqnarray}
   \Gamma_{+} + \Gamma_{-}  &=& \frac{1}{T_{1}}, \\
   \frac{1}{2}\Gamma_{+} + \frac{1}{2}\Gamma_{-} + \Gamma_{\phi} &=& \frac{1}{T_{2}},\\
   \Gamma_{\phi} &=& \frac{1}{T_{\phi}},\\
   \frac{\Gamma_{+}-\Gamma_{-}}{\Gamma_{+}+\Gamma_{-}} &=& p_{\mathrm{th}}.
\end{eqnarray}
In the numerical experiments, we choose $T_1 = 30~\mu \mathrm{s}$, $T_2 = 20~\mu\mathrm{s}$, and $p_{\mathrm{th}} = 0.01$.
Values of the average gate fidelity for each gate, which include both of coherent errors and decoherence, are shown in Table \ref{table:CoherentErrorAGF}. 
The values of the average gate infidelity are shown only the first two digits, and they are in order of $10^{-4}$.
The depolarizing error rates for state and POVM are $0.015$ and $0.10$, respectively.

\begin{table}[bt]
\begin{center}
\begin{tabular}{|c|c|c|c|c|}
\hline
   Gate & $\Delta \omega$ & $A \cdot W$ & $\phi$ & 1 $-$ AGF \\
\hline   
\hline
   $\mathcal{G}_{0}^{\mathrm{true}}$ & 0.01 & 0 & 0 &  $ 2.0 \times 10^{-3}$\\
\hline
   $\mathcal{G}_{1}^{\mathrm{true}}$ & 0.01 & $\pi/2 + 0.1$ & 0.1 & $5.5 \times 10^{-3}$ \\
\hline
   $\mathcal{G}_{2}^{\mathrm{true}}$ & 0.01 & $\pi /2 + 0.1$ & 0.1 &  $5.5 \times 10^{-3}$ \\
\hline
\end{tabular}
\caption{Coherent error parameters and average gate fidelity for $\mathcal{G}^{\mathrm{prepared}}$ in the numerical experiments}
\label{table:CoherentErrorAGF}
\end{center}
\end{table}

The schedule of the experiments consists of sub-experiments.
Every sub-experiments start with the state initialization $\rho^{\mathrm{prepared}}$ and end with the measurement $\bm{\Pi}^{\mathrm{prepared}}$.
Gate sequences between $\rho^{\mathrm{prepared}}$ and $\bm{\Pi}^{\mathrm{prepared}}$ are shown in Table \ref{table:ExperimentSchedule}.
The set of sub-experiments satisfies the SCIC condition.
We choose common number of repetitions, $N$, for each sub-experiment.

\begin{table}[bh]
\begin{center}
\begin{tabular}{|c|c||c|c||c|c|}
\hline
   ID & Gate Sequence & ID & Gate Sequence & ID & Gate Sequence \\
\hline   
\hline
   1 & G0 $\cdot$ G0 & 16 & G0 $\cdot$ G2 $\cdot$ G0 & 31 & G2 $\cdot$ G1 $\cdot$ G0 \\
\hline 
   2 & G0 $\cdot$ G1 & 17  & G0 $\cdot$ G2 $\cdot$ G1 & 32 & G2 $\cdot$ G1 $\cdot$ G1 \\
\hline
   3 & G0 $\cdot$ G2 & 18 & G0 $\cdot$ G2 $\cdot$ G2& 33 & G2 $\cdot$ G1 $\cdot$ G2\\
\hline
   4 & G1 $\cdot$ G0 & 19 & G1 $\cdot$ G0 $\cdot$ G0 & 34 & G2 $\cdot$ G2 $\cdot$ G0 \\
\hline
   5 & G1 $\cdot$ G1 & 20 & G1 $\cdot$ G0 $\cdot$ G1 & 35 & G2 $\cdot$ G2 $\cdot$ G1 \\
\hline
   6 & G1 $\cdot$ G2 & 21 & G1 $\cdot$ G0 $\cdot$ G2 & 36 & G2 $\cdot$ G2 $\cdot$ G2 \\
\hline
   7 & G2 $\cdot$ G0 & 22 & G1 $\cdot$ G1 $\cdot$ G0 & 37 & G1 $\cdot$ G1 $\cdot$ G0 $\cdot$ G0 \\
\hline
  8 & G2 $\cdot$ G1 & 23 & G1 $\cdot$ G1 $\cdot$ G1 & 38 &  G1 $\cdot$ G1 $\cdot$ G0 $\cdot$ G1 \\
\hline
   9 & G2 $\cdot$ G2 & 24 & G1 $\cdot$ G1 $\cdot$ G2 & 39 & G1 $\cdot$ G1 $\cdot$ G0 $\cdot$ G2 \\
\hline
   10 & G0 $\cdot$ G0 $\cdot$ G0 & 25 & G1 $\cdot$ G2 $\cdot$ G0 & 40 & G1 $\cdot$ G1 $\cdot$ G1 $\cdot$ G0 \\
\hline
   11 & G0 $\cdot$ G0 $\cdot$ G1 & 26 & G1 $\cdot$ G2 $\cdot$ G1 & 41 & G1 $\cdot$ G1 $\cdot$ G1 $\cdot$ G1 \\
\hline
   12 & G0 $\cdot$ G0 $\cdot$ G2 & 27 & G1 $\cdot$ G2 $\cdot$ G2 & 42 & G1 $\cdot$ G1 $\cdot$ G1 $\cdot$ G2 \\
\hline
   13 & G0 $\cdot$ G1 $\cdot$ G0 & 28 & G2 $\cdot$ G0 $\cdot$ G0 & 43 & G1 $\cdot$ G1 $\cdot$ G2 $\cdot$ G0 \\
\hline
   14 & G0 $\cdot$ G1 $\cdot$ G1 & 29 & G2 $\cdot$ G0 $\cdot$ G1 & 44 & G1 $\cdot$ G1 $\cdot$ G2 $\cdot$ G1 \\
\hline
   15 & G0 $\cdot$ G1 $\cdot$ G2 & 30 & G2 $\cdot$ G0 $\cdot$ G2 & 45 & G1 $\cdot$ G1 $\cdot$ G2 $\cdot$ G2 \\ 
\hline
\end{tabular}
\caption{List of gate sequences used in the numerical experiments. The operation order is from left to right. G0, G1, G2 correspond to $\mathcal{G}_{0}^{\mathrm{prepared}}$, $\mathcal{G}_{1}^{\mathrm{prepared}}$, $\mathcal{G}_{2}^{\mathrm{prepared}}$, respectively.}
\label{table:ExperimentSchedule}
\end{center}
\end{table}

   \subsection{Optimization Solver and Physicality Constraints}

   We numerically implemented the RSC estimator for 1-qubit systems with IPOPT \cite{IPOPT}.
   IPOPT is implemented by C++ and provides interfaces to convert the objective function and constraints into a standard form of the solver in several programming languages.
   We used C++ to shorten the computation time.
   The information of the first and second derivatives of the objective function and constraints are optionally acceptable with an interface of IPOPT.
   Such optional information is helpful for making the computation time even shorter.
    We provided them to the interface with optional parameters.
    At the interface, we can specify our degree of tolerance of acceptable violation of the constraints, $\delta$.
    The tolerance parameter of $\delta = 0$ means that we do not accept any violation of the constraints, and ideally $\delta = 0$ would be desired.
    However, we chose $\delta=10^{-4}$ for the numerical simulations reported in this manuscript, because computational time for the optimization becomes longer ad we set smaller $\delta$.
    We observed unphysical estimates sometimes in the simulations and confirmed that all violations are controlled to be below $\delta$.
        
   We explain our numerical treatment of the physicality constraints on quantum operations.
   We chose the parametrization of quantum operations by real numbers explained in Appendix~\ref{sec:parametrization}.
   As explained there (Eqs.~(\ref{eq:physicality_state}), (\ref{eq:physicality_povm}), (\ref{eq:TP_HS}), (\ref{eq:CP_CJ})), the physicality constraints on quantum operations are categorized into two types, equality constraints and inequality constraints.
   The equality constraints have been taken into account by the parametrization itself automatically.
   All of the inequality constraints are represented in the form of the positive-semidefiniteness of an Hermitian matrix such as $\rho \succeq 0,\ \Pi_{x} \succeq 0,\ \mathrm{CJ}(\mathcal{G}_{j}) \succeq 0$.
   The positive-semidefiniteness of an Hermitian matrix is rewritten as a set of polynomial inequalities \cite{Kimura2003_PLA,Byrd2003_PRA}.
   We provided the information of the polynomial inequalities for quantum operations with their first and second derivatives to the interface of IPOPT.
   The parametrization of a gate is based on the HS matrix, and we derived and used the following equality to represent the inequality constraint on the gate w.r.t. the HS matrix,
   \begin{eqnarray}
      \CJrep (\mathcal{G}) = \sum_{\alpha, \beta = 0}^{d^2 -1} \mathrm{HS}(\mathcal{G})_{\alpha \beta} B_{\alpha} \otimes \overline{B_{\beta}},\label{eq:HS_CJ_relation}   
   \end{eqnarray} 
   where $\bm{B}=\{ B_{\alpha} \}_{\alpha=0}^{d^2 -1}$ is the matrix basis introduced in Sec.~\ref{sec:parametrization}, and $\overline{B_{\beta}}$ is the complex conjugate of the matrix $B_{\beta}$.\\
   {\bf Proof} (Eq.~(\ref{eq:HS_CJ_relation})):
   Matrix elements of the HS matrix of a linear map $\mathcal{G}$ with respect to the basis $\bm{B}$ is given as
\begin{eqnarray}
   \HSrep (\mathcal{G})_{\alpha \beta} = \Trace \left[ B_{\alpha}^{\dagger} \ \mathcal{G} (B_{\beta})\right], 
\end{eqnarray}   
for $\alpha, \beta = 0, \ldots , d^2 -1$.  
The action of the map is represented with the CJ matrix as
\begin{eqnarray}
   \mathcal{G}(\rho) = \mbox{Tr}_{2} \left[ \left( I_{1} \otimes \rho^{T} \right) \CJrep (\mathcal{G})\right].
\end{eqnarray}
Then
\begin{eqnarray}
   \HSrep (\mathcal{G})_{\alpha \beta}
   &=& \Trace [ B_{\alpha}^{\dagger} \mathcal{G} (B_{\beta})] \notag \\
   &=& \mbox{Tr}_{1} \left[ B_{\alpha}^{\dagger} \mbox{Tr}_{2} \left[ \left( I_{1} \otimes B_{\beta}^{T} \right) \CJrep (\mathcal{G})\right] \right] \\
   &=& \mbox{Tr}_{1, 2} \left[ \left( B_{\alpha}^{\dagger} \otimes B_{\beta}^{T} \right) \CJrep (\mathcal{G}) \right] \\  
   &=& \mbox{Tr}_{1, 2} \left[ \left( B_{\alpha} \otimes \overline{B_{\beta}} \right)^{\dagger} \CJrep (\mathcal{G}) \right]. 
\end{eqnarray}
$\square$\\
Note that the proof holds for any orthonormal matrix basis $\bm{B}$, which is not necessarily Hermitian or $B_{0}=I/\sqrt{d}$.
Therefore Eq.~(\ref{eq:HS_CJ_relation}) holds not only for the (generalized) Pauli basis, but also for the other orthonormal basis including the computational basis.

In our numerical implementation, we chose the Affine parametrization of quantum operations.
There are several parameterizations for quantum operations.
A Cholesky parametrization is a famous one, which has an advantage that a parametrized matrix is automatically positive-semidefinite.
So, in the Cholesky parametrization, the inequality constraints of quantum operations are automatically taken into account by the parametrization itself.
Additionally, the equality constrains of them are treated by quadratic polynomials, which is also simple to treat numerically.
Because of these advantages, we chose the Cholesky parametrization for density matrix, POVM elements, and CJ matrices of gates at the beginning of our numerical implementation.
However the combination of the parametrization and IPOPT did not work well, i.e., the numerical optimization was unstable and results of the solver were far from the optimal point beyond the numerical error.
We switched the parametrization from Cholesky to the Affine one.
Then the instability was not observed, and results of the solver were optimal (with numerical error).
We consider there are two possible reasons of the numerical instability.

The first one is the nonlinearity of the Cholesky parametrization which is a quadratic parametrization.
In general, an optimization problem tends to become harder as its nonlinearity becomes higher.
A probability of a measurement outcome in the SCQT approach is a nonlinear function of quantum operations of interest.
So, the data-fitting in the approach is originally a nonlinear optimization problem, which is independent of use of regularization.
When we use the Cholesky parametrization, quantum operations are perametrized quadratically, and the nonlinearity of the optimization problem becomes higher than that with an Affine parametrization.
This increase of the nonlinearity may be a reason of the instability.
 
 The second one is the non-uniqueness of the Cholesky parametrization for positive-semidefinite matrices with zero-eigenvalues.
 The parametrization is known to be unique for positive definite matrices, but it loses the uniqueness against matrices with zero-eigenvalues.
 In the numerical experiments, we investigated quantum operations close to a target set of operations.
 In QIP experiments, target operations are typically pure states, projective measurements, and unitary operations.
 These targets correspond to rank-1 positive semidefinite matrices which has many zero-eigenvalues.
 When the true set of quantum operations is close to the singular point, the solver searches parameter regions close to the singular point, and Cholesky parametrization can become unstable there during optimization.
 The non-uniqueness of the parametrization may be another reason of the instability.  
 
   \subsection{Other results of numerical simulation}
   
   In this subsection, we show numerical results of depolarizing, amplitude damping, and rotation error models.
   We also show results of the realistic model with parameters different from those shown in the main text.

\begin{figure*}[tb]
   \begin{center}
      \includegraphics[width=\linewidth]{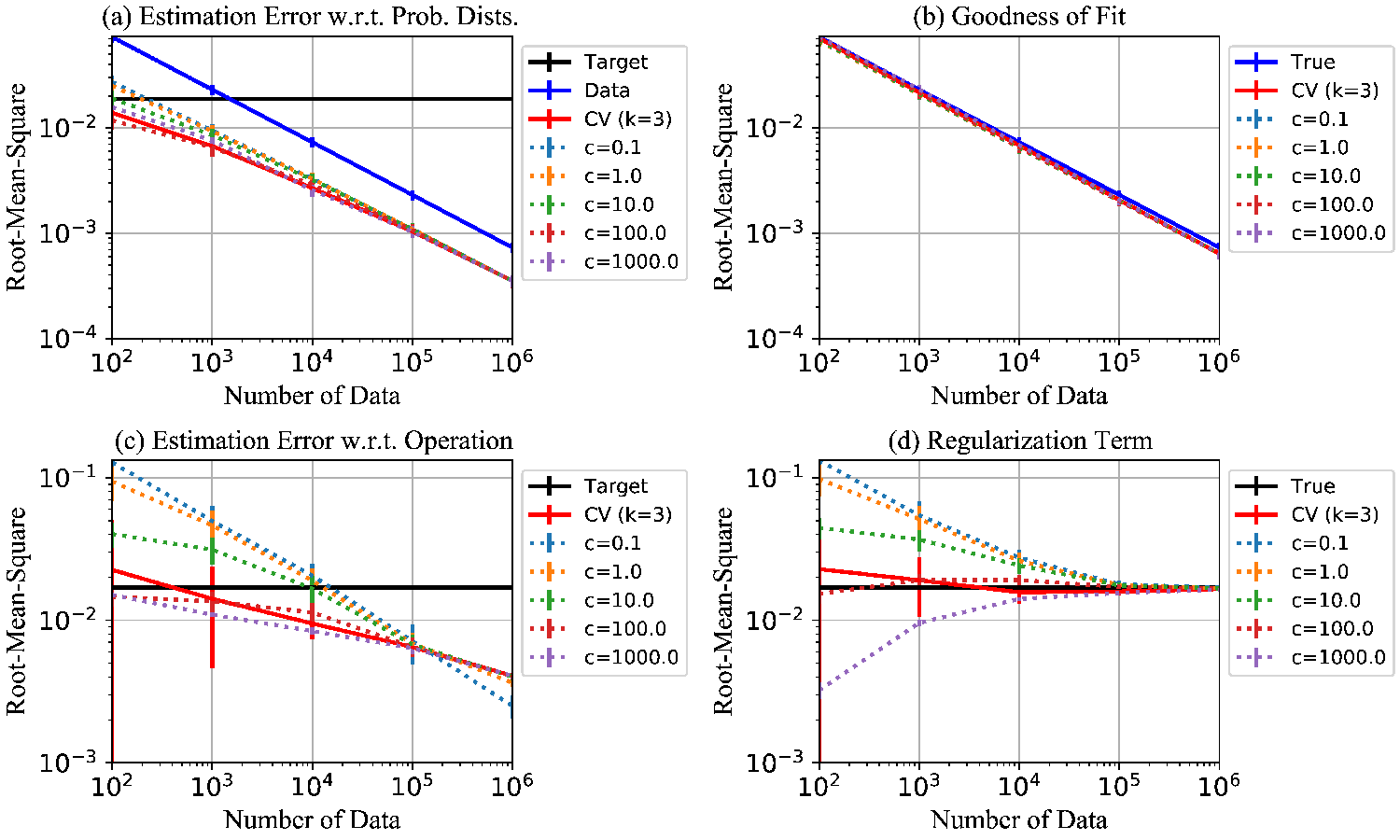}
      \caption{Plots of Root-Mean-Squares (RMS) of loss and regularization functions for the RSC estimator against number of data $N$ when the error model is a depolarizing. Panel (a) is for the estimation error in the space of the probability distributions. Panel (b) is for the goodness of fit to data. Panel (c) is for the estimation error in the space of quantum operations. Panel (d) is for the regularization term to the target set.}
      \label{fig:type0_sub_depolarizing}
   \end{center}
   \end{figure*}

\begin{figure*}[tb]
   \begin{center}
      \includegraphics[width=\linewidth]{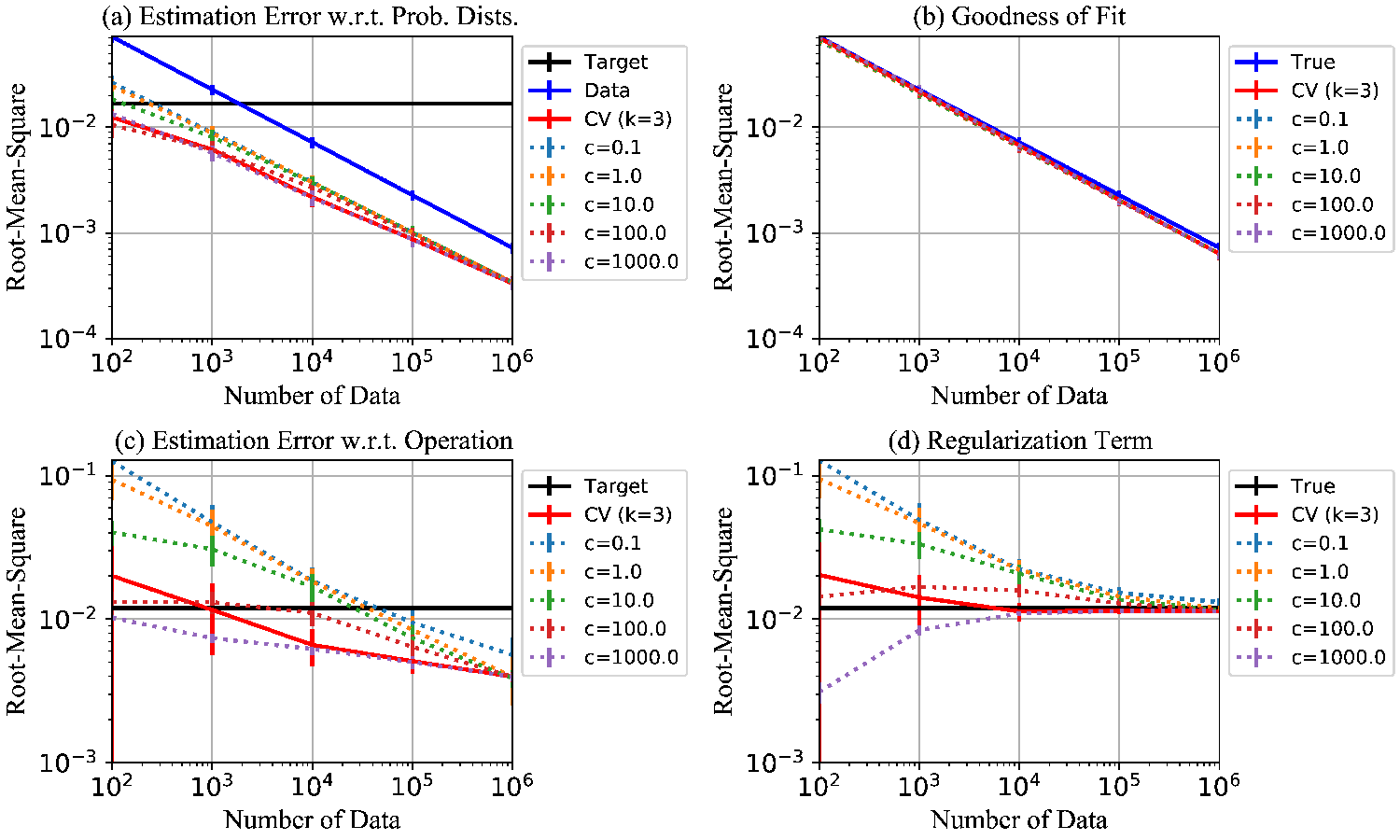}
      \caption{Plots of Root-Mean-Squares (RMS) of loss and regularization functions for the RSC estimator against number of data $N$ when the error model is an amplitude damping. Panel (a) is for the estimation error in the space of the probability distributions. Panel (b) is for the goodness of fit to data. Panel (c) is for the estimation error in the space of quantum operations. Panel (d) is for the regularization term to the target set.}
      \label{fig:type0_sub_amplitudedamping}
   \end{center}
   \end{figure*}

   Figures \ref{fig:type0_sub_depolarizing} and \ref{fig:type0_sub_amplitudedamping} are for the root-mean-squares of quantities related to the loss and regularization functions of the depolarizing and amplitude damping errors, respectively.
   The depolarizing error rate $p$ and the amplitude damping rate $\gamma$ are chosen to be $0.01$, which are common for all operations.
   These figures correspond to Fig.~\ref{fig:type0_main_lindblad01} in the main text.
   Detailed behaviors of each line changes along with the choice of the error model and parameter values.
   The existence of the gap between $\sest_N$ and $\bm{f}_N$ in Panels (a) of Figures \ref{fig:type0_sub_depolarizing} and \ref{fig:type0_sub_amplitudedamping} are common.
   Hence we consider that the RSC estimator's predictability of probability distributions higher than the empirical distributions is a feature independent of the details of the choice of error models and parameter values.
    
      \begin{table}[bt]
\begin{center}
\begin{tabular}{|c|c|c|c|c|}
\hline
   Gate & $\Delta \omega$ & $A \cdot W$ & $\phi$ & 1 $-$ AGF \\
\hline   
\hline
   $\mathcal{G}_{0}^{\mathrm{true}}$ & 0.01 & 0 & 0 &  $2.0 \times 10^{-3}$\\
\hline
   $\mathcal{G}_{1}^{\mathrm{true}}$ & 0.01 & $\pi/2 + 0.1$ & 0.15 & $9.8 \times 10^{-3}$ \\
\hline
   $\mathcal{G}_{2}^{\mathrm{true}}$ & 0.01 & $\pi /2 + 0.1$ & -0.20 &  $1.6 \times 10^{-2}$ \\
\hline
\end{tabular}
\caption{Coherent error parameters and average gate fidelity for $\mathcal{G}^{\mathrm{prepared}}$ in the additional numerical experiments}
\label{table:CoherentErrorAGF_sub}
\end{center}
\end{table}      

\begin{figure*}[tb]
   \begin{center}
      \includegraphics[width=\linewidth]{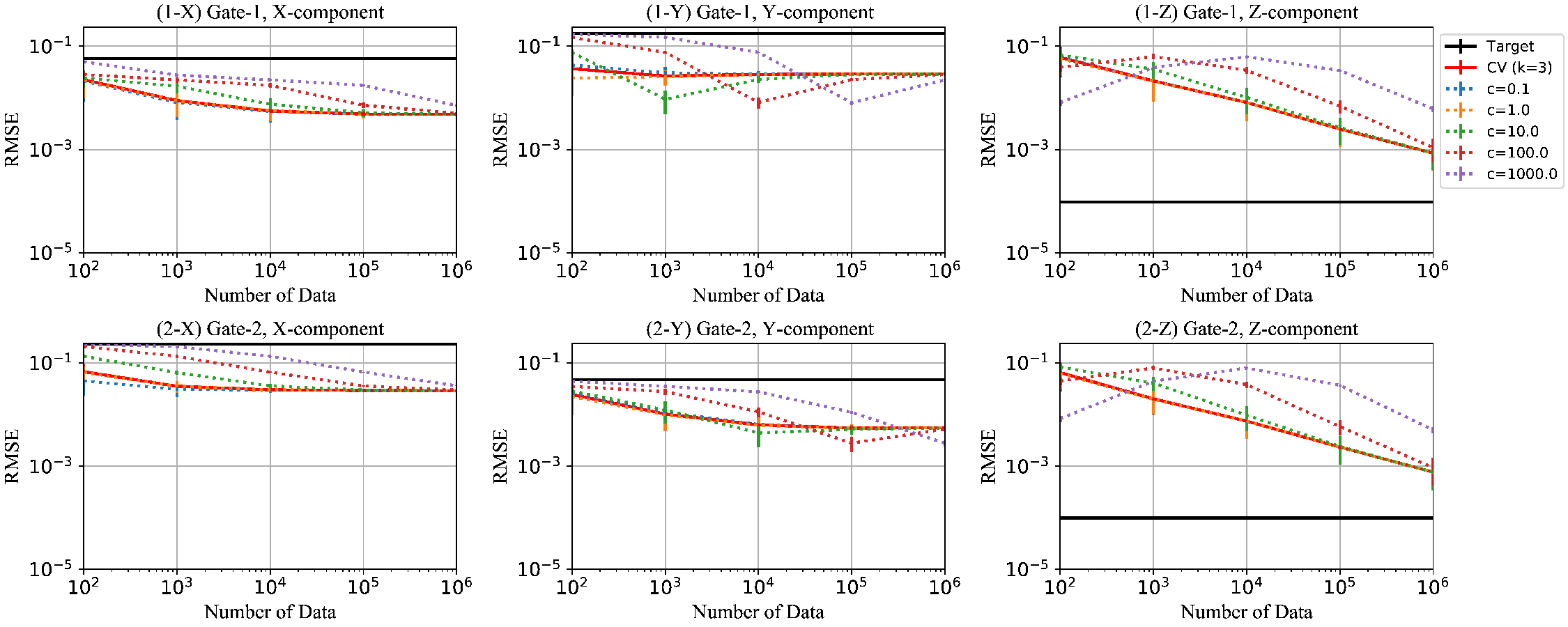}
      \caption{Root-Mean-Squared Error (RMSE) of the estimated Hamiltonian components against the number of data $N$ when the error is generated by the GKLS master equation with parameters different from Fig.~\ref{fig:type2_main_lindblad01}. The number and letter at each panel label corresponds to the gate number (0, 1, 2) and component. All of horizontal and vertical axes are log-scale.}
      \label{fig:type2_sub_lindblad01}
   \end{center}
   \end{figure*}

   Figure \ref{fig:type2_sub_lindblad01} is the result of estimation errors of Hamiltonian component at a case that we choose an error model obeying the GKLS master equation with different parameter values from Fig. \ref{fig:type2_main_lindblad01}.
   The parameter values are summarized at Table \ref{table:CoherentErrorAGF_sub}.
   This corresponds to Fig.~\ref{fig:type2_main_lindblad01} in the main text. 
   Compared to Fig.~\ref{fig:type2_main_lindblad01}, estimation errors of $Y$-component of gate-1 (Panel (1-Y)) and $X$-component of gate-2 (Panel (2-X)) are relatively small.
   The strength of the gauge generator, $ \|a \|$, of Fig. \ref{fig:type2_sub_lindblad01} is considered to be much smaller than that of Fig. \ref{fig:type2_main_lindblad01}. 

\end{document}